\documentclass[11pt]{article}
\usepackage{amsfonts}
\def \sech{\mathop{\mathrm sech}\nolimits}

\usepackage{amssymb,amsmath,mathrsfs,mathdots}
\usepackage{graphicx,subfigure}
\usepackage{indentfirst}
\usepackage{cite}

\begin{document}
\title{Nonintegrable Spatial Discrete Nonlocal Nonlinear Schr\"odinger Equation}
\author{Jia-Liang Ji$\dagger$, Zong-Wei Xu$\ddagger$ and Zuo-Nong Zhu$\ddagger$\footnote{Corresponding author. Email: znzhu@sjtu.edu.cn} \\
$\dagger$ School of Mathematics, Physics and Statistics,\\ Shanghai University of Engineering Science,\\ 333 Longteng Road, Shanghai, 201620, P. R. China\\
$\ddagger$ School of Mathematical Sciences, Shanghai Jiao Tong University,\\ 800 Dongchuan Road, Shanghai, 200240, P. R. China}
\date{}
\maketitle
\begin{abstract}
Integrable and nonintegrable discrete nonlinear Schr\"odinger equations (NLS) are significant models to describe many phenomena in physics. Recently, Ablowitz and Musslimani introduced a class of reverse space, reverse time and reverse space-time nonlocal integrable equations, including nonlocal NLS, nonlocal sine-Gordon equation and nonlocal Davey-Stewartson equation etc. And, the integrable nonlocal discrete NLS has been exactly solved by inverse scattering transform. In this paper, we study a nonintegrable discrete nonlocal NLS which is direct discretization version of the reverse space nonlocal NLS. By applying discrete Fourier transform and modified Neumann iteration, we present its stationary solutions numerically. The linear stability of the stationary solutions is examined. Finally, we study the Cauchy problem for nonlocal NLS equation numerically and find some different and new properties on the numerical solutions comparing with the numerical solutions of the Cauchy problem for NLS equation.
\end{abstract}

\textbf{The nonlinear Schr\"odinger equation (NLS) is an important integrable equation which has widely physical applications. Its spatial discrete version includes integrable and nonintegrable discrete NLS equations which are also significant models to describe many discrete phenomena in physics. Recently, a new integrable system---nonlocal NLS equation introduced by Ablowitz and Musslimani has attracted many researchers. The Cauchy problem of this equation is solved under special initial values by inverse scattering transform (IST). But for general initial values, the solutions of Cauchy problem of nonlocal NLS equation can not be obtained by using IST. In our study, we investigate a nonintegrable space discrete nonlocal NLS equation. We obtain its stationary solitary wave solutions by discrete Fourier transform. The linear stability of these stationary solitary wave solutions are examined. Then, numerical simulations for Cauchy problem of nonlocal NLS equation with periodic boundary conditions are performed, in which we use the integrable scheme and the nonintegrable scheme as a spatial discrete model of nonlocal NLS equation. We find some different and new properties on the numerical solutions for Cauchy problem of nonlocal NLS equation comparing with the numerical solutions of the Cauchy problem for the NLS equation.}
\section{Introduction}
It is well known that the nonlinear Schr\"{o}dinger equation(NLS)
\begin{equation}
iu_{t}+u_{xx}+u^{2}u^{*}=0,\label{eq:LCNLS}
\end{equation}
is one of the most significant integrable equations, which has been investigated extensely \cite{SS07139_,RRT85PD16D_339,RR86PS33_97,RR86PS33_498,AB98OL23_1668,APT04302_,Pel06N19_2695,Yan1016_}. Eq.(\ref{eq:LCNLS}) has various applications such as nonlinear optical fibers \cite{HT73APL23_142,HT73APL23_171}, magneto-static spin waves \cite{Zve83SPJ_}, deep water waves \cite{Per83JAMSSB25_16,BN67JMP46_133,Zak68JAMTP9_190,BR69SAM48_377}, plasma physics \cite{Zak72SPJ35_908,ZS72SPJ34_62,ZS73SPJ37_823}, and so on.
The spatial discrete version of the NLS equation (1) includes the following two models:
\begin{equation}
i{u}_{n,t}+\frac{1}{h^{2}}\left(u_{n+1}-2u_{n}+u_{n-1}\right)+\left|u_{n}\right|^{2}u_{n}=0,\label{eq:LNNLS}
\end{equation}
and
\begin{equation}
i{u}_{n,t}+\frac{1}{h^{2}}\left(u_{n+1}-2u_{n}+u_{n-1}\right)+\left|u_{n}\right|^{2}\frac{u_{n+1}+u_{n-1}}{2}=0\label{eq:LINLS}
\end{equation}
which are nonintegrable and integrable discrete NLS, respectively \cite{APT04302_,AL75JMP16_598,AL76JMP17_1011,CM83LMP7_17,HSGT95PRE52_255,KRB01IJMPB15_2833,HA89PRL62_2065}. Discrete NLS has also wide range of applications such as biomolecular chains \cite{KAT01PRL87_165501}, optic fibers \cite{FCS+03PRL90_23902,FSEC03N422_147,KA03_} etc. Ablowitz and Ladik obtained soliton solutions of Eq.(\ref{eq:LINLS}) through IST\cite{AL75JMP16_598,AL76JMP17_1011}.
Herbst and Ablowitz found numerically induced chaos for the Cauchy problem of NLS equation by using model (\ref{eq:LNNLS})  as a spatial discrete scheme \cite{HA89PRL62_2065}.
Ablowitz, Musslimani and Biondini investigated stationary and traveling-wave solitons of Eq.(\ref{eq:LNNLS}) \cite{AMB02PRE65_26602}.

Very recently, Ablowitz and Musslimani introduced and investigated the nonlocal NLS equation,
\begin{equation}
iu_{t}+u_{xx}+u^{2}u^{*}(-x,t)=0,\label{eq:NNLS}
\end{equation}
which is parity-time-symmetric(PT-symmetric) since it keeps invariant under the transform $x\to-x$, $t\to-t$, $u(x,t)\to u^{*}\left(x,t\right)$. They solved Eq.(\ref{eq:NNLS}) through IST\cite{AM13PRL110_64105,AM16N29_915,AM16SAM_}. Eq.(\ref{eq:NNLS}) has attracted many researchers' attention. Many results of Eq.(\ref{eq:NNLS}) were given in the literatures, such as interactions between dark and antidark soliton via the $N$-fold Darboux transform \cite{LX15PRE91_33202}, rational-soliton solutions \cite{LXM16JPSJ85_124001}, higher-order rational solitons \cite{WYY16C26_63123}, the dynamics of the peregrine rogue wave\cite{GS16CNSNS36_141}, the gauge equivalence between Eq.(\ref{eq:NNLS}) and the unconventional system of coupled Landau-Lifshitz equations \cite{MZ16JMP57_83507,GA16PRA93_62124}. Furthermore, Fokas investigated multi-dimensional version of Eq.\eqref{eq:NNLS} \cite{Fok16N29_319}.

Ablowitz and Musslimani studied the integrable discretization of Eq.(\ref{eq:NNLS}),
\begin{equation}
i{u}_{n,t}+\frac{1}{h^{2}}\left(u_{n+1}-2u_{n}+u_{n-1}\right)+u_{n}u_{-n}^{*}\frac{u_{n+1}+u_{n-1}}{2}=0,\label{eq:ddkeji}
\end{equation}
and solved it through IST\cite{AM14PRE90_32912}. In Ref.\cite{MZ16AML59_115}, $N$-soliton solutions of Eq.(\ref{eq:ddkeji}) are obtained by the Hirota bilinear method.

In this paper, our main purpose is to study the following equation,
\begin{equation}
i{u}_{n,t}+\frac{1}{h^{2}}\left(u_{n+1}-2u_{n}+u_{n-1}\right)+u_{n}^{2}u_{-n}^{*}=0.\label{eq:ddnoke}
\end{equation}
This is a nonintegrable spatial discrete version of Eq.(\ref{eq:NNLS}) and is also PT-symmetric. We will discuss  its stationary-solitary-wave solutions numerically and analyze their linear stability. In Ref.\cite{HA89PRL62_2065}, Herbst and Ablowitz investigated the following Cauchy problem
numerically,
\begin{equation}\left\{\begin{array}{l}
iq_t+q_{xx}+\gamma|q|^2q=0,\\
q(x,0)=q(x),
\end{array}\right.\end{equation}
with a periodic boundary condition, $q(x+L,t)=q(x,t)$. For an appropriate initial data, they found that the nonintegrable discrete scheme for NLS equation (7),
\begin{equation}
iq_{n,t}+\frac{q_{n+1}+q_{n-1}-2q_n}{h^2}+\gamma|q_n|^2q_n=0,
\end{equation}
produces chaotic solution for intermediate levels of mesh refinement. But chaos disappears when the discretization is fine enough, and the convergence to a quasiperiodic solution is obtained. Motivated by Herbst and Ablowitz's
work, we will study the following Cauchy problem,
\begin{equation}\begin{cases}
iu_{t}(x,t)+u_{xx}(x,t)+u^{2}(x,t)u^{*}\left(-x,t\right)=0,\\
u(x,0)=a\left[1+\epsilon\cos\left(\mu x+\varphi_{0}\right)\right], & -\dfrac{L}{2}\leq x\leq\dfrac{L}{2},\\
u\left(-\frac{L}{2},t\right)=u\left(\frac{L}{2},t\right),
\end{cases}\label{eq:CPNL}\end{equation}
where $a,\epsilon,\mu,\varphi_{0}$ are real parameters. It is impossible that this Cauchy problem could be solved by IST. We will use the spatial discrete schemes (\ref{eq:ddkeji}) and (\ref{eq:ddnoke}) to solve the Cauchy problem numerically, respectively. We find that the numerical solution of Cauchy problem \eqref{eq:CPNL} yields the blow-up phenomenon and the initial data has big influence on the blow-up time and position. This is quite different from the case of the Cauchy problem for NLS equation.
\section{Stationary solitary waves of nonintegrable spatial discrete nonlocal NLS equation}
In this section, we consider the discrete nonlocal NLS equation \eqref{eq:ddnoke} and seek for its localized soliton solutions as follows
\begin{equation}\label{b2}
u_n(t)=\phi(nh)e^{i\omega t}\equiv[F(nh)+iG(nh)]e^{i\omega t},
\end{equation}
with $F,G$ being real. It is obvious that Eq.\eqref{eq:ddnoke} becomes
\begin{equation}\label{b3}\begin{gathered}
\mathcal{D}F+(F^2-G^2)\cdot\mathcal{P}F+2FG\cdot\mathcal{P}G=\omega F,\\
\mathcal{D}G-(F^2-G^2)\cdot\mathcal{P}G+2FG\cdot\mathcal{P}F=\omega G,
\end{gathered}\end{equation}
where
\begin{equation}\label{b4}\begin{gathered}
\mathcal{D}X(nh)=\dfrac{1}{h^2}(E_++E_--2)X(nh),\\ \mathcal{P}X(nh)=X(-nh),
\end{gathered}\end{equation}
with $E_\pm X(nh)=X(nh\pm h)$. System \eqref{b3} is a differential advance-delay system, which is difficult to solve. To solve \eqref{b3}, we decompose $F,G$ into a sum of an even function and an odd function,
\begin{equation*}
F=F_{\textrm{e}}+F_{\textrm{o}},\qquad G=G_{\textrm{e}}+G_{\textrm{o}}.
\end{equation*}
So, system \eqref{b3} has the form,
\begin{equation}\label{b5}\begin{gathered}
\mathcal{D}(F_{\textrm{e}}+F_{\textrm{o}})+[(F_{\textrm{e}}+F_{\textrm{o}})^2-(G_{\textrm{e}}+G_{\textrm{o}})^2](F_{\textrm{e}}-F_{\textrm{o}}) +2(F_{\textrm{e}}+F_{\textrm{o}})(G_{\textrm{e}}^2-G_{\textrm{o}}^2)=\omega (F_{\textrm{e}}+F_{\textrm{o}}),\\
\mathcal{D}(G_{\textrm{e}}+G_{\textrm{o}})-[(F_{\textrm{e}}+F_{\textrm{o}})^2-(G_{\textrm{e}}+G_{\textrm{o}})^2](G_{\textrm{e}}-G_{\textrm{o}}) +2(F_{\textrm{e}}^2-F_{\textrm{o}}^2)(G_{\textrm{e}}+G_{\textrm{o}})=\omega (G_{\textrm{e}}+G_{\textrm{o}}).
\end{gathered}\end{equation}
Then, applying discrete Fourier transform on \eqref{b5}, which is defined as
\begin{gather}
\hat{F}(q)=\underset{m=-\infty}{\overset{+\infty}{\sum}}F(mh)e^{-iqmh},\label{b6}\\ F(nh)=\dfrac{h}{2\pi}\int_{-\pi/h}^{\pi/h}\hat{F}(q)e^{iqnh}\textrm{d}q,\label{b11}
\end{gather}
and separating the real and imaginary parts, we get
\begin{equation}\label{b7}\begin{aligned}
\hat{F}_{\textrm{e}}(q) & =\dfrac{1}{\Omega(q)}\{Q_1[\hat{F}_{\textrm{e}},\tilde{F}_{\textrm{o}},\hat{G}_{\textrm{e}},\tilde{G}_{\textrm{o}}](q)\ast\hat{F}_{\textrm{e}}(q) +2Q_2[\hat{F}_{\textrm{e}},\tilde{F}_{\textrm{o}},\hat{G}_{\textrm{e}},\tilde{G}_{\textrm{o}}](q)\ast\tilde{F}_{\textrm{o}}(q)\\ & \qquad +2Q_3[\hat{G}_{\textrm{e}},\tilde{G}_{\textrm{o}}](q)\ast\hat{F}_{\textrm{e}}(q)\}\\ & \equiv\mathcal{K}_1[\omega;\hat{F}_{\textrm{e}}(q),\tilde{F}_{\textrm{o}}(q),\hat{G}_{\textrm{e}}(q),\tilde{G}_{\textrm{o}}(q)],\\
\tilde{F}_{\textrm{o}}(q) & =\dfrac{1}{\Omega(q)}\{-Q_1[\hat{F}_{\textrm{e}},\tilde{F}_{\textrm{o}},\hat{G}_{\textrm{e}},\tilde{G}_{\textrm{o}}](q)\ast\tilde{F}_{\textrm{o}}(q) +2Q_2[\hat{F}_{\textrm{e}},\tilde{F}_{\textrm{o}},\hat{G}_{\textrm{e}},\tilde{G}_{\textrm{o}}](q)\ast\hat{F}_{\textrm{e}}(q)\\ & \qquad +2Q_3[\hat{G}_{\textrm{e}},\tilde{G}_{\textrm{o}}](q)\ast\tilde{F}_{\textrm{o}}(q)\}\\ & \equiv\mathcal{K}_2[\omega;\hat{F}_{\textrm{e}}(q),\tilde{F}_{\textrm{o}}(q),\hat{G}_{\textrm{e}}(q),\tilde{G}_{\textrm{o}}(q)],\\
\hat{G}_{\textrm{e}}(q) & =\dfrac{1}{\Omega(q)}\{-Q_1[\hat{F}_{\textrm{e}},\tilde{F}_{\textrm{o}},\hat{G}_{\textrm{e}},\tilde{G}_{\textrm{o}}](q)\ast\hat{G}_{\textrm{e}}(q) -2Q_2[\hat{F}_{\textrm{e}},\tilde{F}_{\textrm{o}},\hat{G}_{\textrm{e}},\tilde{G}_{\textrm{o}}](q)\ast\tilde{G}_{\textrm{o}}(q)\\ & \qquad +2Q_4[\hat{F}_{\textrm{e}},\tilde{F}_{\textrm{o}}](q)\ast\hat{G}_{\textrm{e}}(q)\}\\ & \equiv\mathcal{K}_3[\omega;\hat{F}_{\textrm{e}}(q),\tilde{F}_{\textrm{o}}(q),\hat{G}_{\textrm{e}}(q),\tilde{G}_{\textrm{o}}(q)],\\
\tilde{G}_{\textrm{o}}(q) & =\dfrac{1}{\Omega(q)}\{Q_1[\hat{F}_{\textrm{e}},\tilde{F}_{\textrm{o}},\hat{G}_{\textrm{e}},\tilde{G}_{\textrm{o}}](q)\ast\tilde{G}_{\textrm{o}}(q) -2Q_2[\hat{F}_{\textrm{e}},\tilde{F}_{\textrm{o}},\hat{G}_{\textrm{e}},\tilde{G}_{\textrm{o}}](q)\ast\hat{G}_{\textrm{e}}(q)\\ & \qquad +2Q_4[\hat{F}_{\textrm{e}},\tilde{F}_{\textrm{o}}](q)\ast\tilde{G}_{\textrm{o}}(q)\}\\ & \equiv\mathcal{K}_4[\omega;\hat{F}_{\textrm{e}}(q),\tilde{F}_{\textrm{o}}(q),\hat{G}_{\textrm{e}}(q),\tilde{G}_{\textrm{o}}(q)],
\end{aligned}\end{equation}
where $\tilde{F}_{\textrm{o}}(q)=i\hat{F}_{\textrm{o}}(q)$, $\tilde{G}_{\textrm{o}}(q)=i\hat{G}_{\textrm{o}}(q)$, and
\begin{equation}\label{b8}\begin{aligned}
& \Omega(q)=\omega+\dfrac{2}{h^2}[1-\cos(hq)],\\
& Q_1[\hat{F}_{\textrm{e}},\tilde{F}_{\textrm{o}},\hat{G}_{\textrm{e}},\tilde{G}_{\textrm{o}}](q)=\dfrac{h^2}{4\pi^2} (\hat{F}_{\textrm{e}}\ast\hat{F}_{\textrm{e}}-\tilde{F}_{\textrm{o}}\ast\tilde{F}_{\textrm{o}} -\hat{G}_{\textrm{e}}\ast\hat{G}_{\textrm{e}}+\tilde{G}_{\textrm{o}}\ast\tilde{G}_{\textrm{o}})(q),\\
& Q_2[\hat{F}_{\textrm{e}},\tilde{F}_{\textrm{o}},\hat{G}_{\textrm{e}},\tilde{G}_{\textrm{o}}](q)=\dfrac{h^2}{4\pi^2} (\hat{F}_{\textrm{e}}\ast\tilde{F}_{\textrm{o}}-\hat{G}_{\textrm{e}}\ast\tilde{G}_{\textrm{o}})(q),\\
& Q_3[\hat{G}_{\textrm{e}},\tilde{G}_{\textrm{o}}](q)=\dfrac{h^2}{4\pi^2}(\hat{G}_{\textrm{e}}\ast\hat{G}_{\textrm{e}} +\tilde{G}_{\textrm{o}}\ast\tilde{G}_{\textrm{o}})(q),\\
& Q_4[\hat{F}_{\textrm{e}},\tilde{F}_{\textrm{o}}](q)=\dfrac{h^2}{4\pi^2}(\hat{F}_{\textrm{e}}\ast\hat{F}_{\textrm{e}} +\tilde{F}_{\textrm{o}}\ast\tilde{F}_{\textrm{o}})(q),
\end{aligned}\end{equation} with
\begin{equation*}
(\hat{L}\ast\hat{M}\ast\hat{N})(q)=\iint\hat{L}(q_1)\hat{M}(q_2)\hat{N}(q-q_1-q_2)~\textrm{d}q_1\textrm{d}q_2.
\end{equation*}
Through modified Neumann iterative algorithms\cite{Pet76SJPP2_257,AB98OL23_1668,Yan1016_}, we construct the following iteration schemes,
\begin{equation}\label{b9}\begin{aligned}
\hat{F}_{\textrm{e},n+1}(q) & =\left|\dfrac{\alpha} {\beta}\right|^{3/2} \mathcal{K}_1[\omega;\hat{F}_{\textrm{e},n}(q),\tilde{F}_{\textrm{o},n}(q),\hat{G}_{\textrm{e},n}(q),\tilde{G}_{\textrm{o},n}(q)],\\ \tilde{F}_{\textrm{o},n+1}(q) & =\left|\dfrac{\alpha} {\beta}\right|^{3/2} \mathcal{K}_2[\omega;\hat{F}_{\textrm{e},n}(q),\tilde{F}_{\textrm{o},n}(q),\hat{G}_{\textrm{e},n}(q),\tilde{G}_{\textrm{o},n}(q)],\\ \hat{G}_{\textrm{e},n+1}(q) & =\left|\dfrac{\alpha} {\beta}\right|^{3/2} \mathcal{K}_3[\omega;\hat{F}_{\textrm{e},n}(q),\tilde{F}_{\textrm{o},n}(q),\hat{G}_{\textrm{e},n}(q),\tilde{G}_{\textrm{o},n}(q)],\\ \tilde{G}_{\textrm{o},n+1}(q) & =\left|\dfrac{\alpha} {\beta}\right|^{3/2} \mathcal{K}_4[\omega;\hat{F}_{\textrm{e},n}(q),\tilde{F}_{\textrm{o},n}(q),\hat{G}_{\textrm{e},n}(q),\tilde{G}_{\textrm{o},n}(q)],
\end{aligned}\end{equation}
where $\tfrac\alpha\beta$ is a stabilizing factor to ensure the convergence of the iteration scheme,
\begin{equation}\label{b10}\begin{aligned}
\alpha & =\int\left(\hat{F}_{\textrm{e},n}^2(q)+\tilde{F}_{\textrm{o},n}^2(q)+\hat{G}_{\textrm{e},n}^2(q)+\tilde{G}_{\textrm{o},n}^2(q)\right)\textrm{d}q,
\\ \beta & =\int\left(\hat{F}_{\textrm{e},n}(q) \mathcal{K}_1[\omega;\hat{F}_{\textrm{e},n}(q),\tilde{F}_{\textrm{o},n}(q),\hat{G}_{\textrm{e},n}(q),\tilde{G}_{\textrm{o},n}(q)]\right.\\ & \qquad+\tilde{F}_{\textrm{o},n}(q) \mathcal{K}_2[\omega;\hat{F}_{\textrm{e},n}(q),\tilde{F}_{\textrm{o},n}(q),\hat{G}_{\textrm{e},n}(q),\tilde{G}_{\textrm{o},n}(q)]\\ & \qquad+\hat{G}_{\textrm{e},n}(q) \mathcal{K}_3[\omega;\hat{F}_{\textrm{e},n}(q),\tilde{F}_{\textrm{o},n}(q),\hat{G}_{\textrm{e},n}(q),\tilde{G}_{\textrm{o},n}(q)]\\ & \qquad\left.+\tilde{G}_{\textrm{o},n}(q) \mathcal{K}_4[\omega;\hat{F}_{\textrm{e},n}(q),\tilde{F}_{\textrm{o},n}(q),\hat{G}_{\textrm{e},n}(q),\tilde{G}_{\textrm{o},n}(q)]\right)\textrm{d}q.
\end{aligned}\end{equation}
Numerical simulation shows that under appropriate values of the parameters, when $n\rightarrow\infty$,
\(\left|\dfrac{\alpha} {\beta}\right|\rightarrow 1,\)
and all function sequences $\{\hat{F}_{\textrm{e},n}(q)\}$, $\{\tilde{F}_{\textrm{o},n}(q)\}$, $\{\hat{G}_{\textrm{e},n}(q)\}$ and $\{\tilde{G}_{\textrm{o},n}(q)\}$ are convergent. Denote
\begin{equation*}\begin{array}{ll}
\hat{F}_{\textrm{e},n}(q)\rightarrow\hat{F}_{\textrm{e},s}(q), & \tilde{F}_{\textrm{o},n}(q)\rightarrow\tilde{F}_{\textrm{o},s}(q),\\ \hat{G}_{\textrm{e},n}(q)\rightarrow\hat{G}_{\textrm{e},s}(q), & \tilde{G}_{\textrm{o},n}(q)\rightarrow\tilde{G}_{\textrm{o},s}(q), \end{array} \qquad n\rightarrow\infty.
\end{equation*}
Thus, $\{\hat{F}_{\textrm{e},s}(q),\tilde{F}_{\textrm{o},s}(q),\hat{G}_{\textrm{e},s}(q),\tilde{G}_{\textrm{o},s}(q)\}$ is a fixed point of nonlinear integral system \eqref{b9}. Applying inverse Fourier transform \eqref{b11}, we get the approximate solution of discrete nonlocal NLS equation \eqref{eq:ddnoke}.

\begin{figure}[htbp]
\centering
\subfigure[]{\includegraphics[width=0.25\textwidth]{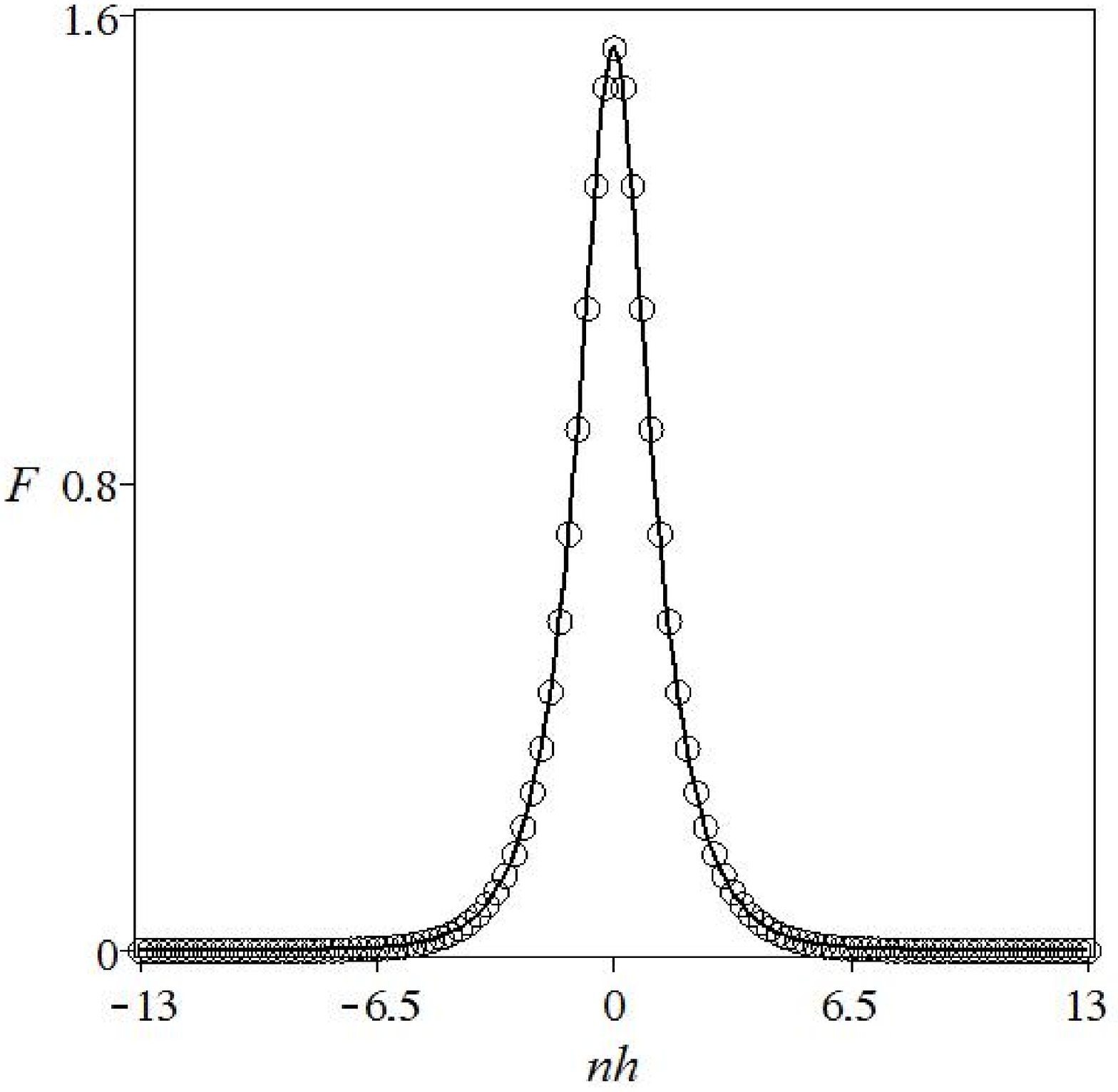}}\qquad
\subfigure[]{\includegraphics[width=0.25\textwidth]{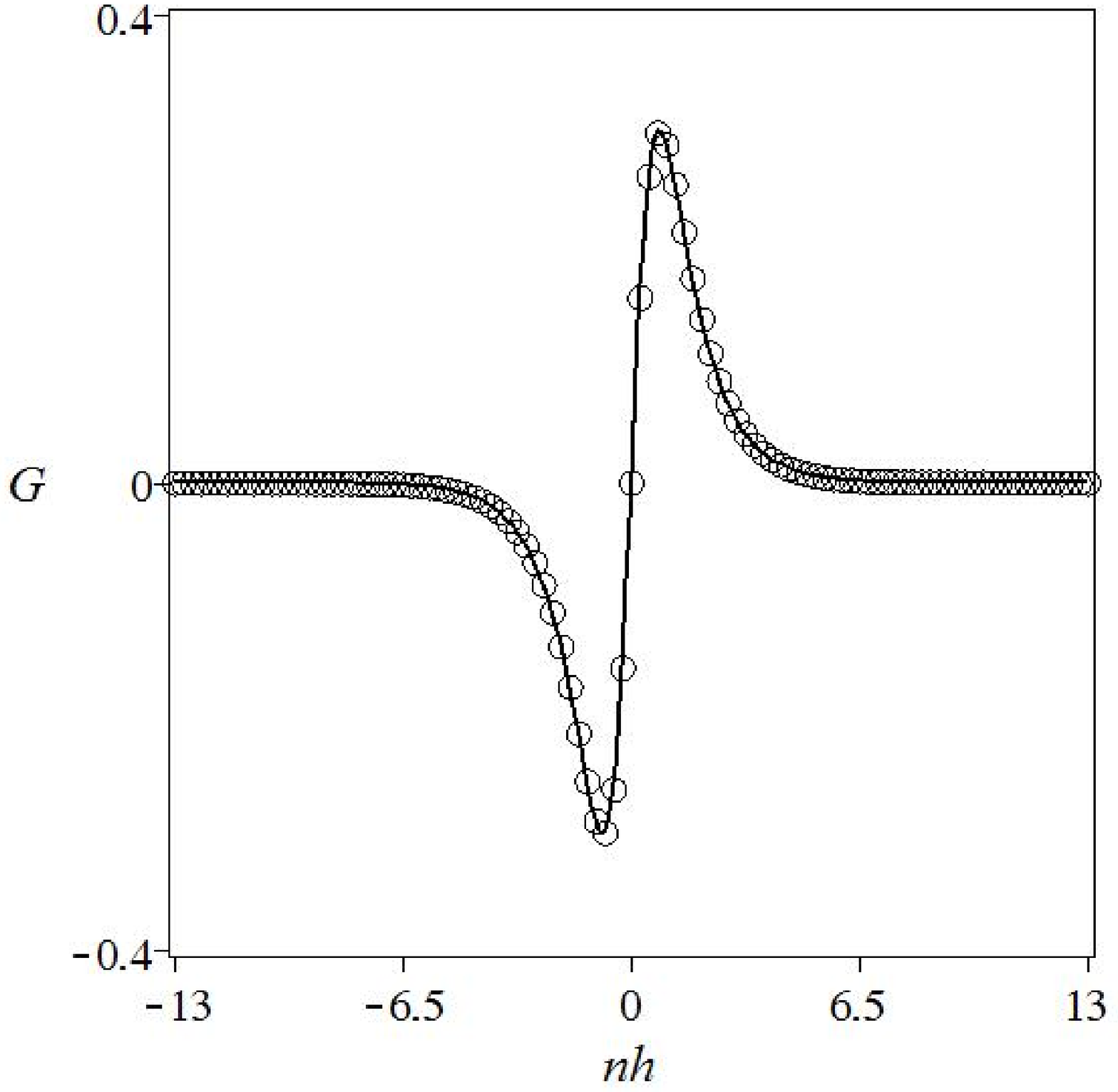}}\qquad
\subfigure[]{\includegraphics[width=0.25\textwidth]{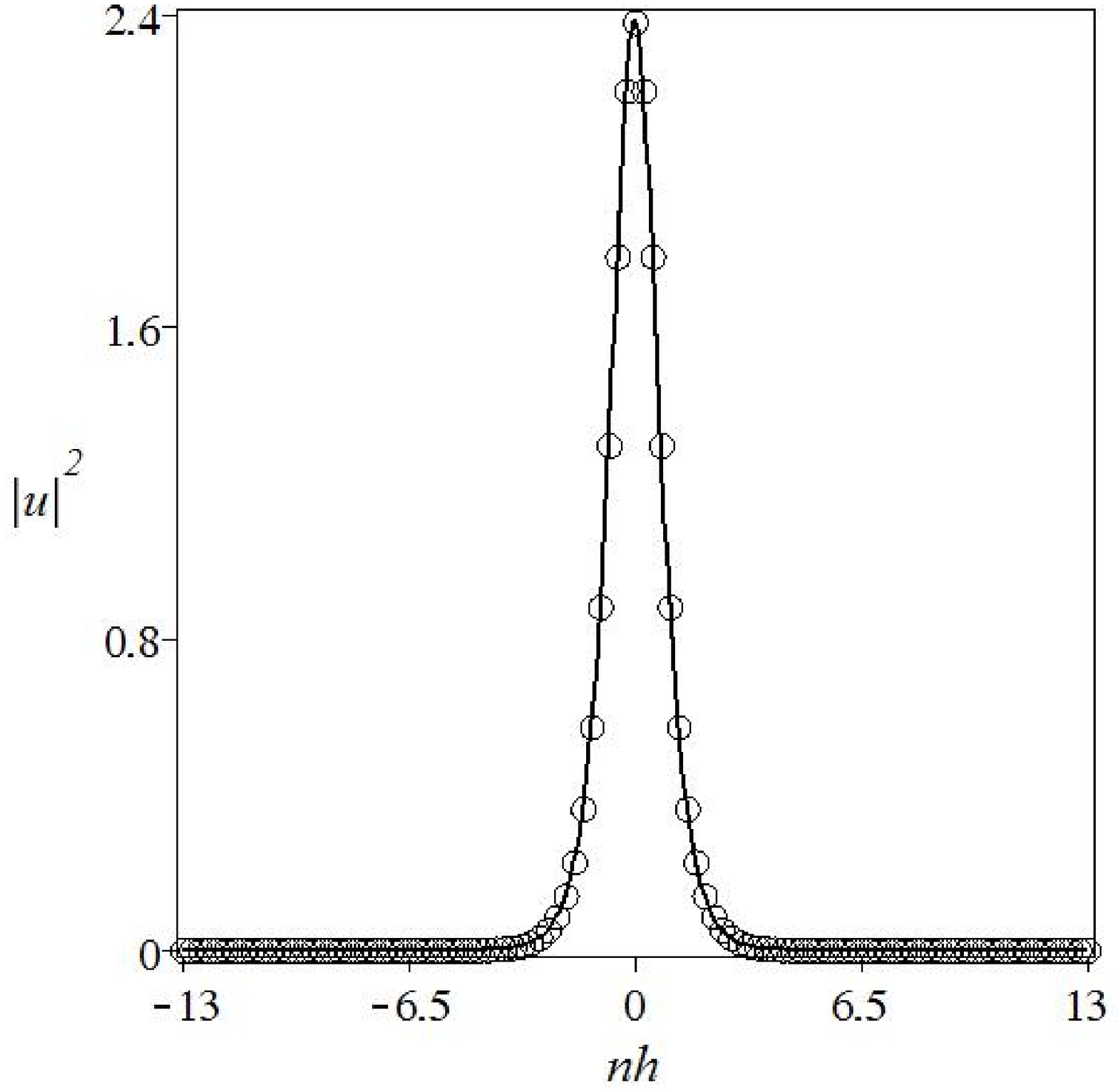}}
\caption{\small{Modes of $u_n(t)$ with $h=0.25$ and $\omega=1$, which initial data is given by \eqref{b12}. All these modes are depicted in physical space.}}\label{fg2-1}
\end{figure}

\begin{figure}[htbp]
\centering
\subfigure[]{\includegraphics[width=0.25\textwidth]{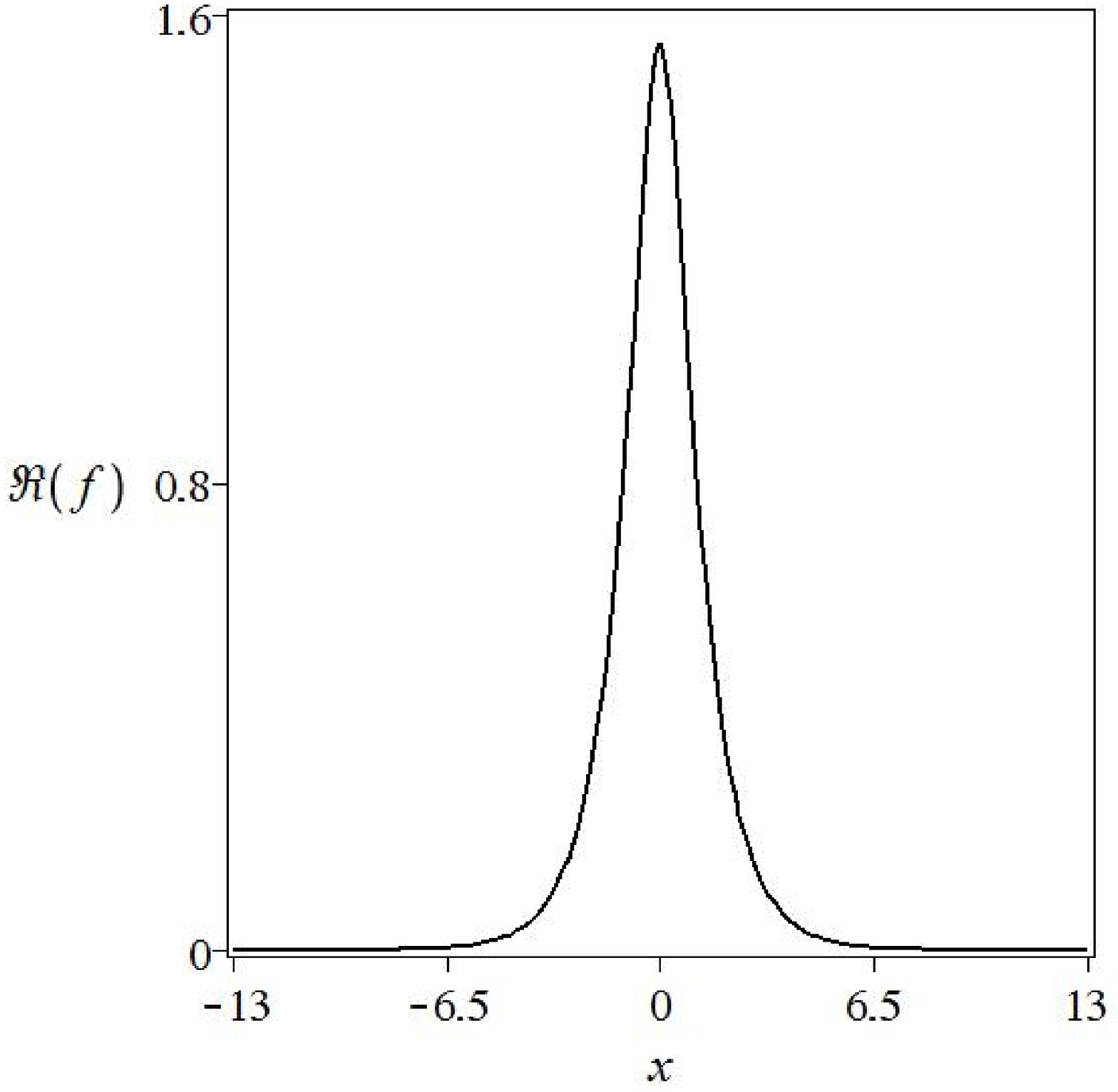}}\qquad
\subfigure[]{\includegraphics[width=0.25\textwidth]{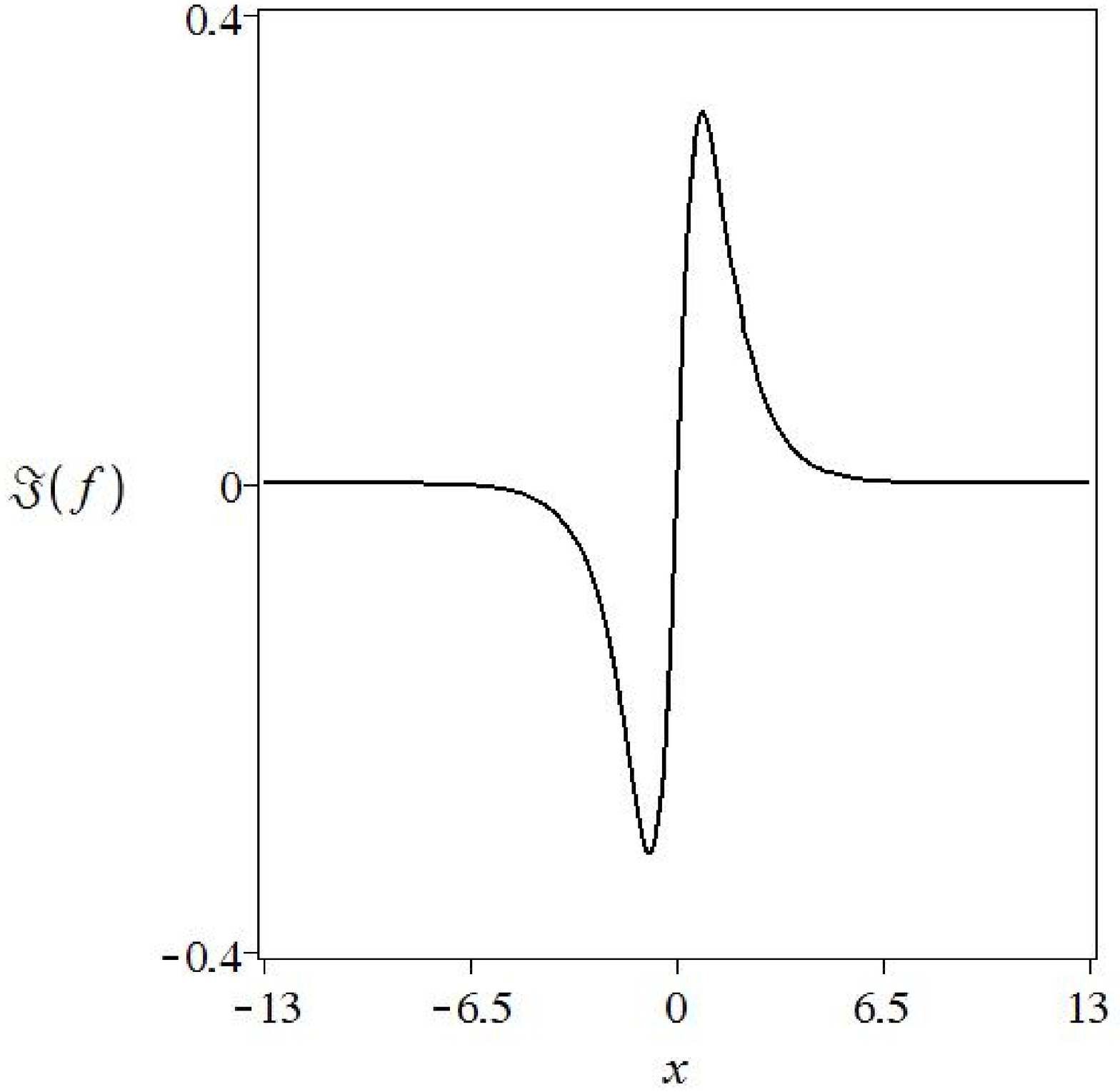}}\qquad
\subfigure[]{\includegraphics[width=0.25\textwidth]{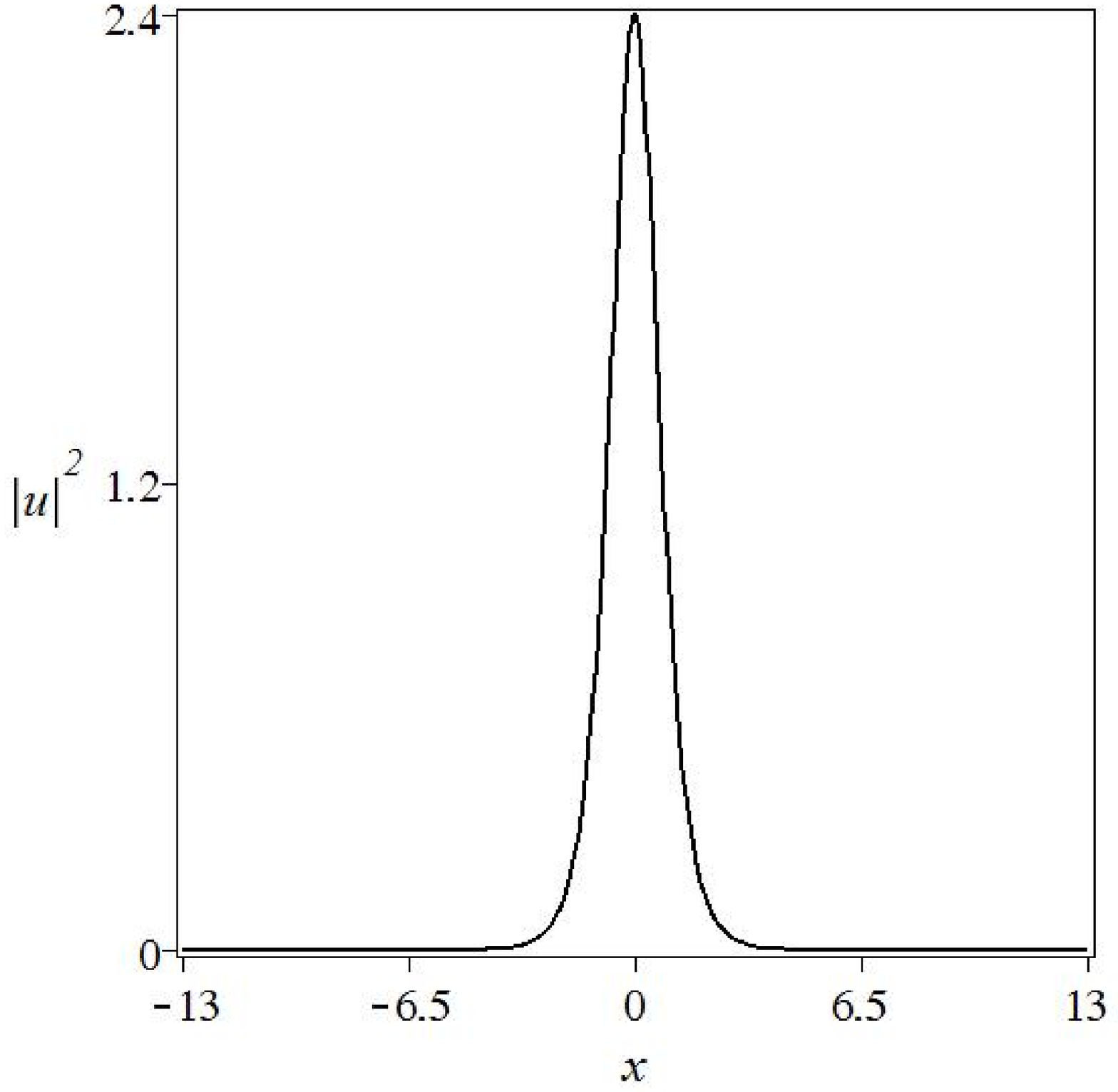}}
\caption{\small{Stationary solution \eqref{4.b13} to Eq.\eqref{4.a4}, where the parameters are $a=1/\sqrt{2}$, $\theta_1=3\pi/2$ and $\theta_2=-\arccos{\sqrt{\frac{5}{6}}}$.}}\label{4.fg2-2}
\end{figure}

Let us give an example. Set $h=0.25$, $\omega=1$ and the following initial data:
\begin{equation}\label{b12}
\hat{F}_{\textrm{e},0}(q)=\sech(q),\quad\tilde{F}_{\textrm{o},0}(q)=0,\quad\hat{G}_{\textrm{e},0}(q)=0,\quad\tilde{G}_{\textrm{o},0}(q)=\sin(q)\sech(q).
\end{equation}
The result of the numerical simulation is given by the following table:
\begin{center}\begin{tabular}{|c|c|c|c|}\hline
$n$ & $|\alpha/\beta|$ & $\||\hat{F}_{\textrm{e},n}|-|\hat{F}_{\textrm{e},s}|\|_{L^2}$ & $\||\tilde{G}_{\textrm{o},n}|-|\tilde{G}_{\textrm{o},s}|\|_{L^2}$ \\ \hline
8 & 1.001518064 & 0.08597422926 & 0.04659383926 \\ \hline
9 & 1.000796248 & 0.04361879844 & 0.02356541023 \\ \hline
10 & 1.000411536 & 0.02200440477 & 0.01187203961 \\ \hline
11 & 1.000210991 & 0.01099636203 & 0.005930276334 \\ \hline
12 & 1.000107681 & 0.005397427744 & 0.002910885638 \\ \hline
13 & 1.000054805 & 0.002552027827 & 0.001376742451 \\ \hline
14 & 1.000027842 & 0.001106701695 & 0.0005973233238 \\ \hline
15 & 1.000014119 & 0.0003727100677 & 0.0002013074880 \\ \hline
16 & 1.000007149 & / & / \\ \hline
\end{tabular}\end{center}
\begin{center}{\small {Table: the errors under $L^2$-norm between the approximate solution $u_n$ and solution $u_s$}}\end{center}
Here, we take $\hat{F}_{\textrm{e},s}(q)=\hat{F}_{\textrm{e},16}(q)$ and $\tilde{G}_{\textrm{o},s}(q)=\tilde{G}_{\textrm{o},16}(q)$. The errors under $L^2$-norm are described in the table. The modes of the solution $u_n$ is shown in Fig.\ref{fg2-1}. Let us consider the stationary solutions to nonlocal NLS equation,
\begin{equation}\label{4.a4}
iu_{t}(x,t)+u_{xx}(x,t)+u^{2}(x,t)u^\ast(-x,t)=0.
\end{equation}
Note that the following solution of Eq.\eqref{4.a4} is obtained in Ref.\cite{plus1}:
\begin{equation}\label{4.b13}
u(x,t)=2iae^{2ia^2t+i\theta_1}\sech(\sqrt{2}ax+i\theta_2)\triangleq f(x)e^{2ia^2t},
\end{equation}
wherw $a,\theta_1,\theta_2$ are arbitrary real numbers. From Fig.\ref{4.fg2-2}, we can see that the numerical solution $u_n$ obtained from the initial values \eqref{b12} fits this stationary solution $u$ very well.

\section{Linear stability of stationary solitary waves\label{sec:Linear-stability-of}}
\begin{figure*}[htbp]
\centering
\subfigure[]{\includegraphics[width=0.25\textwidth]{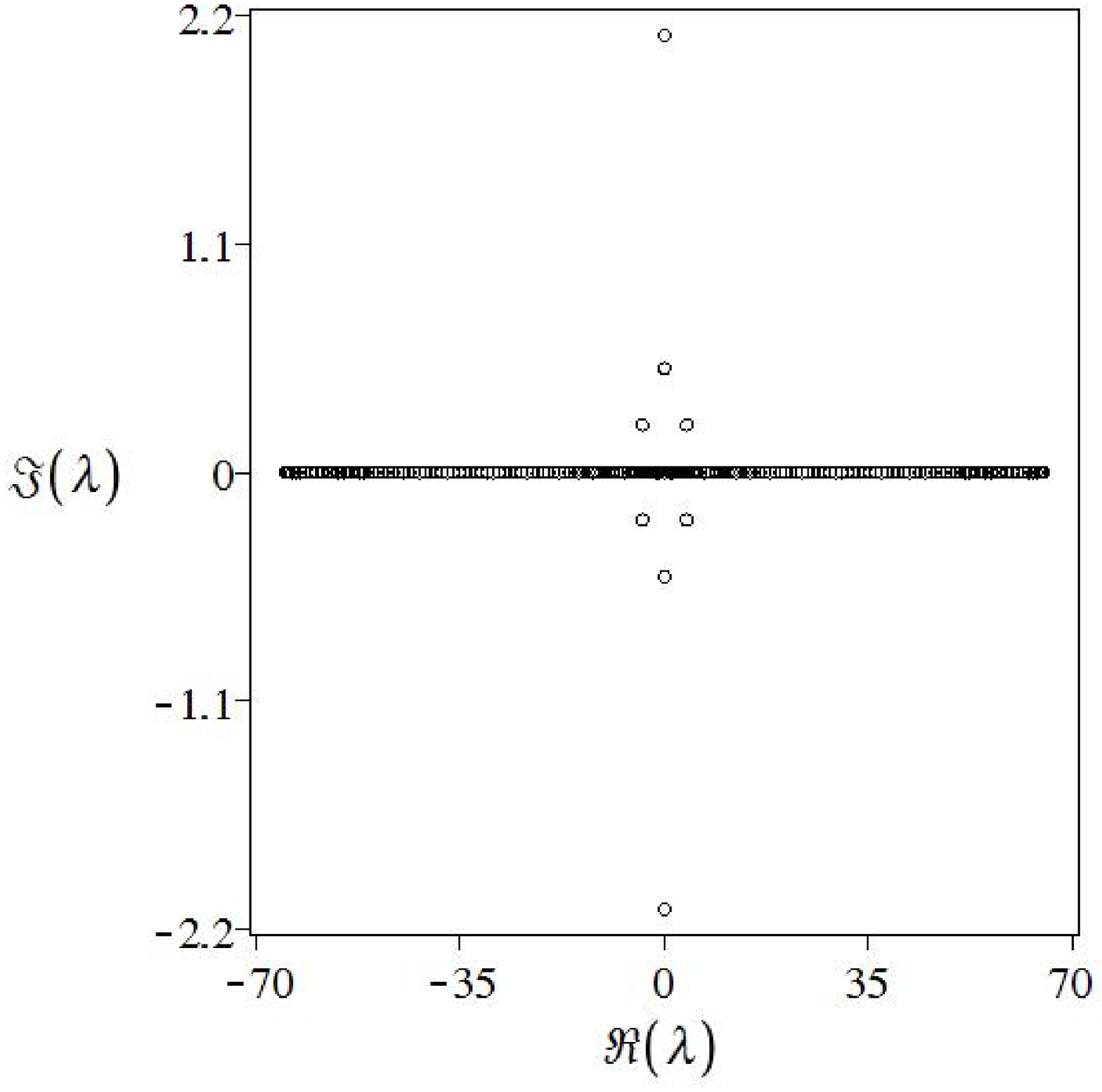
}}\qquad
\subfigure[]{\includegraphics[width=0.25\textwidth]{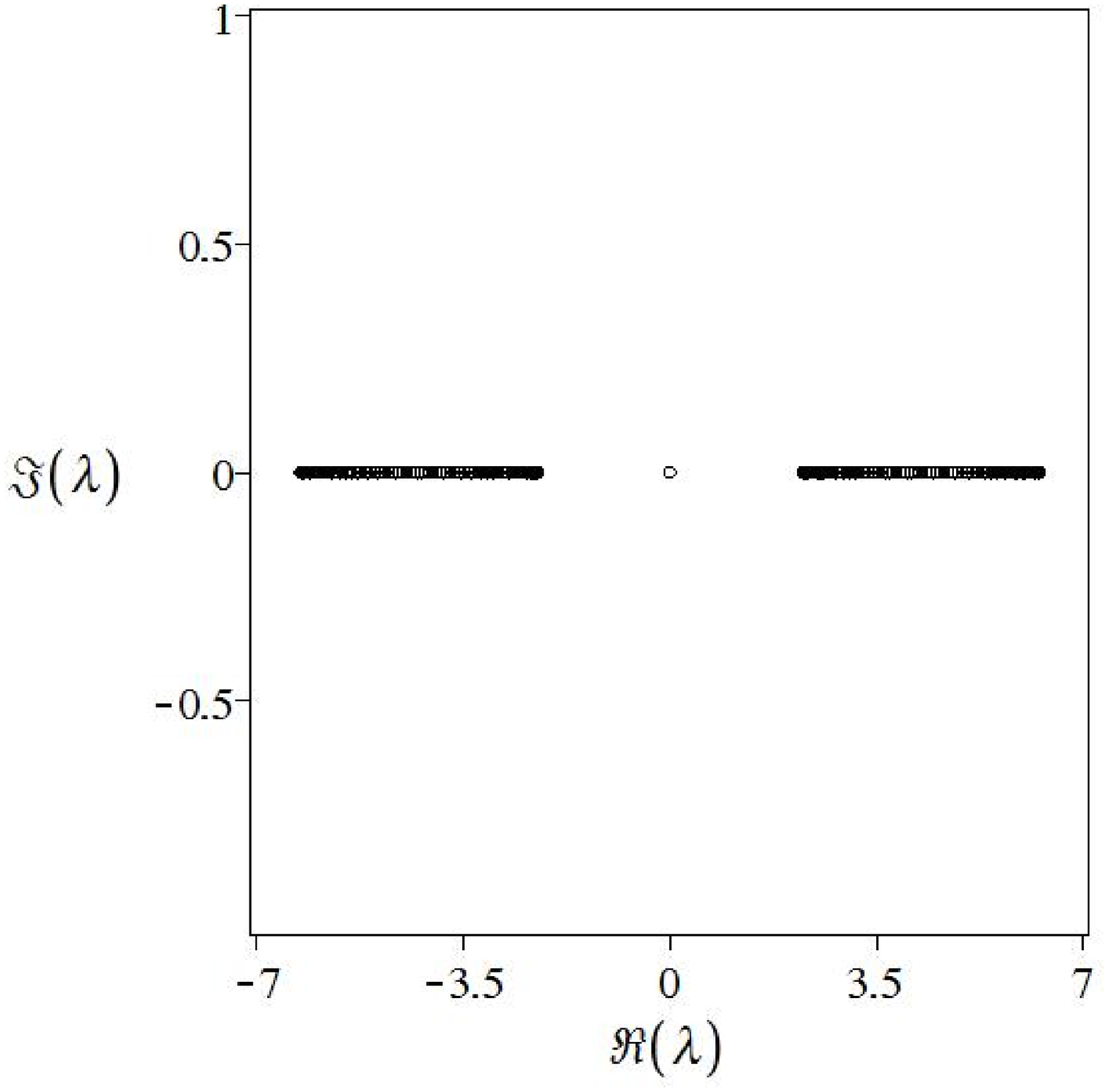}}\qquad
\subfigure[]{\includegraphics[width=0.25\textwidth]{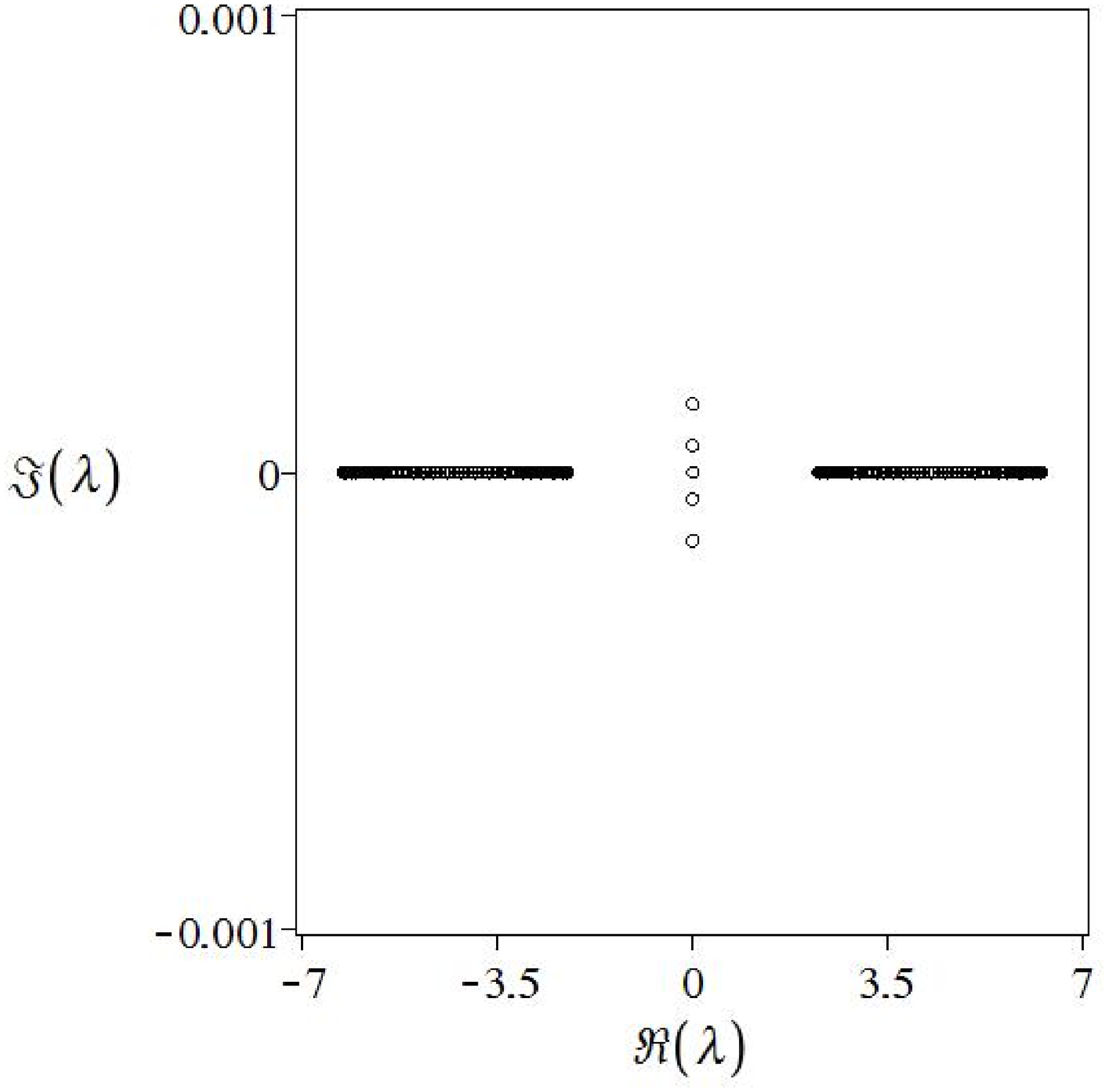}}
\caption{\small{(a) Eigenvalues of stability problem \eqref{c6} of stationary solution to nonlocal NLS \eqref{eq:ddnoke}, which initial data is given by \eqref{b12}; (b) Eigenvalues for $Q_n^{[1],s}$, as $z=0.5$, $\varphi_1=0$ and $\varphi_2=0$; (c) Eigenvalues for $Q_n^{[1],s}$, as $z=0.5$, $\varphi_1=0$ and $\varphi_2=\frac{3}{4}\pi$.}}\label{fg3-1}
\end{figure*}

In this section, we will numerically determine the linear stability of stationary solutions of discrete nonlocal NLS equation \eqref{eq:ddnoke}. Assume that $u_n(t)$ is a localized solution to Eq.\eqref{eq:ddnoke}  and has the form as \eqref{b2}. Thus, $\phi_n$ holds for a difference equation:
\begin{equation}\label{c1}
-\omega\phi_n+\dfrac{1}{h^2}(\phi_{n+1}-2\phi_n+\phi_{n-1})+\phi_n^2\phi_{-n}^\ast=0.
\end{equation}
Give a small perturbation to $u_n$ as follows
\begin{equation}\label{c2}
u_n(t)=[\phi_n+v_n(t)]e^{i\omega t},
\end{equation}
where $|v_n|\ll|\phi_n|$. Substituting \eqref{c2} into \eqref{eq:ddnoke} and neglecting the nonlinear terms of $v_n(t)$ yields a linear equation with respective to $v_n(t)$,
\begin{equation}\label{c3}
i{v}_{n,t}-\omega v_n+\dfrac{1}{h^2}(v_{n+1}-2v_n+v_{n-1})+\phi_n^2v_{-n}^\ast+2\phi_n\phi_{-n}^\ast v_n=0.
\end{equation}
Assume that the perturbation $v_n(t)$ can be expressed as
\begin{equation}\label{c4}
v_n(t)=a_ne^{-i\lambda t}+b_n^\ast e^{i\lambda^\ast t}.
\end{equation}
Then, Eq.\eqref{c3} becomes to the linear eigenvalue equations
\begin{equation}\label{c5}\begin{aligned}
\lambda a_n-\omega a_n+\dfrac{1}{h^2}(a_{n+1}-2a_n+a_{n-1})+\phi_n^2b_{-n}+2\phi_n\phi_{-n}^\ast a_n & =0,\\
-\lambda b_n-\omega b_n+\dfrac{1}{h^2}(b_{n+1}-2b_n+b_{n-1})+\phi_n^{\ast2}a_{-n}+2\phi_n^\ast\phi_{-n}b_n & =0.
\end{aligned}\end{equation}
This implies a matrix eigenvalue problem for eigenvalues $\lambda$ and the corresponding eigenfunctions $\{a_n,b_n\}$. To solve the problem numerically, we shall cut off the infinite lattice into finite length as $-N\leq n\leq N$. The eigenvalue problem is as follows:
\begin{equation}\label{c6}
\left(\begin{array}{cc}-\textbf{A} & \textbf{B}\\ -\textbf{B}^\ast & \textbf{A}^\ast\end{array}\right)\left(\begin{array}{c}\vec{a}\\ -\vec{b}\end{array}\right) =\lambda\left(\begin{array}{c}\vec{a}\\ -\vec{b}\end{array}\right),
\end{equation}
where $\vec{a}=(a_{-N},a_{-N+1},\cdots,a_N)^{\textrm{T}}$, $\vec{b}=(b_{-N},b_{-N+1},\cdots,b_N)^{\textrm{T}}$ and
\begin{align*}
& \textbf{A}=\left(\begin{array}{cccccc} A_{-N} & 1/h^2 & ~ & ~ & ~ & ~\\ 1/h^2 & A_{-N+1} & 1/h^2 & ~ & ~ & ~\\ ~ & 1/h^2 & A_{-N+2} & \ddots & ~ & ~\\ ~ & ~ & \ddots & \ddots & \ddots & ~\\ ~ & ~ & ~ & \ddots & A_{N-1} & 1/h^2\\ ~ & ~ & ~ & ~ & 1/h^2 & A_N \end{array}\right),\\
& \textbf{B}=\left(\begin{array}{cccc} ~ & ~ & ~ & \phi_{-N}^2\\ ~ & ~ & \iddots & ~\\ ~ & \phi_{N-1}^2 & ~ & ~\\ \phi_N^2 & ~ & ~ & ~ \end{array}\right), \end{align*}
with $A_n=-\omega-\frac{2}{h^2}+2\phi_n\phi_{-n}^\ast~(n=-N,-N+1,\cdots,N)$. Corresponding to the stationary solution gained in the above section, we give its stability analysis. It is shown in Fig.\ref{fg3-1}(a) that this stability problem contains eight eigenvalues with nonzero imaginary part, which means that the stationary solution is linear unstable.\par

\begin{figure*}[htbp]
\centering
\subfigure[]{\includegraphics[width=0.25\textwidth]{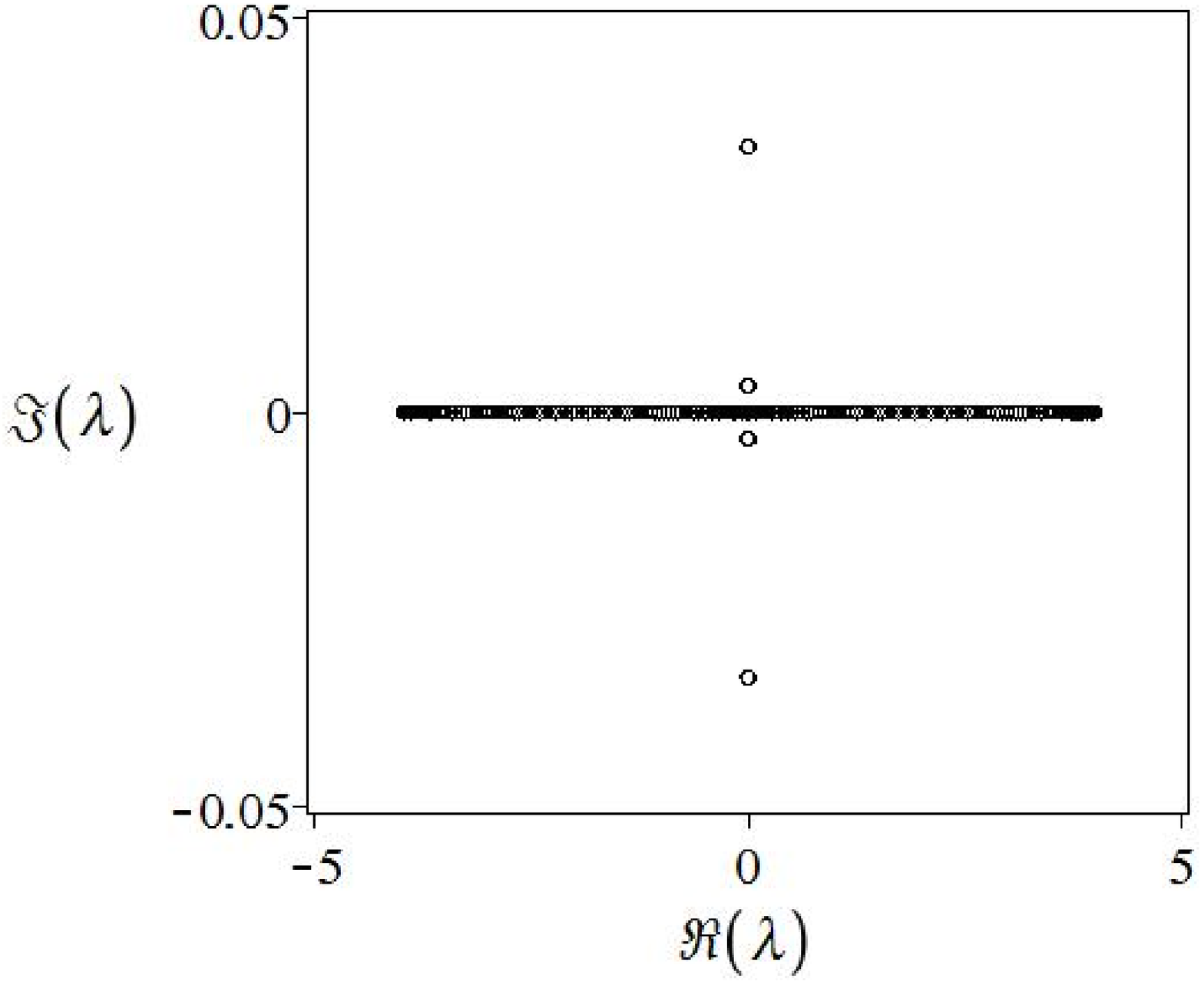}}\qquad
\subfigure[]{\includegraphics[width=0.25\textwidth]{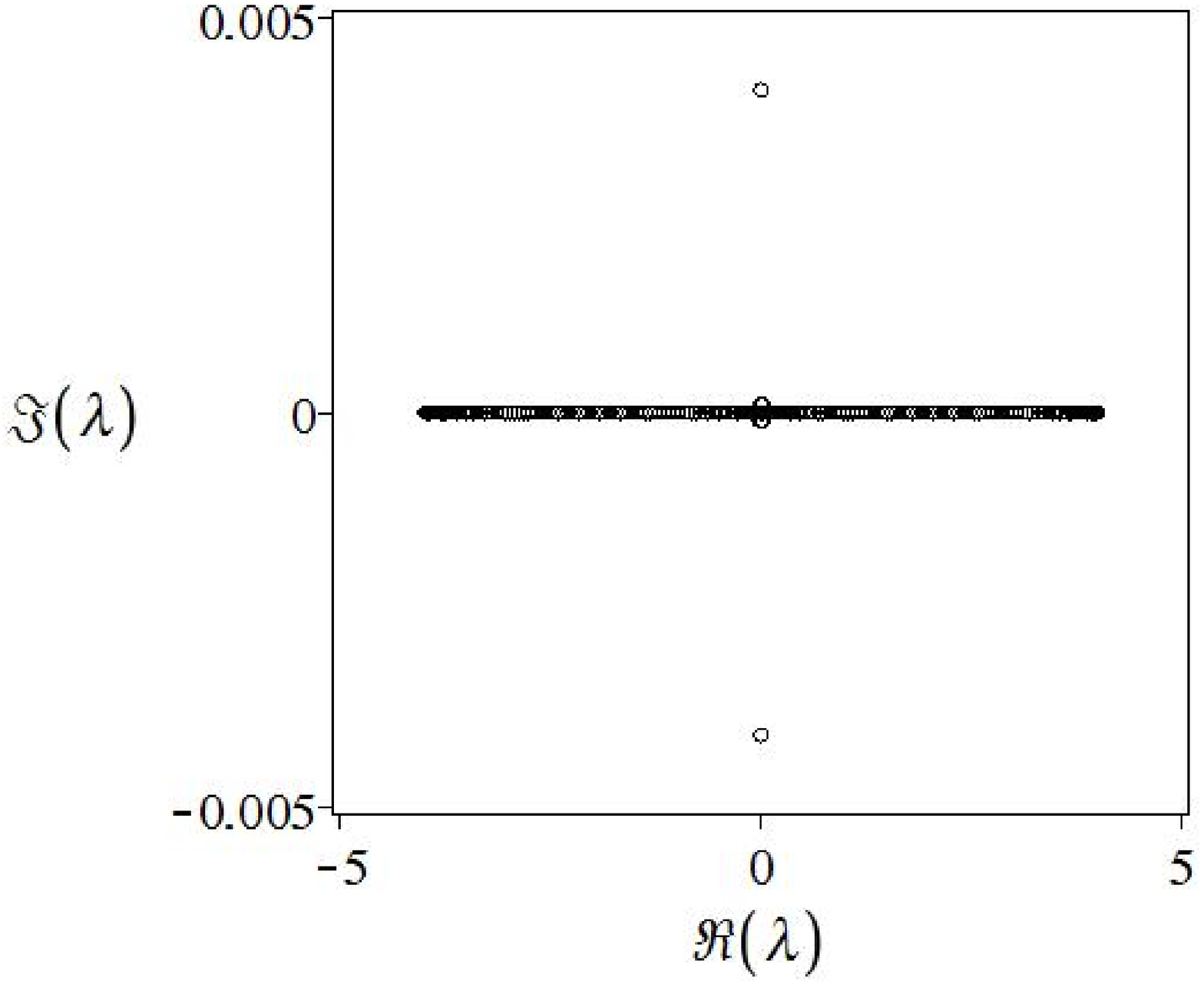}}
\caption{\small{Eigenvalues for $Q_n^{[2],s}$, as $b=3$. In (a), $\varepsilon=0.25$. In (b), $\varepsilon=0.05$.}}\label{fg3-3}
\end{figure*}

To compare with this result, let us we discuss the stability of stationary solitary solution for the integrable spatial discrete nonlocal NLS equation\cite{AM14PRE90_32912},
\begin{equation}\label{c7}
i{Q}_{n,t}=Q_{n+1}-2Q_n+Q_{n-1}+Q_nQ_{-n}^\ast(Q_{n+1}+Q_{n-1}).
\end{equation}
The one-soliton solution to Eq.\eqref{c7} reads
\begin{equation}\label{c8} Q_n^{[1]}=-\dfrac{(z_1z_2)^{-1}(z_1^2-z_2^2)e^{i\varphi_2}e^{-2iw_2t}z_2^{2n}}{1+e^{i(\varphi_1+\varphi_2)}e^{2i(w_1-w_2)t}z_1^{-2n}z_2^{2n}},
\end{equation}
where $w_j=(z_j-z_j^{-1})^2/2~(j=1,2)$, $z_1, z_2$ are positive and $\varphi_1,\varphi_2 $ are arbitrary real numbers. Set $z_1z_2=1$ and denote $z_2$ as $z$. The stationary solution derived from Eq.\eqref{c8} is
\begin{equation}\label{c9}
Q_n^{[1],s}=\dfrac{(z^2-z^{-2})e^{i\varphi_2}z^{2n}}{1+e^{i(\varphi_1+\varphi_2)}z^{4n}}e^{-i(z-z^{-1})^2t}\triangleq\phi_n^{[1]}e^{i\omega^{[1]}t}.
\end{equation}
Moreover, Eq.\eqref{c7} is studied by Hirota bilinear technique in Ref.\cite{MZ16AML59_115}. The following soliton solution to Eq.\eqref{c7} is given \cite{MZ16AML59_115},
\begin{equation}\label{c10}
Q_n^{[2]}=\dfrac{\varepsilon e^{an+2t\sinh a\sin b}e^{i(bn+2t(1-\cosh a\cos b))}}{1-\frac{\varepsilon^2}{4\sin^2b}e^{2ibn+4t\sinh a\sin b}},
\end{equation}
where $a,b$ are constants. Let $a=0$. Another stationary solution,
\begin{equation}\label{c11}
Q_n^{[2],s}=\dfrac{\varepsilon e^{ibn}}{1-\frac{\varepsilon^2}{4\sin^2b}e^{2ibn}}e^{2it(1-\cos b)}\triangleq\phi_n^{[2]}e^{i\omega^{[2]}t}.
\end{equation}
to Eq.\eqref{c7} is derived. We use the same method to analyze the linear stability of stationary solutions \eqref{c9} and \eqref{c11}. For Eq.\eqref{c7}, the final eigenvalue problem reaches at the same form of Eq.\eqref{c6}, but every entry in $\textbf{A}$ and $\textbf{B}$ is changed to
\begin{align*}
& \textbf{A}=\left(\begin{array}{cccccc} A_{-N}^{[j]} & \tilde{A}_{-N}^{[j]} & ~ & ~ & ~ & ~\\ \tilde{A}_{-N}^{[j]} & A_{-N+1}^{[j]} & \tilde{A}_{-N+1}^{[j]} & ~ & ~ & ~\\ ~ & \tilde{A}_{-N+1}^{[j]} & A_{-N+2}^{[j]} & \ddots & ~ & ~\\ ~ & ~ & \ddots & \ddots & \ddots & ~\\ ~ & ~ & ~ & \ddots & A_{N-1}^{[j]} & \tilde{A}_{N-1}^{[j]}\\ ~ & ~ & ~ & ~ & \tilde{A}_{N-1}^{[j]} & A_N^{[j]} \end{array}\right),\\
& \textbf{B}=\left(\begin{array}{cccc} ~ & ~ & ~ & \phi_{-N}^{[j]}(\phi_{-N+1}^{[j]}+\phi_{-N-1}^{[j]})\\ ~ & ~ & \iddots & ~\\ ~ & \phi_{N-1}^{[j]}(\phi_N^{[j]}+\phi_{N-2}^{[j]}) & ~ & ~\\ \phi_N^{[j]}(\phi_{N+1}^{[j]}+\phi_{N-1}^{[j]}) & ~ & ~ & ~ \end{array}\right),
\end{align*}
with
\begin{align*}
& A_n^{[j]}=\omega^{[j]}-2+\phi_{-n}^{[j]*}(\phi_{n+1}^{[j]}+\phi_{n-1}^{[j]}),\quad (n=-N,-N+1,\cdots,N,\quad j=1,2),\\
& \tilde{A}_n^{[j]}=\dfrac{1}{h^2}+\phi_n^{[j]}\phi_{-n}^{[j]*},\quad (n=-N,-N+1,\cdots,N-1,\quad j=1,2),
\end{align*}
in which $\phi_{-N-1}^{[j]}=\phi_{N+1}^{[j]}\equiv 0,\quad (j=1,2)$. If $Q_n^{[1],s}$ is also the solution to the classical discrete NLS, which requires $\varphi_1=2k_1\pi$ and $\varphi_2=2k_2\pi$ with two integers $k_{1,2}$, then $Q_n^{[1],s}$ is linear stable, as is shown in Fig.\ref{fg3-1}(b). From Fig.\ref{fg3-1}(c) and Fig.\ref{fg3-3}, we can see that both of two stationary solutions contain eigenvalues with nonzero imaginary part. This also implies that these two stationary solutions are both linear unstable. But for $Q_n^{[2],s}$, when $\varepsilon$ become smaller, the maximum absolute value of the imaginary parts of the eigenvalues also becomes smaller.

\section{Numerical Solution of Cauchy Problem \eqref{eq:CPNL}}
In this section, we solve Cauchy problem \eqref{eq:CPNL} numerically in periodic boundary condition. We use spatial discrete schemes (\ref{eq:ddkeji}) and (\ref{eq:ddnoke}) as a discrete model of nonlocal NLS equation \eqref{eq:NNLS},respectively. We first divide $[-\frac{L}{2},\frac{L}{2}]$ into $N$ intervals with step $h=\frac{L}{N}$, and then we solve the following Cauchy problems numerically:
\begin{align}
  &\begin{cases}
  i{u}_{n,t}+\frac{1}{h^{2}}\left(u_{n+1}-2u_{n}+u_{n-1}\right)+u_{n}^{2}u_{-n}^{*}=0,\\
  u_{n}\left(0\right)=a[1+\epsilon\cos\left(\mu nh+\varphi_{0}\right)],
\end{cases}
\label{eq:CPDNN}
\\
  &\begin{cases}
  i{u}_{n,t}+\frac{1}{h^{2}}\left(u_{n+1}-2u_{n}+u_{n-1}\right)+u_{n}u_{-n}^{*}\frac{u_{n+1}+u_{n-1}}{2}=0, \\
  u_{n}\left(0\right)=a[1+\epsilon\cos\left(\mu nh+\varphi_{0}\right)],
\end{cases}
\label{eq:CPDNI}
\end{align}
where $-\frac{N}{2}\leq n\leq\frac{N}2$. We use a six-order self-adaptive Runge-Kutta method \cite{EHJ76_}, to simulate the evolution of the Cauchy problems \eqref{eq:CPDNN} and \eqref{eq:CPDNI} by setting $L=2\sqrt{2}\pi$, $\epsilon=0.1$,  $a=1$ and $\mu=\frac{2\pi}L$, respectively.

In this performance, we first would like to check that for evolution of the solution to the Cauchy problems, what can be obtained when the space step $h$ changes for fixed initial data. Then, we discuss the influence of initial data on the solutions to the Cauchy problem for fixed step $h$. We also give a special observation for the difference among Cauchy problems \eqref{eq:CPDNN},  \eqref{eq:CPDNI} and the following Cauchy problem of discrete NLS equation,
\begin{align}
  &\begin{cases}
  i{u}_{n,t}+\frac{1}{h^{2}}\left(u_{n+1}-2u_{n}+u_{n-1}\right)+|u_{n}|^{2}u_{n}=0,\\
  u_{n}\left(0\right)=a[1+\epsilon\cos\left(\mu nh+\varphi_{0}\right)].
\end{cases}
\label{eq:CPDLN}
\end{align}

\begin{figure}[htbp]
\centering
\subfigure[$t=3.2$]
{\includegraphics[width=0.32\textwidth]{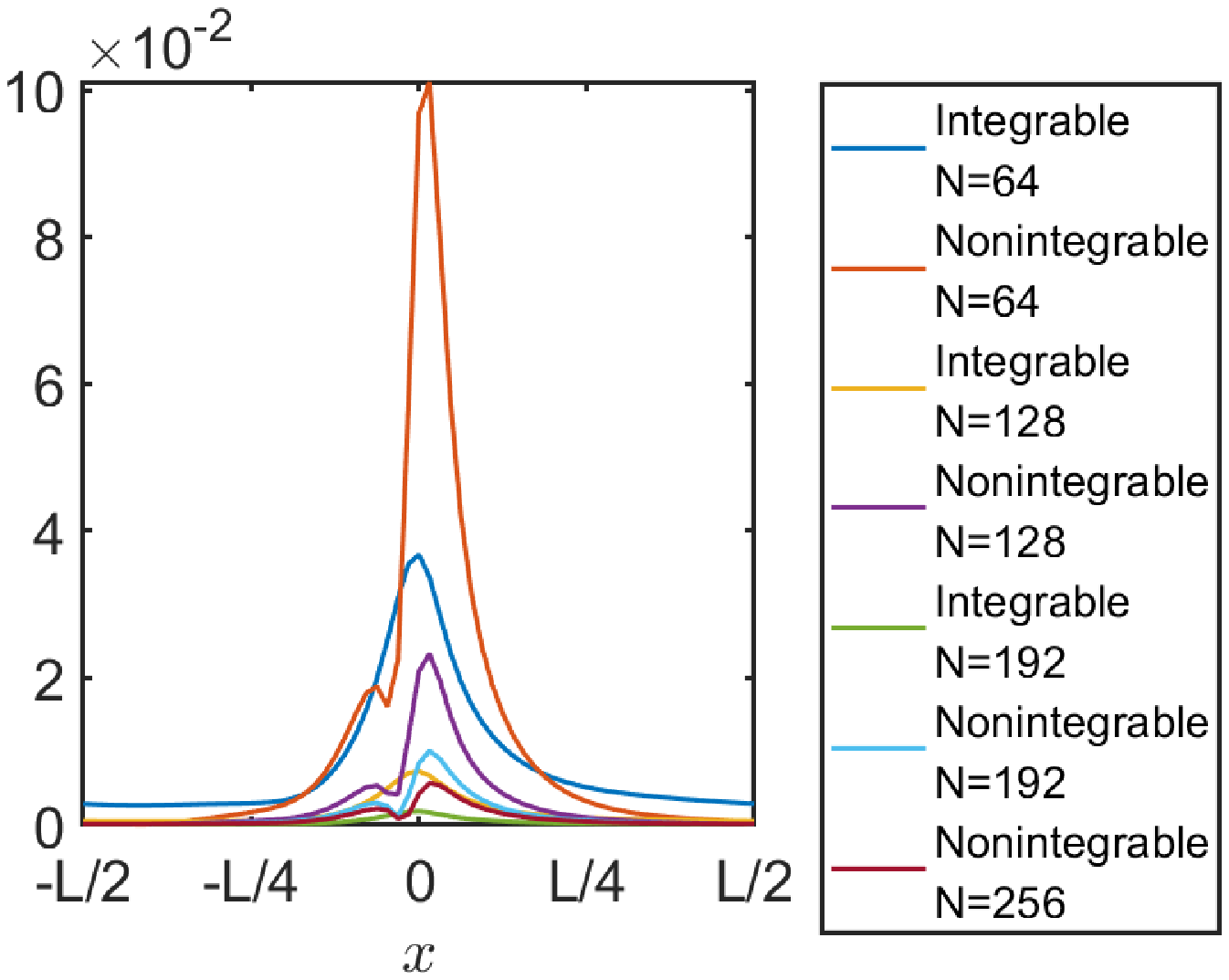}}\,
\subfigure[$t=6.4$]
{\includegraphics[width=0.32\textwidth]{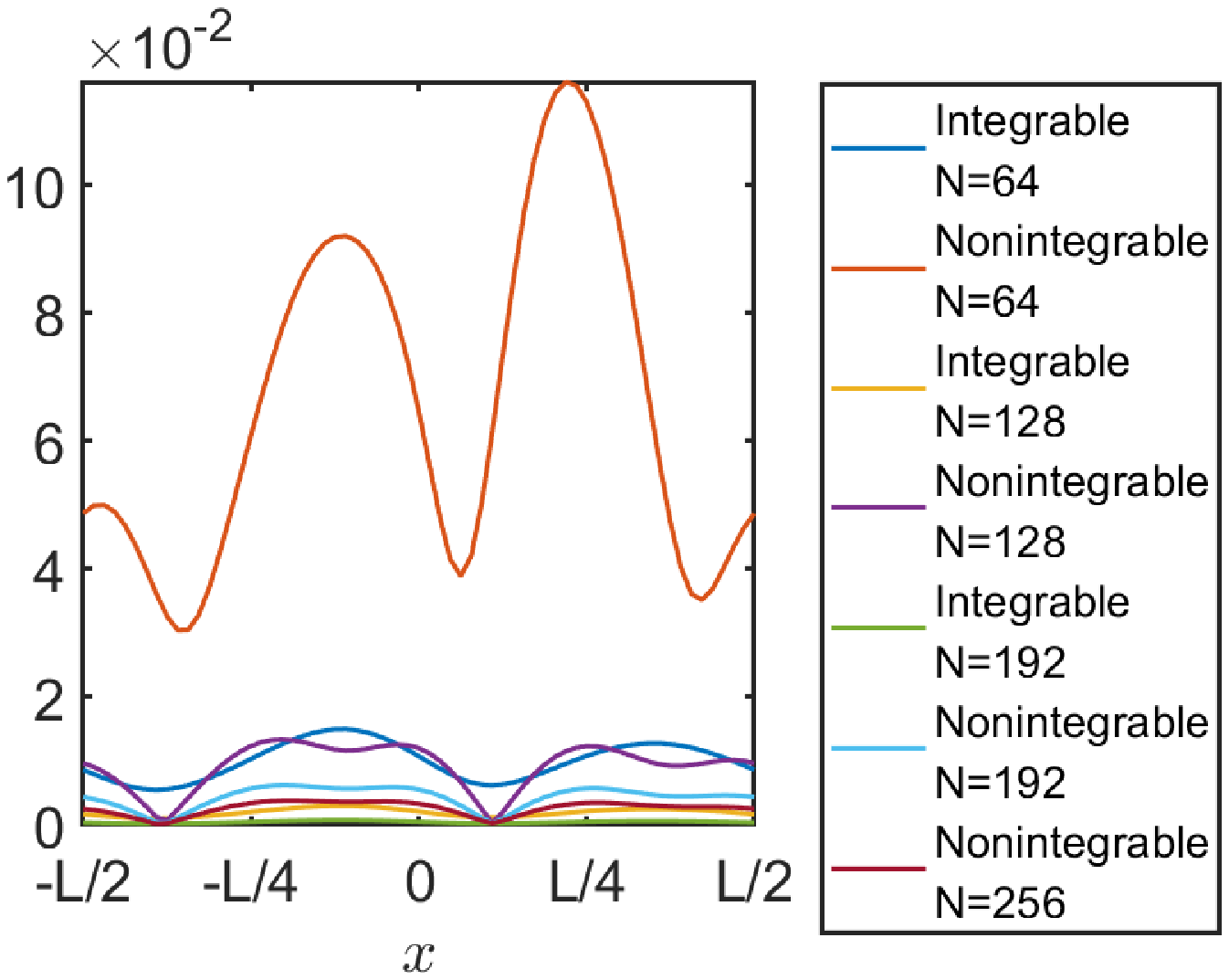}}\,
\subfigure[$t=9.6$]
{\includegraphics[width=0.32\textwidth]{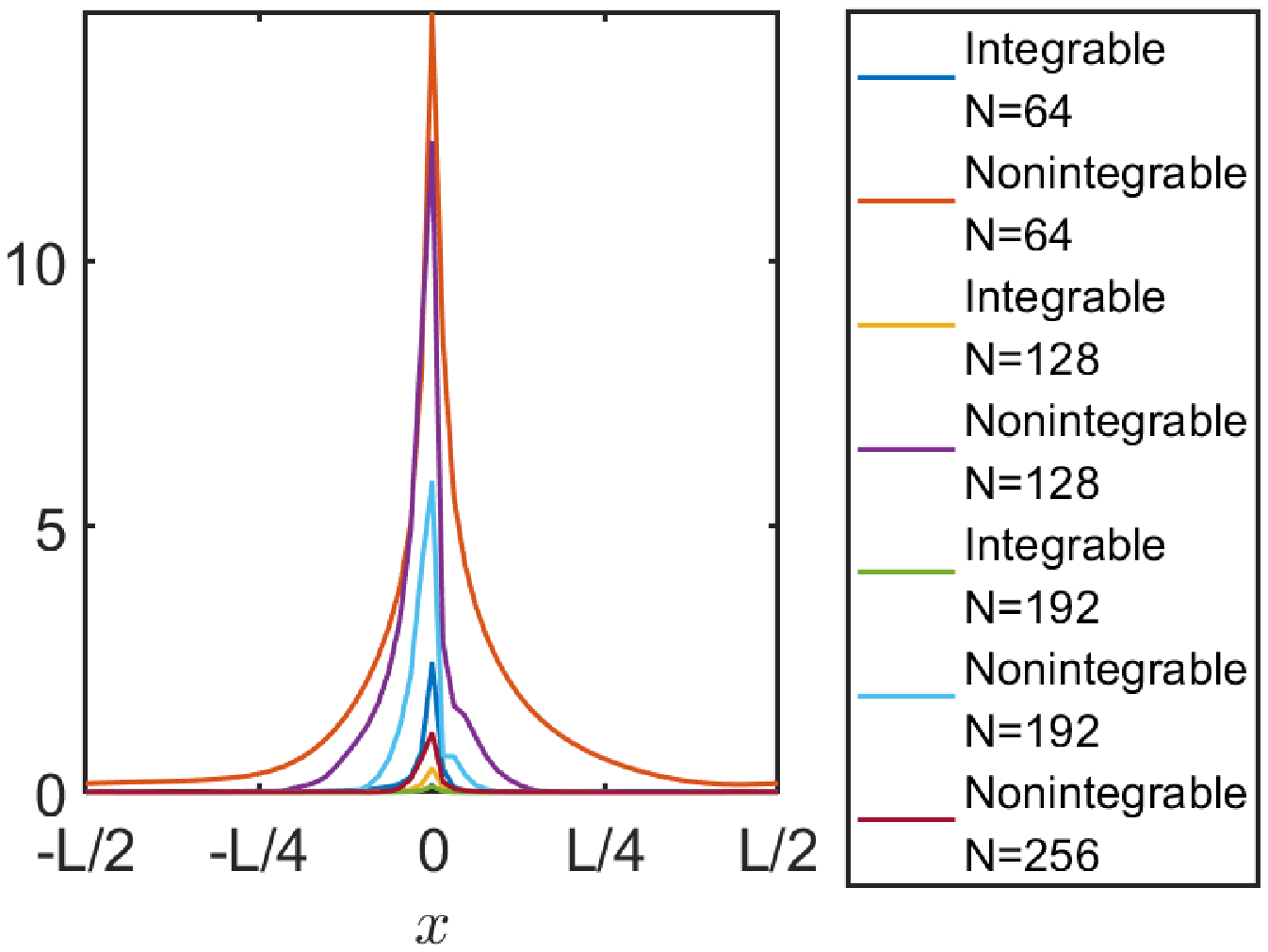}}\\
\caption{Error compared to the numerical solution of the
integrable case with $N=256$.
\label{fig:chatu}}
\end{figure}

\begin{figure}[htbp]
\centering
\subfigure[$N=64$]
{\includegraphics[width=0.32\textwidth]{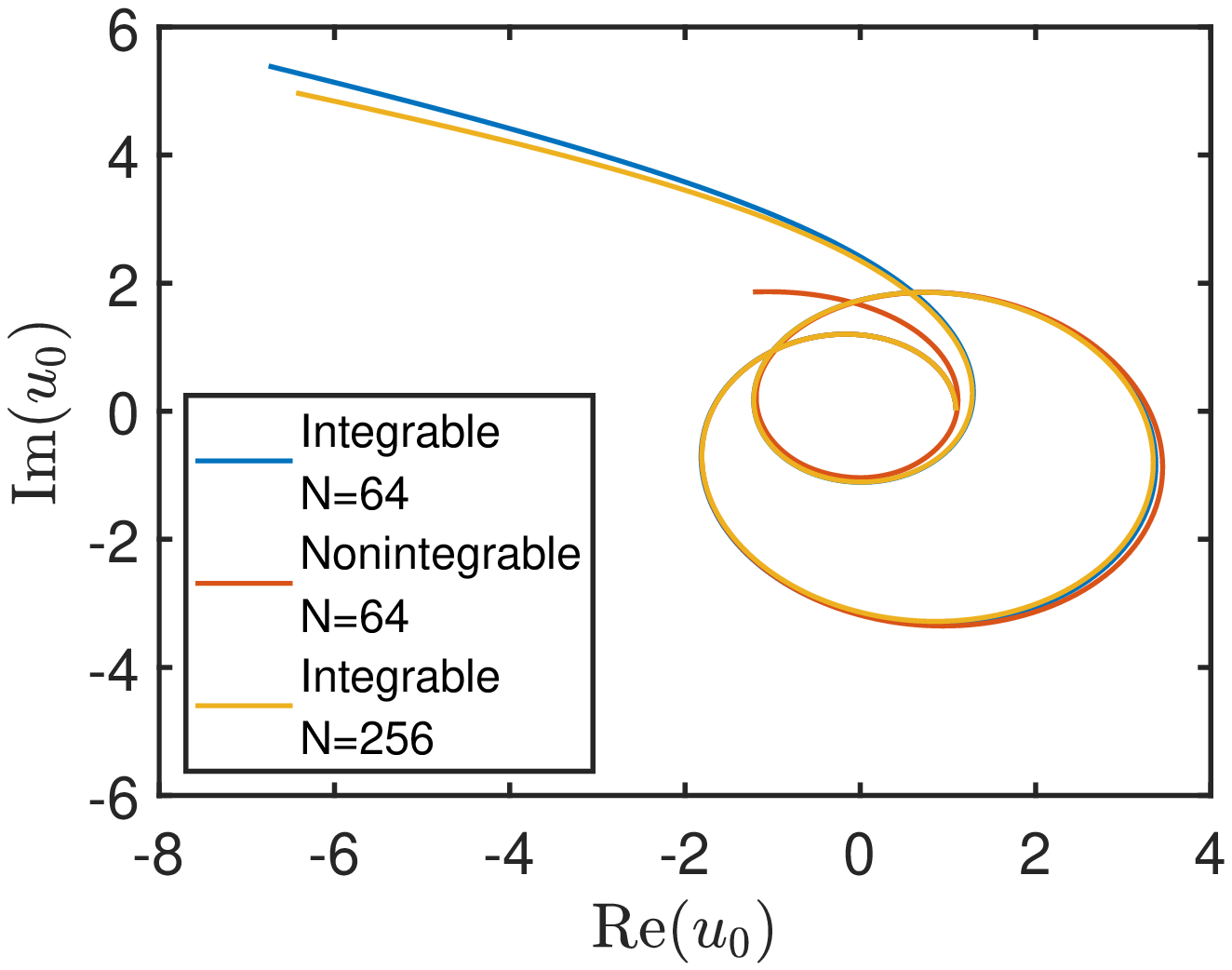}}\qquad
\subfigure[$N=128$]
{\includegraphics[width=0.32\textwidth]{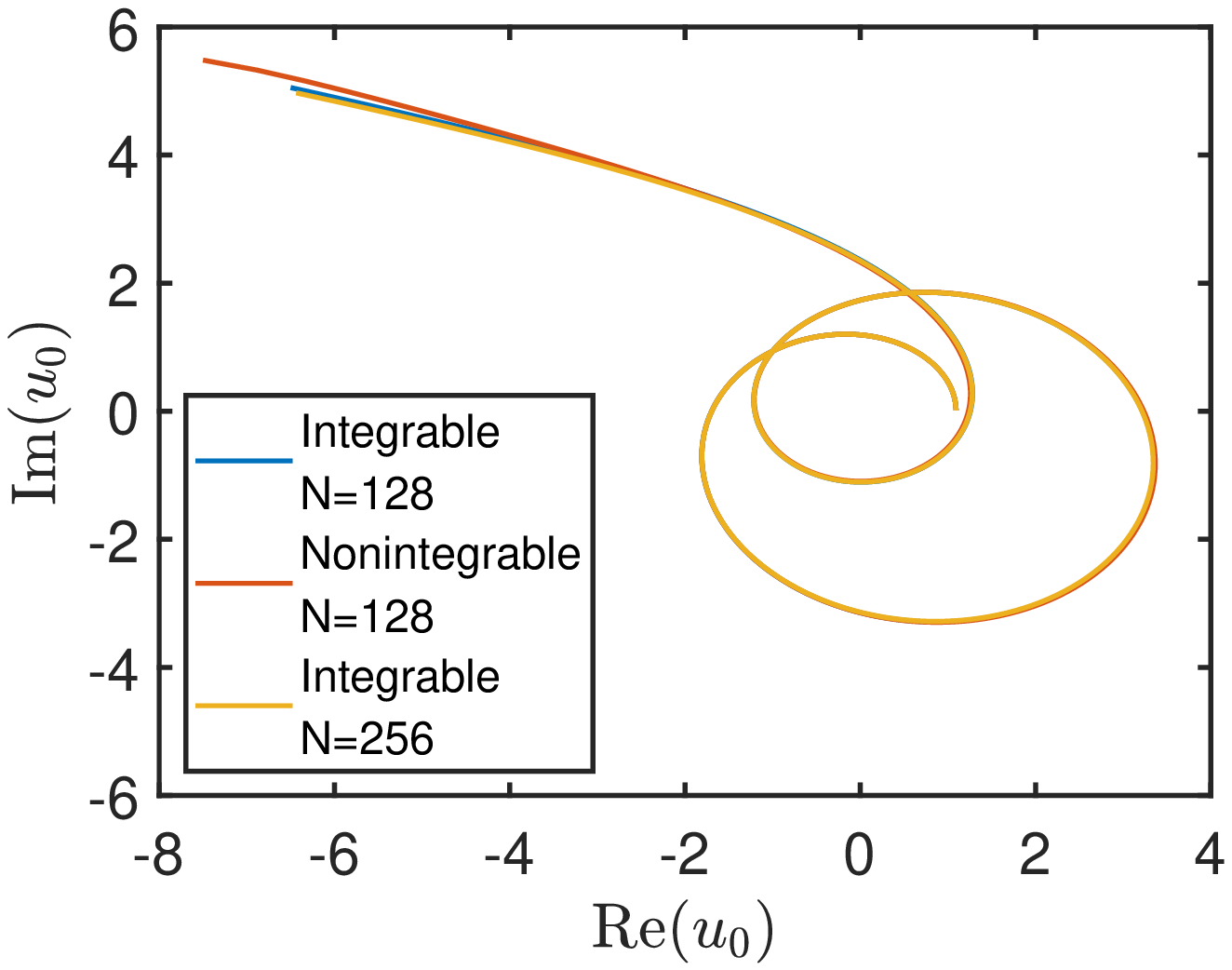}}
\caption{Trajectory of the numerical solution $u(0,t),0\leq t\leq9.5$ compared to that of the
integrable case with $N=256$.
\label{fig:chaxiang}}
\end{figure}

Next, let us consider some specific examples. We set $\varphi_0=\frac{1}{8}\pi$. As we known, the discrete solution of Cauchy problem \eqref{eq:CPDNI} converges to the one of Cauchy problem \eqref{eq:CPNL}. We denote $\hat{u}^{256}_n(t)$ as the approximate solution to the Cauchy problem \eqref{eq:CPDNI} with $N=256$. We simulate the envolution of solutions to Cauchy problems \eqref{eq:CPDNN} and \eqref{eq:CPDNI} with $N=64,\,128,\,192$, and we consider the value of $|\hat{u}_n^{256}(t)-u_n^N(t)|$. Fig.\ref{fig:chatu} describes the value. In Fig.\ref{fig:chaxiang}, we show the trajectory of the numerical solution at $n=0$ corresponding to the case of $N=64,128,256$. From Fig.\ref{fig:chatu} and Fig.\ref{fig:chaxiang}, we can see that the solutions of Cauchy problem \eqref{eq:CPDNI} converges faster than those of  Cauchy problem \eqref{eq:CPDNN}.

\begin{figure}[htbp]
\centering
\subfigure[$\left|u_{0}(t)\right|$]{\includegraphics[width=0.3390\textwidth]{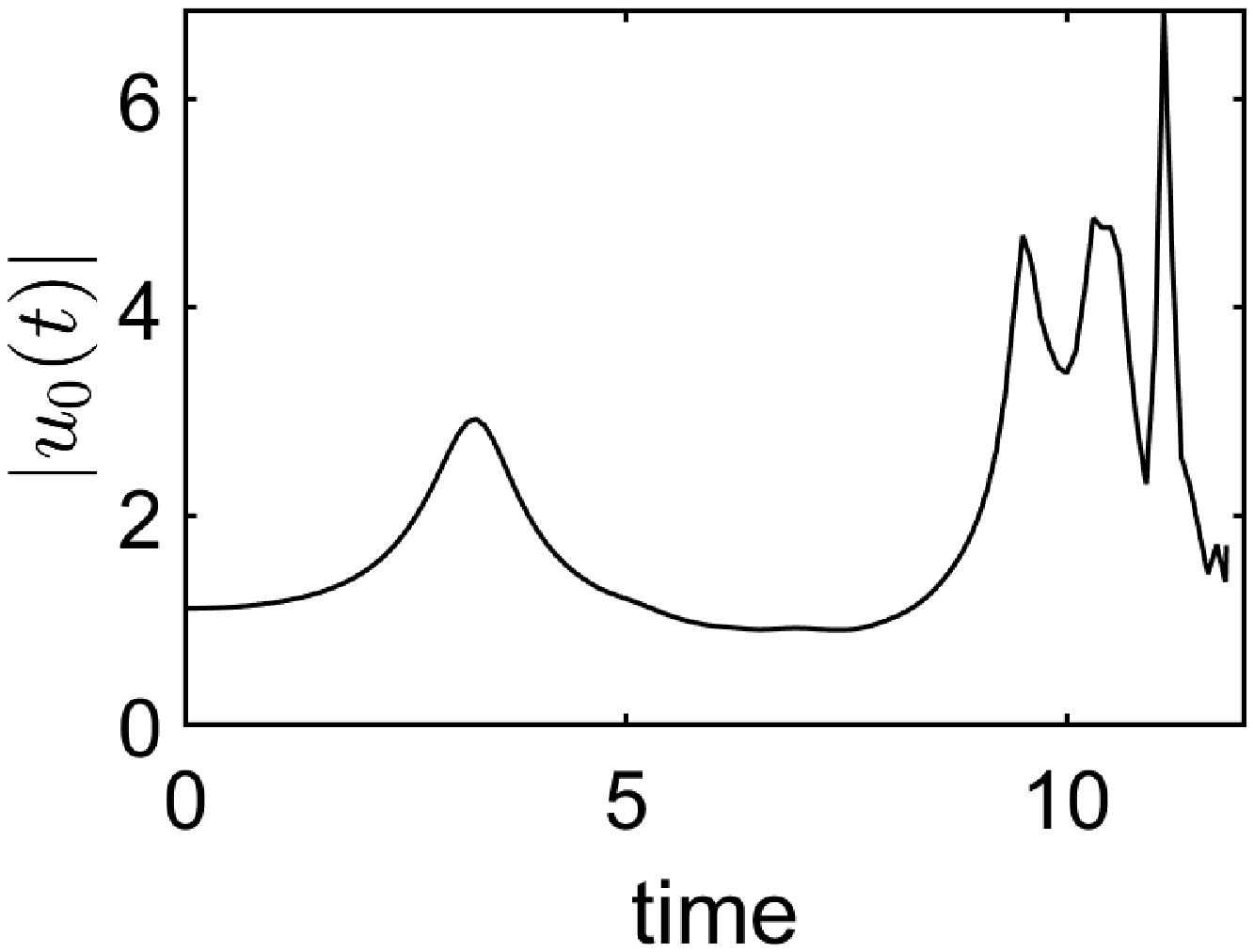
}}\qquad
\subfigure[$\left|u_{n}\left(11.8075\right)\right|$]{\includegraphics[width=0.3390\textwidth]{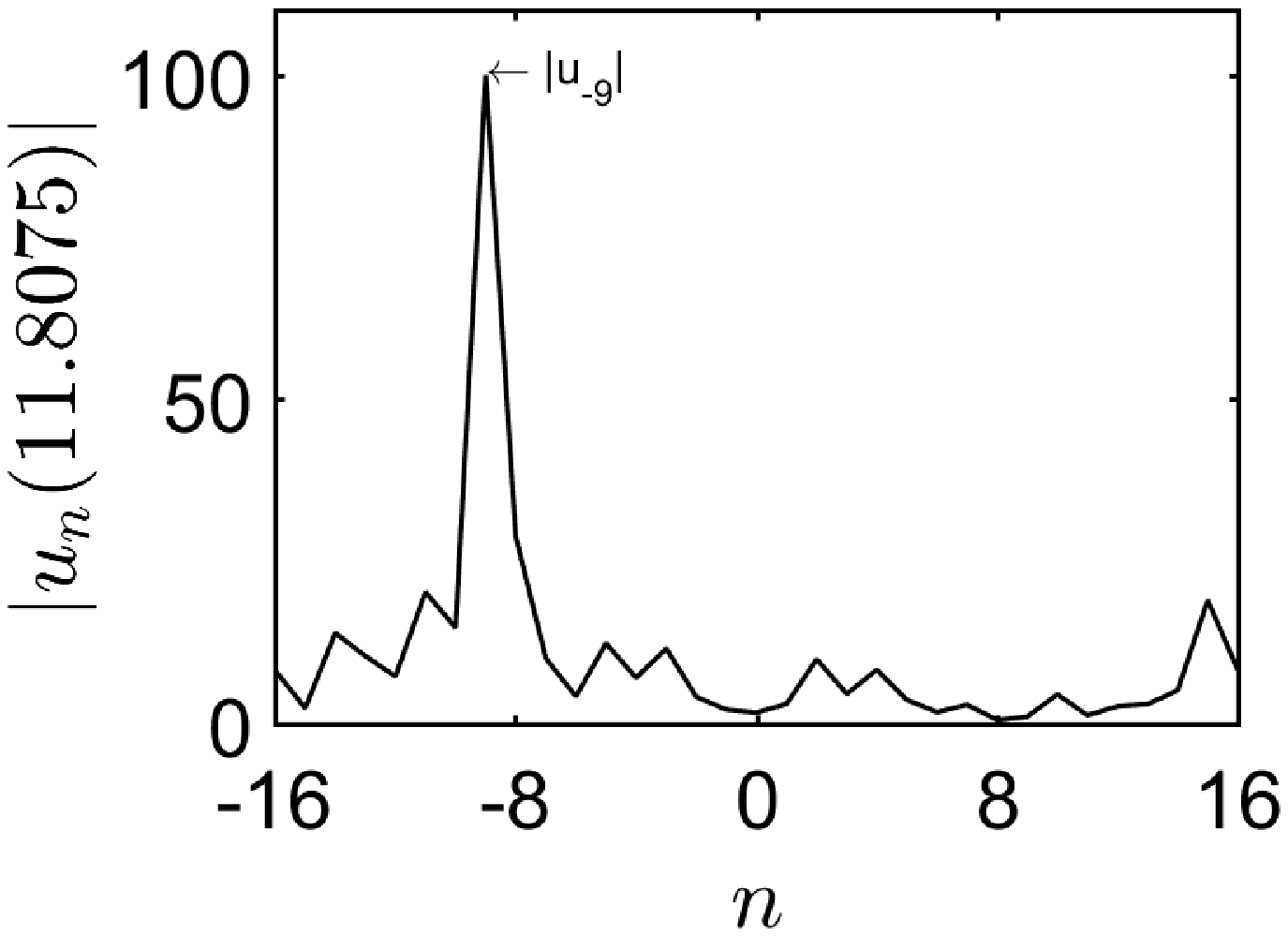}}\\
\subfigure[trajectory of $u_{0}\left(t\right)$]{\includegraphics[width=0.3390\textwidth]{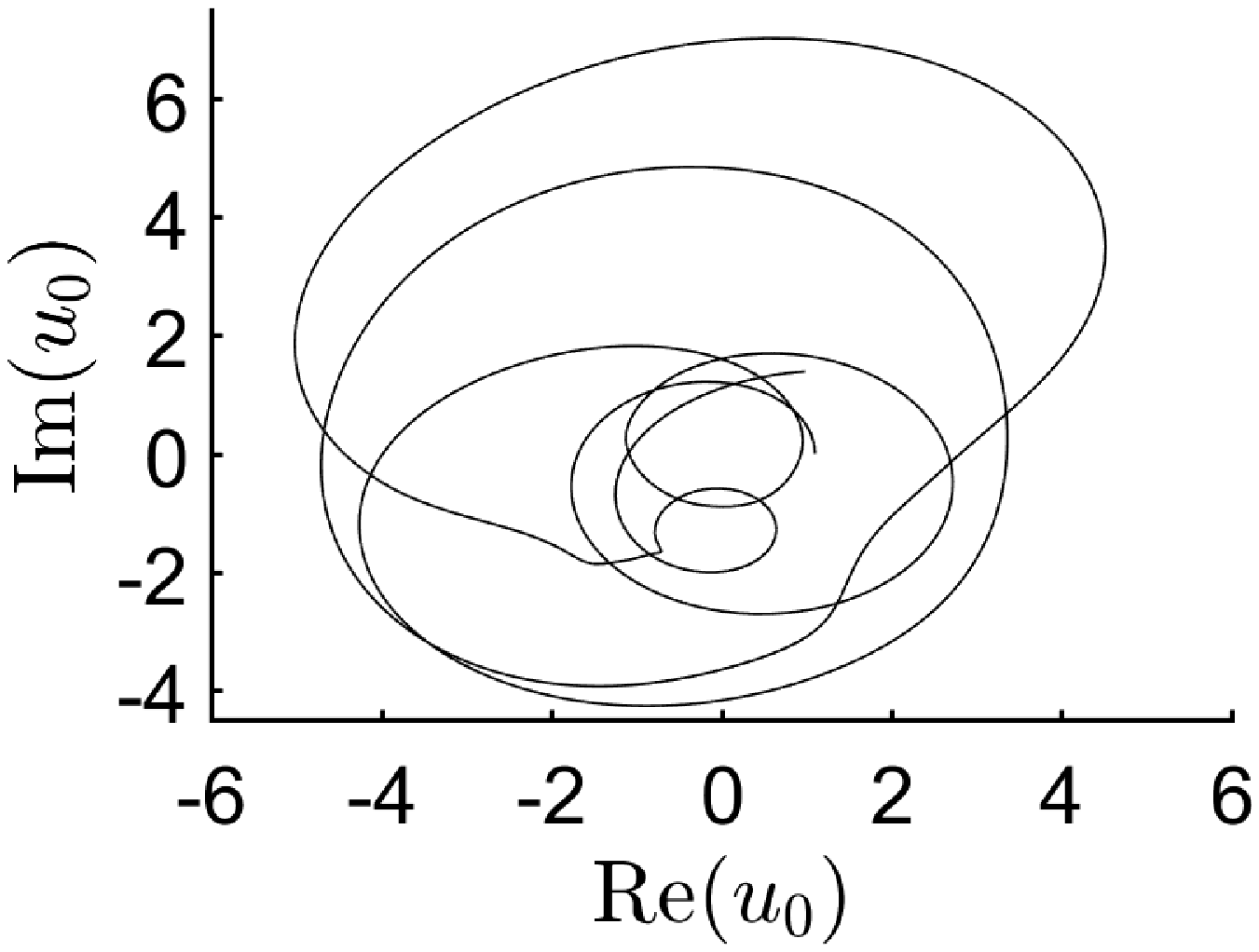
}}\qquad
\subfigure[$\left|u_{0}\left(t\right)\right|$ of the Cauchy problem \eqref{eq:CPDLN}]{\includegraphics[width=0.3390\textwidth]{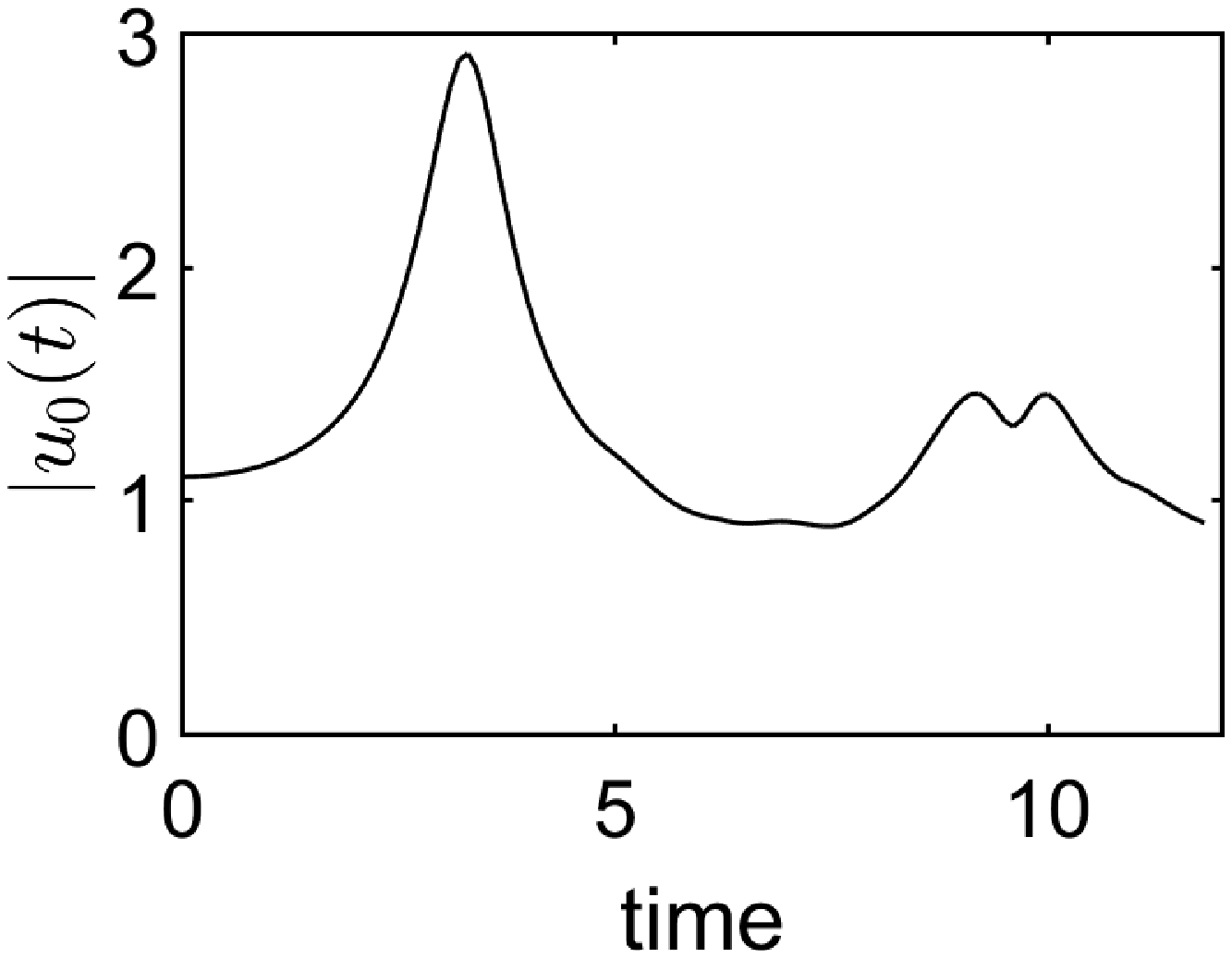
}}
\caption{The numerical solution to Cauchy problem \eqref{eq:CPDNN} with $N=32$ and $\varphi_{0}=\frac{1}{256}\pi$. Subfigures (a) and (c) describe the evolution of $u_0(t)$, where  $0\leq t\leq11.8075$. Subfigure (b) shows the mode of the solution when the maximum module reaches 100 at $t=11.8075$.\label{fig:noke32}}
\end{figure}

Let us consider the case of $\varphi_0=\frac\pi{256}$. We simulate and compare the evolution of the solution to Cauchy problems \eqref{eq:CPDNN} and (36) with $N=32,64,128$.
In the case of $N=32$, Fig.\ref{fig:noke32} shows that the maximum module of the solution of Cauchy problem \eqref{eq:CPDNN} reaches 100 at grid $n={-9}$ and $t=11.8075$.

\begin{figure}[htbp]
\centering
\subfigure[$\left|u_{0}(t)\right|$]{\includegraphics[width=0.3390\textwidth]{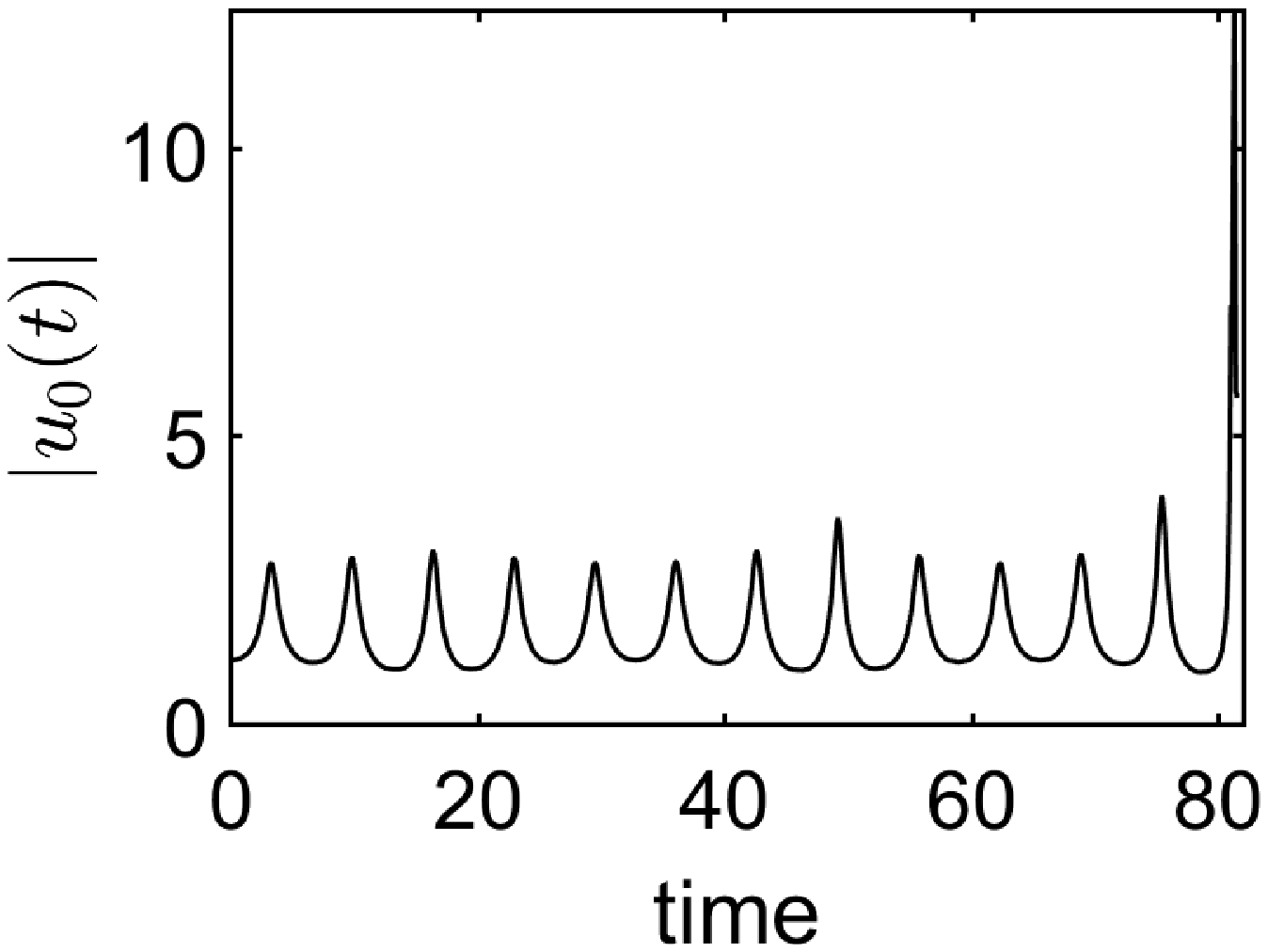
}}\qquad
\subfigure[$\left|u_{n}\left(81.5007\right)\right|$]{\includegraphics[width=0.3390\textwidth]{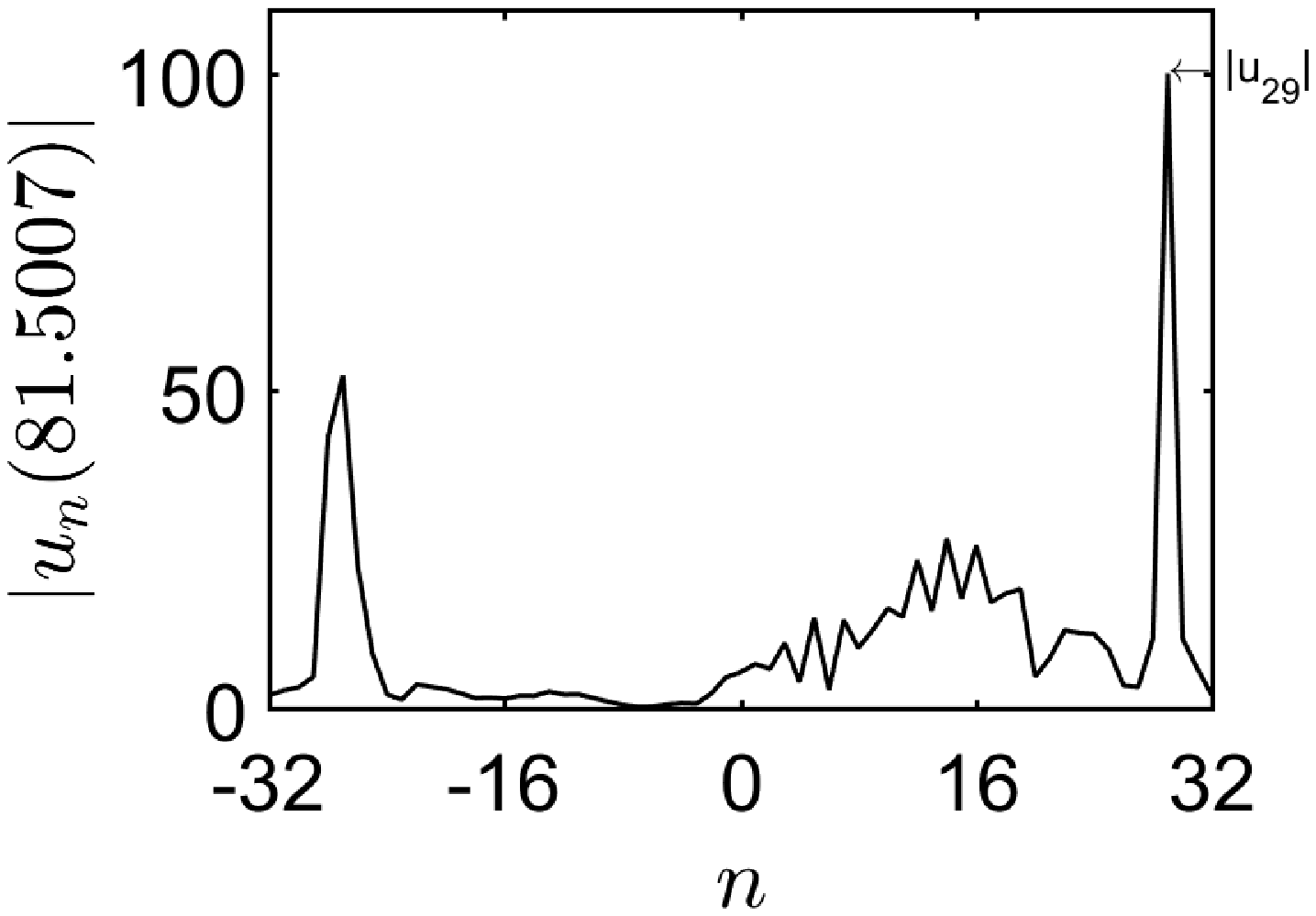
}}\\
\subfigure[trajectory of $u_{0}\left(t\right)$]{\includegraphics[width=0.3390\textwidth]{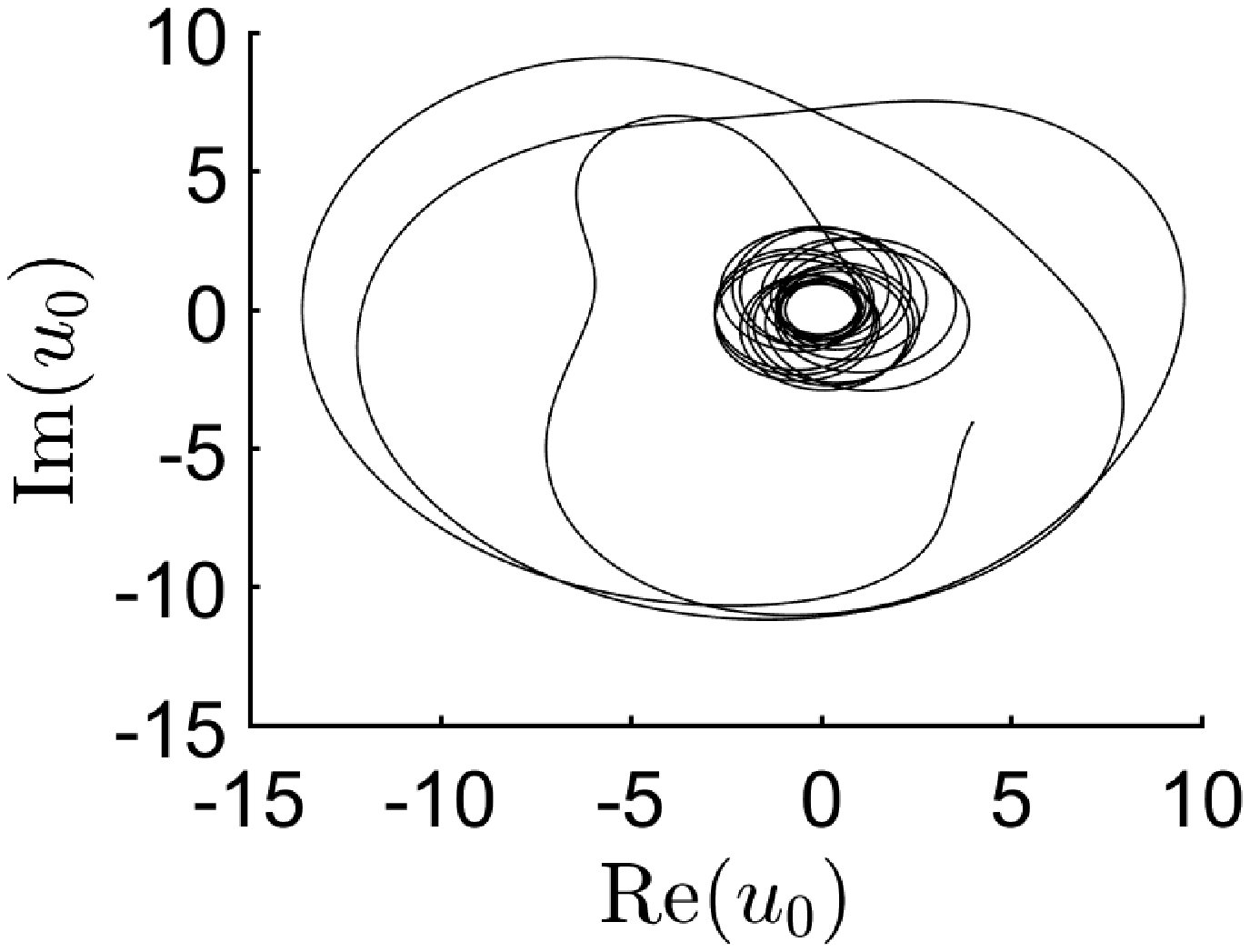
}}\qquad
\subfigure[$\left|u_{0}\left(t\right)\right|$ of the Cauchy problem \eqref{eq:CPDLN}]{\includegraphics[width=0.3390\textwidth]{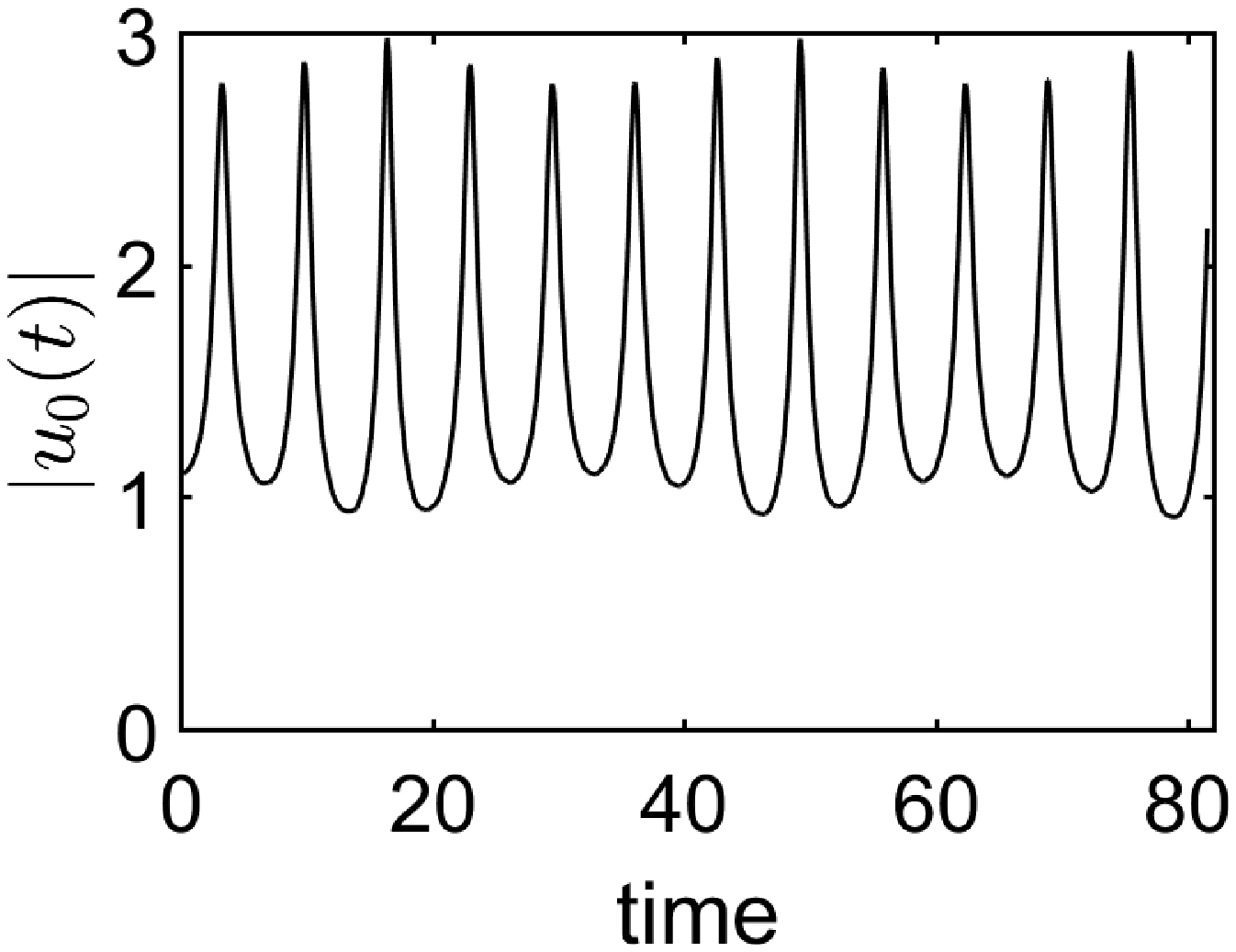
}}
\caption{The numerical solution to Cauchy problem \eqref{eq:CPDNN} with $N=64$ and $\varphi_{0}=\frac{1}{256}\pi$. Subfigures (a) and (c) describe the evolution of $u_0(t)$, where $0\leq t\leq 81.5007$. Subfigure (b) shows the mode of the solution when the maximum module reaches 100 at $t=81.5007$.
\label{fig:noke32-1}}
\end{figure}

We refine the grid to $N=64$. Fig.\ref{fig:noke32-1} shows that the module of the solution to the Cauchy problem \eqref{eq:CPDNN} at $n=0$ oscillates with the time evolution before a sudden growth at $t=81.5007$. Solution to Cauchy problem \eqref{eq:CPDLN} is bounded for the case of $N=32$ and $N=64$.

\begin{figure}[htbp]
\centering
\subfigure[$\left|u_{0}(t)\right|$]{\includegraphics[width=0.3390\textwidth]{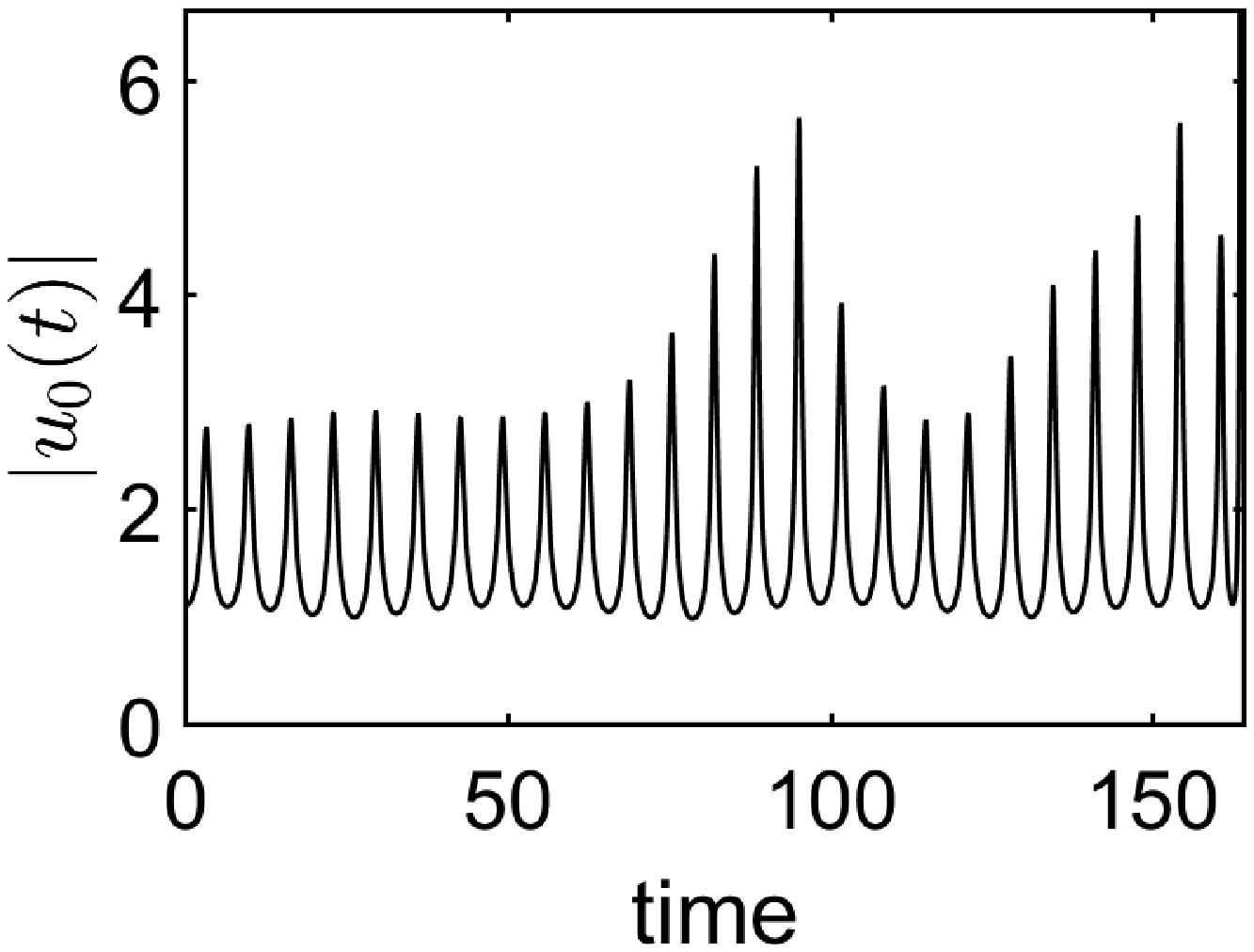
}}\qquad
\subfigure[$\left|u_{n}\left(163.4591\right)\right|$]{\includegraphics[width=0.3390\textwidth]{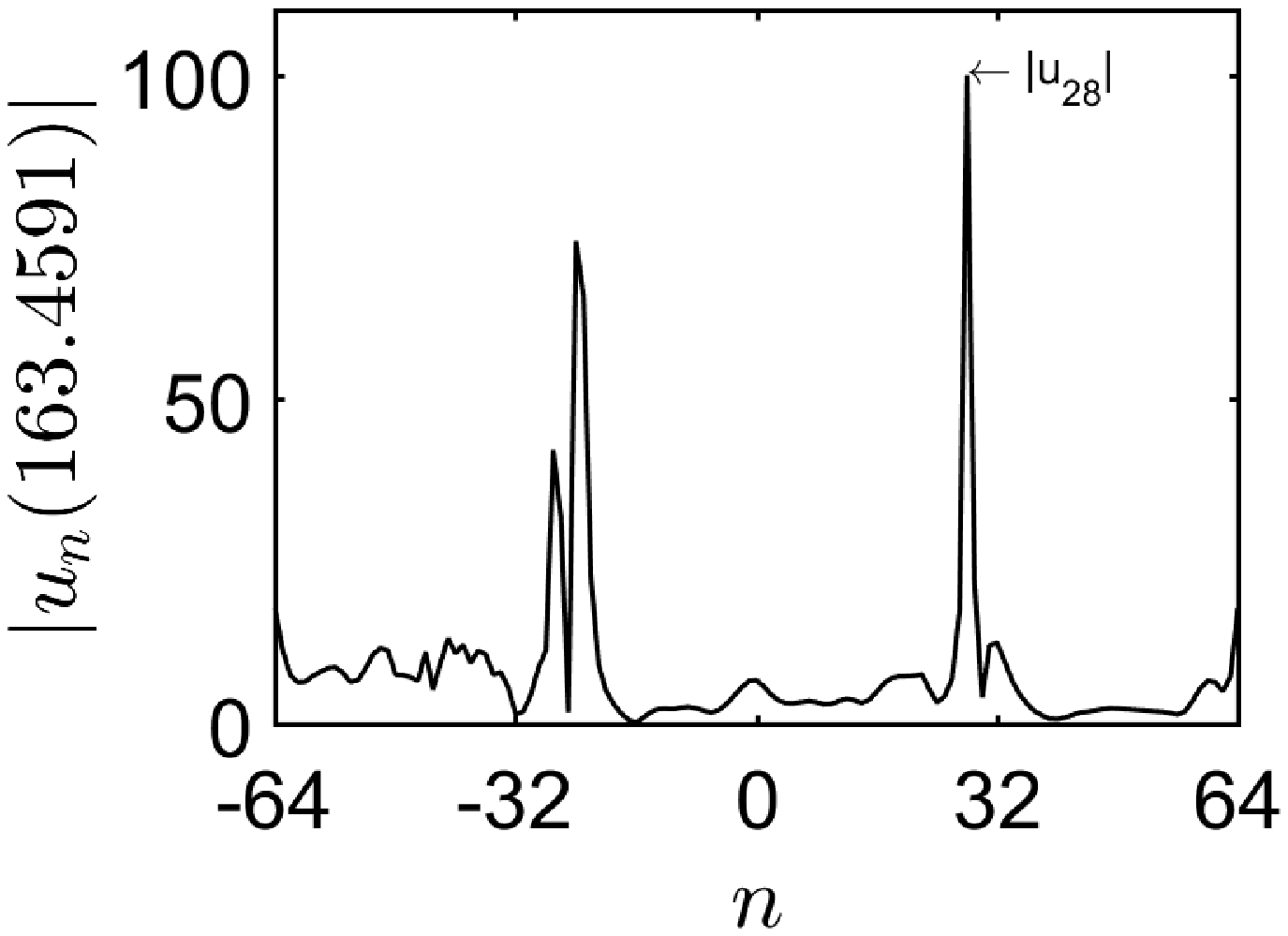
}}\\
\subfigure[trajectory of $u_{0}\left(t\right)$]{\includegraphics[width=0.3390\textwidth]{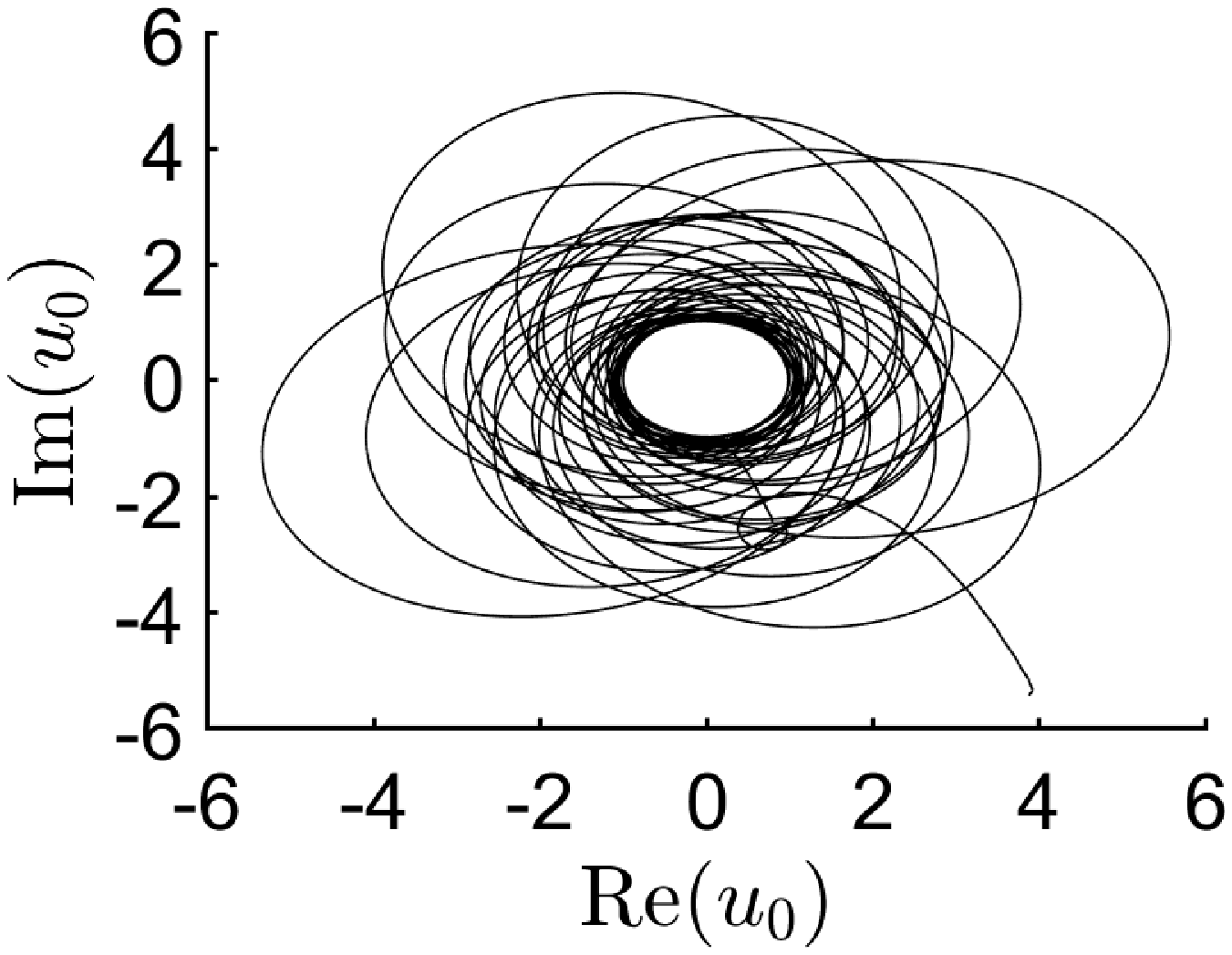
}}\qquad
\subfigure[$\left|u_{0}\left(t\right)\right|$ of the Cauchy problem \eqref{eq:CPDLN}]{\includegraphics[width=0.3390\textwidth]{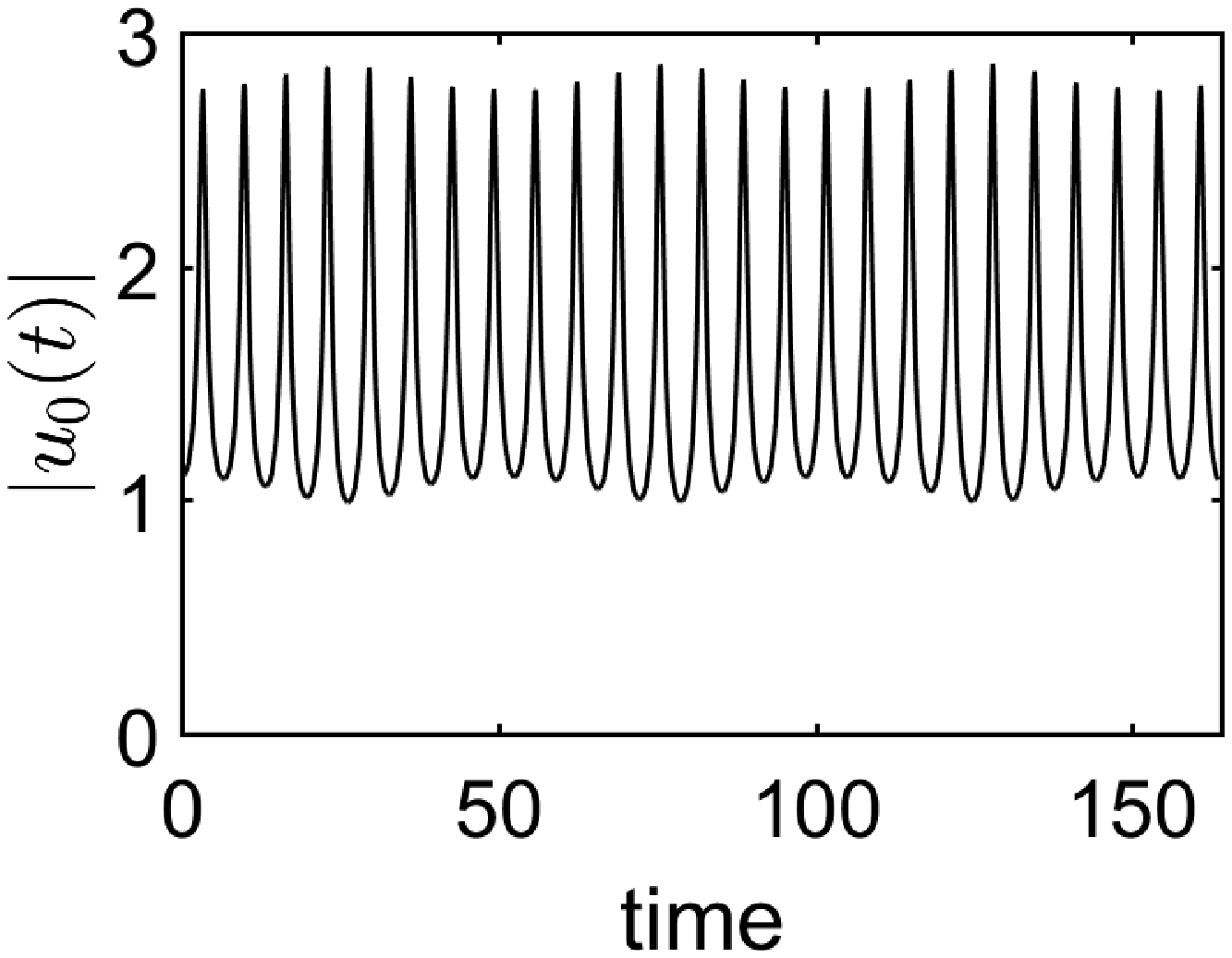
}}
\caption{The numerical solution to Cauchy problem \eqref{eq:CPDNN} with $N=128$ and $\varphi_{0}=\frac{1}{256}\pi$. Subfigures (a) and (c) describe the evolution of $u_0(t)$, where $0\leq t\leq 163.4591$. Subfigure (b) shows the mode of the solution when the maximum module reaches 100 at $t=163.4591$.
\label{fig:noke32-2}}
\end{figure}

We refine the grid once again to $N=128$. The computational results are given in Fig.\ref{fig:noke32-2}. The module of the numerical solution of Cauchy problem \eqref{eq:CPDNN} at $n=0$ oscillates longer than the case of $N=64$. For the same initial condition, we still obtain bounded solutions to Cauchy problem \eqref{eq:CPDLN}.

From these examples, we find that for nonlocal nonintegrable scheme \eqref{eq:ddnoke}, the points at which the maximum modules of the solutions first to reach 100 are different and the time consumed is also different. We also find that the numerical solutions become quasiperiodic before the solution blows up with the mesh refinement increasing.

\begin{figure}[htbp]
\centering
\subfigure[$N=32$, $\left|u_{0}(t)\right|$]{\includegraphics[width=0.2871\textwidth]{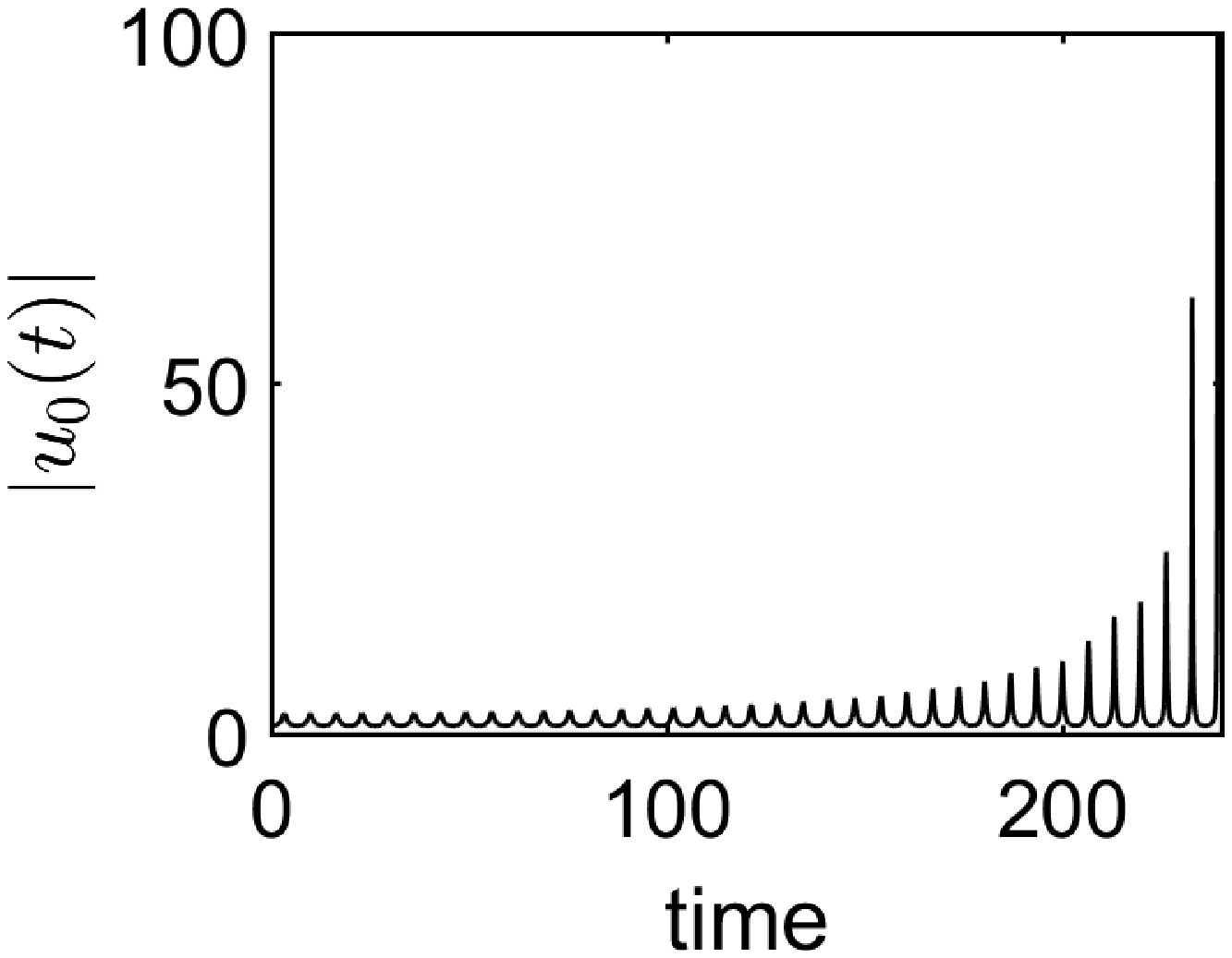
}}~
\subfigure[$N=32$, $\left|u_{n}\left(239.0951\right)\right|$]{\includegraphics[width=0.2871\textwidth]{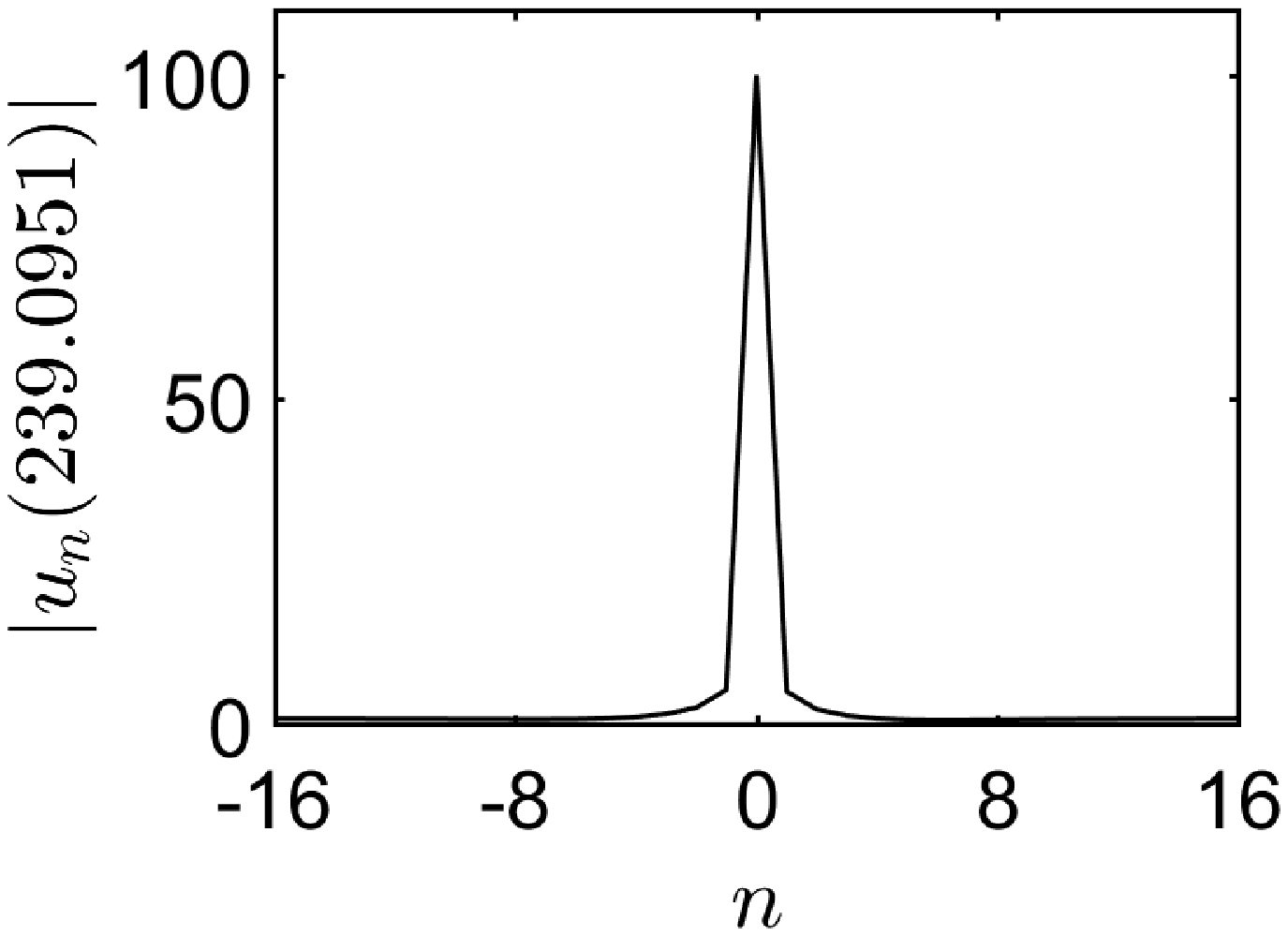
}}~
\subfigure[$N=32$, trajectory of $u_{0}\left(t\right)$, $0\leq t\leq239.0951$]{\includegraphics[width=0.2871\textwidth]{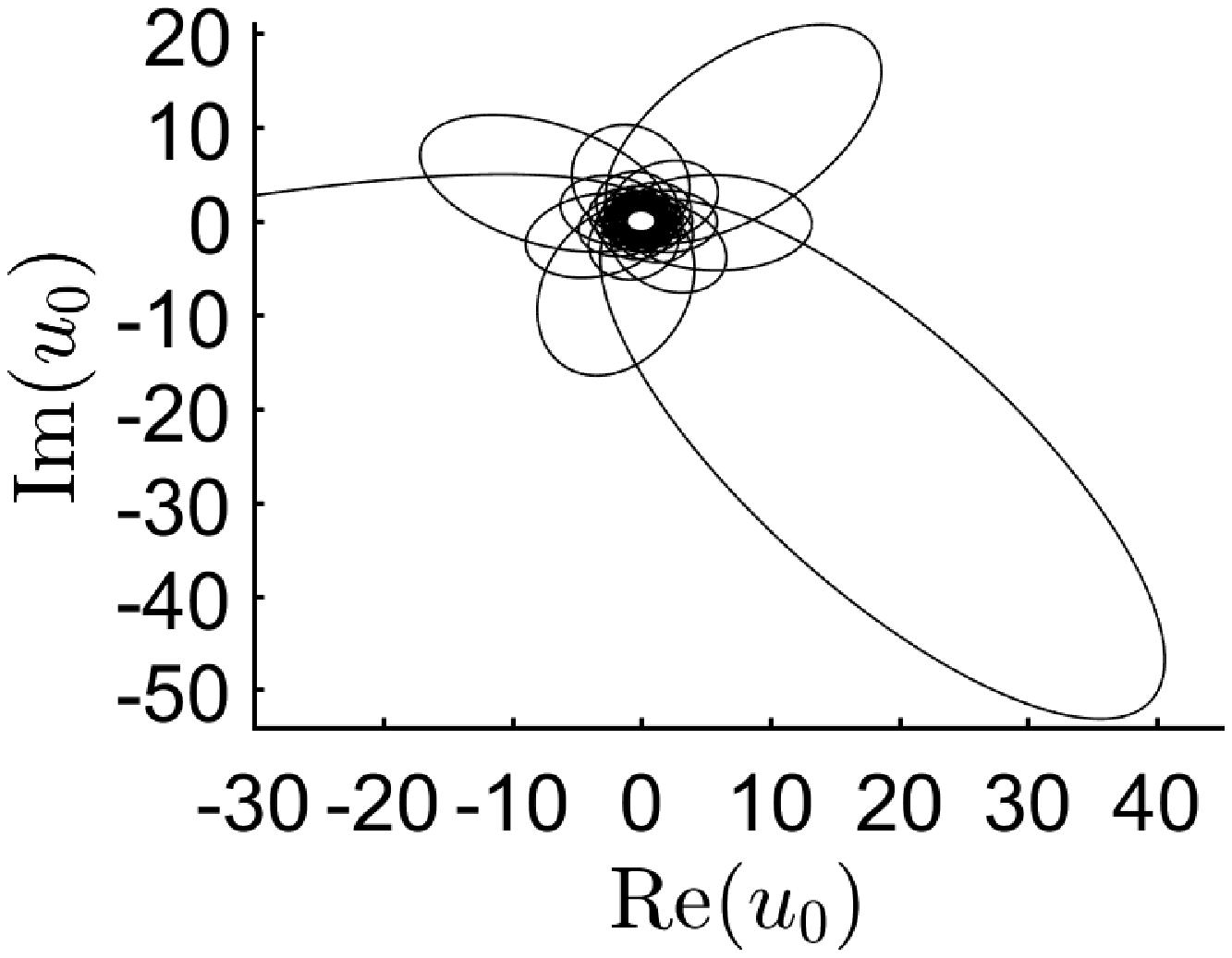
}}\\
\subfigure[$N=64$, $\left|u_{0}(t)\right|$]{\includegraphics[width=0.2871\textwidth]{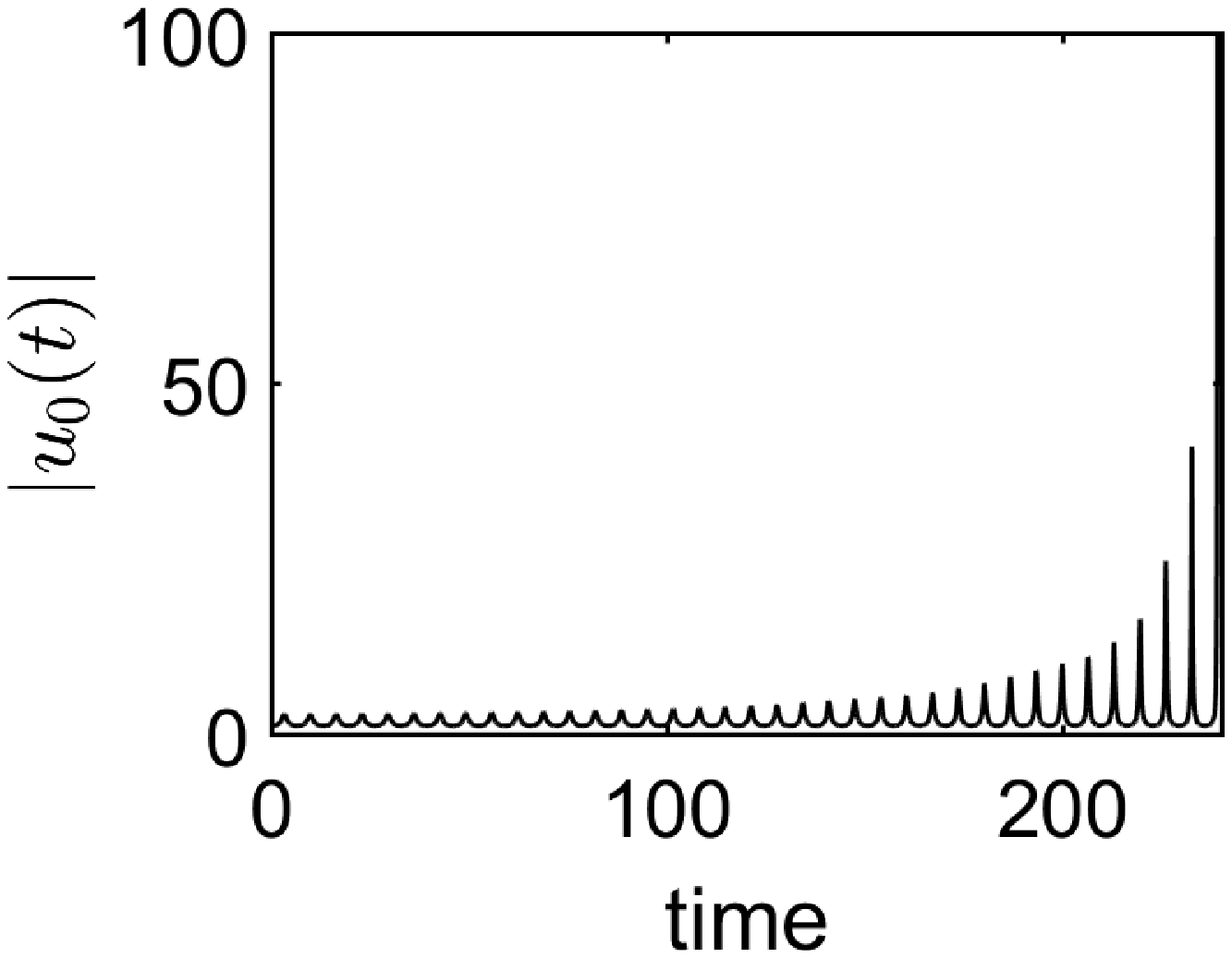
}}~
\subfigure[$N=64$, $\left|u_{n}\left(239.0620\right)\right|$]{\includegraphics[width=0.2871\textwidth]{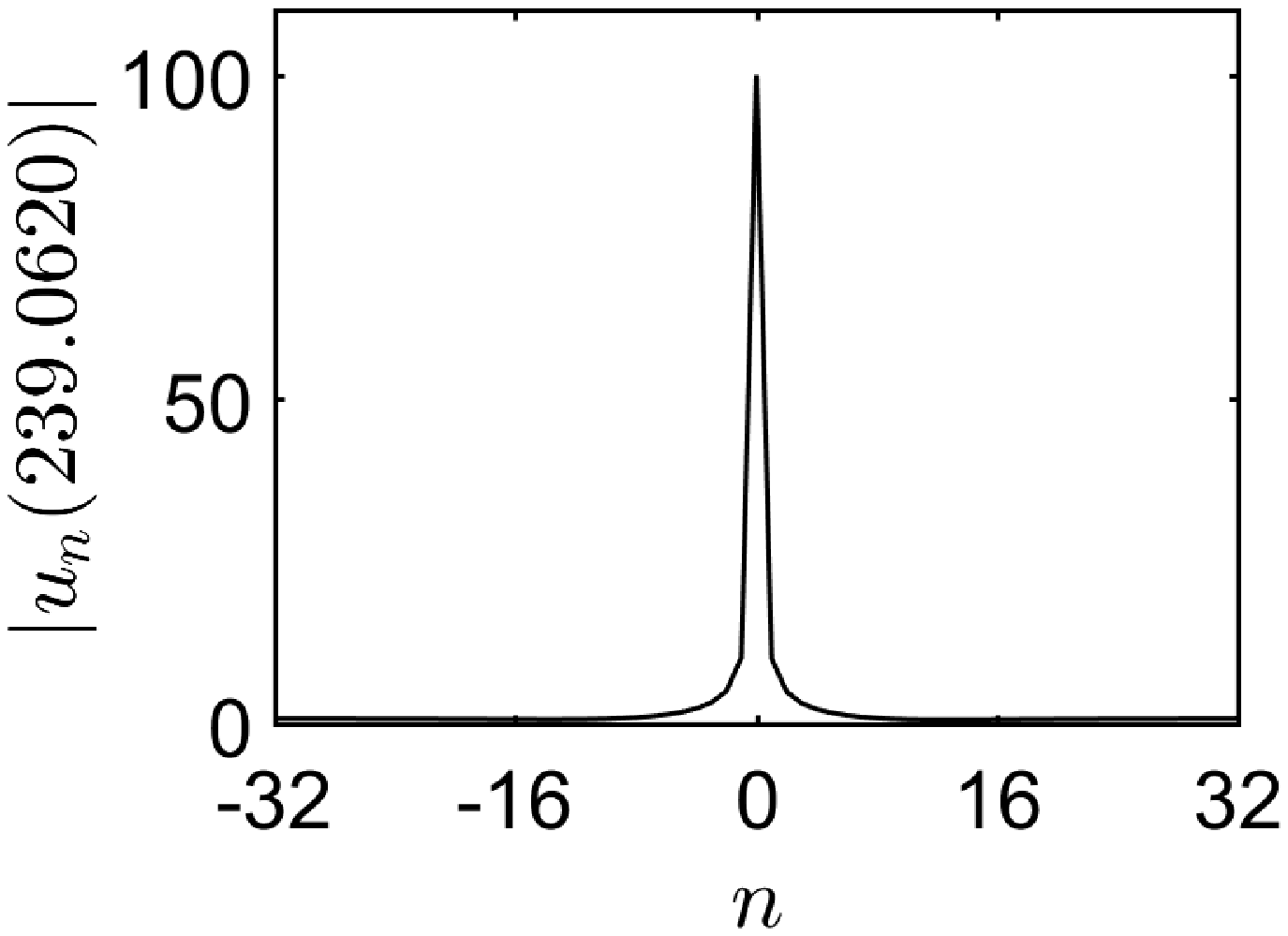
}}~
\subfigure[$N=64$, trajectory of $u_{0}\left(t\right)$, $0\leq t\leq239.0620$]{\includegraphics[width=0.2871\textwidth]{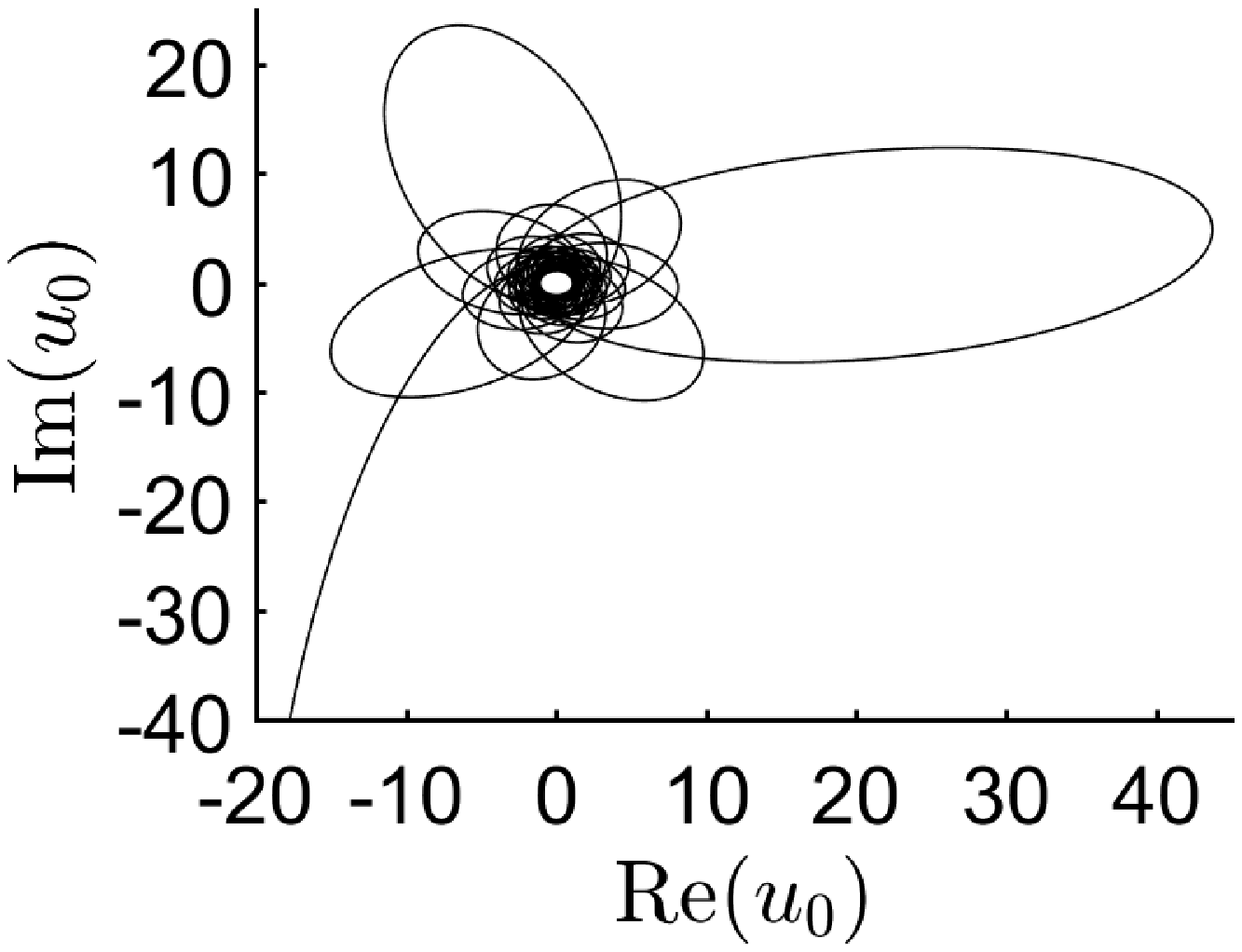
}}\\
\subfigure[$N=128$, $\left|u_{0}(t)\right|$]{\includegraphics[width=0.2871\textwidth]{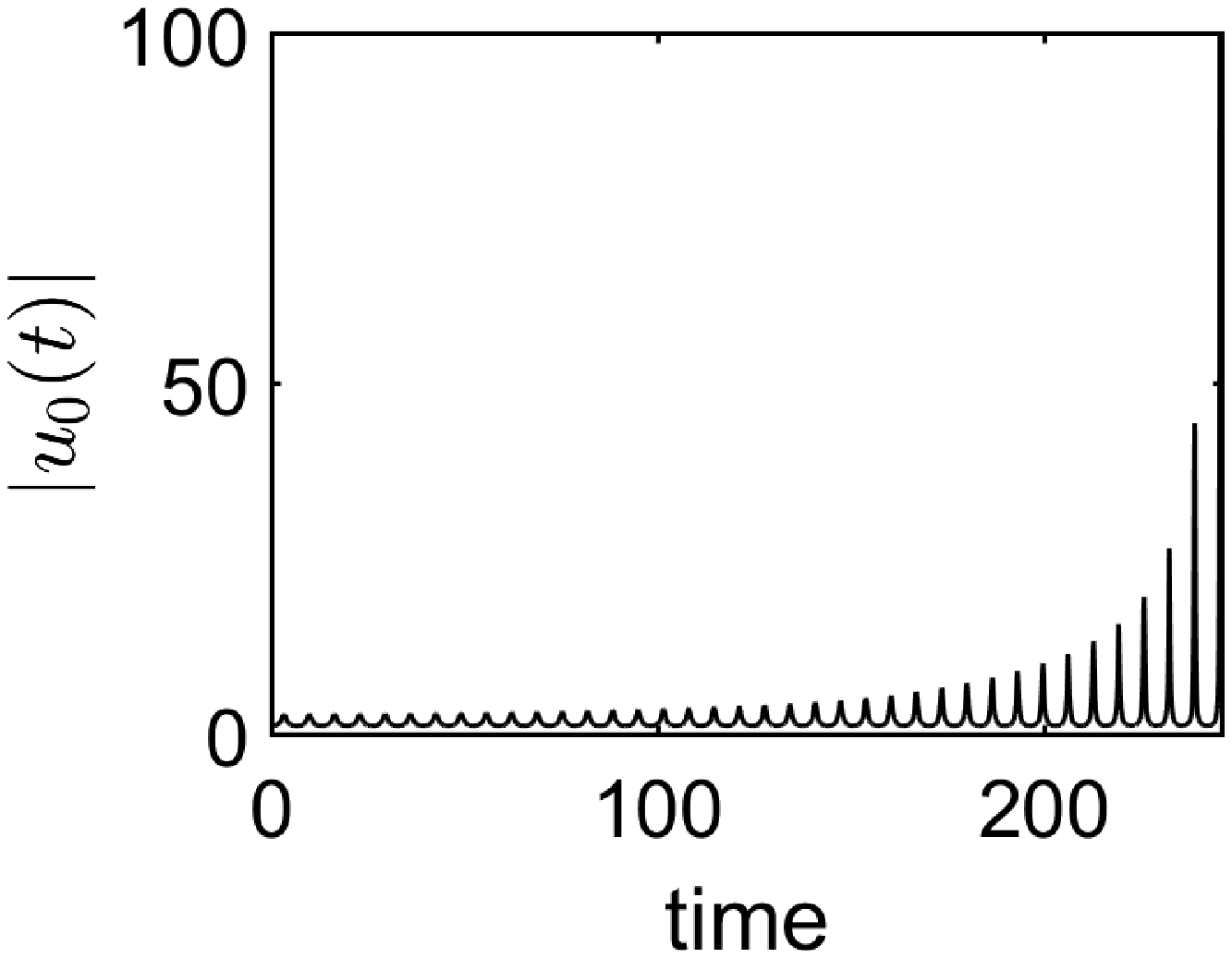
}}~
\subfigure[$N=128$, $\left|u_{n}\left(245.5561\right)\right|$]{\includegraphics[width=0.2871\textwidth]{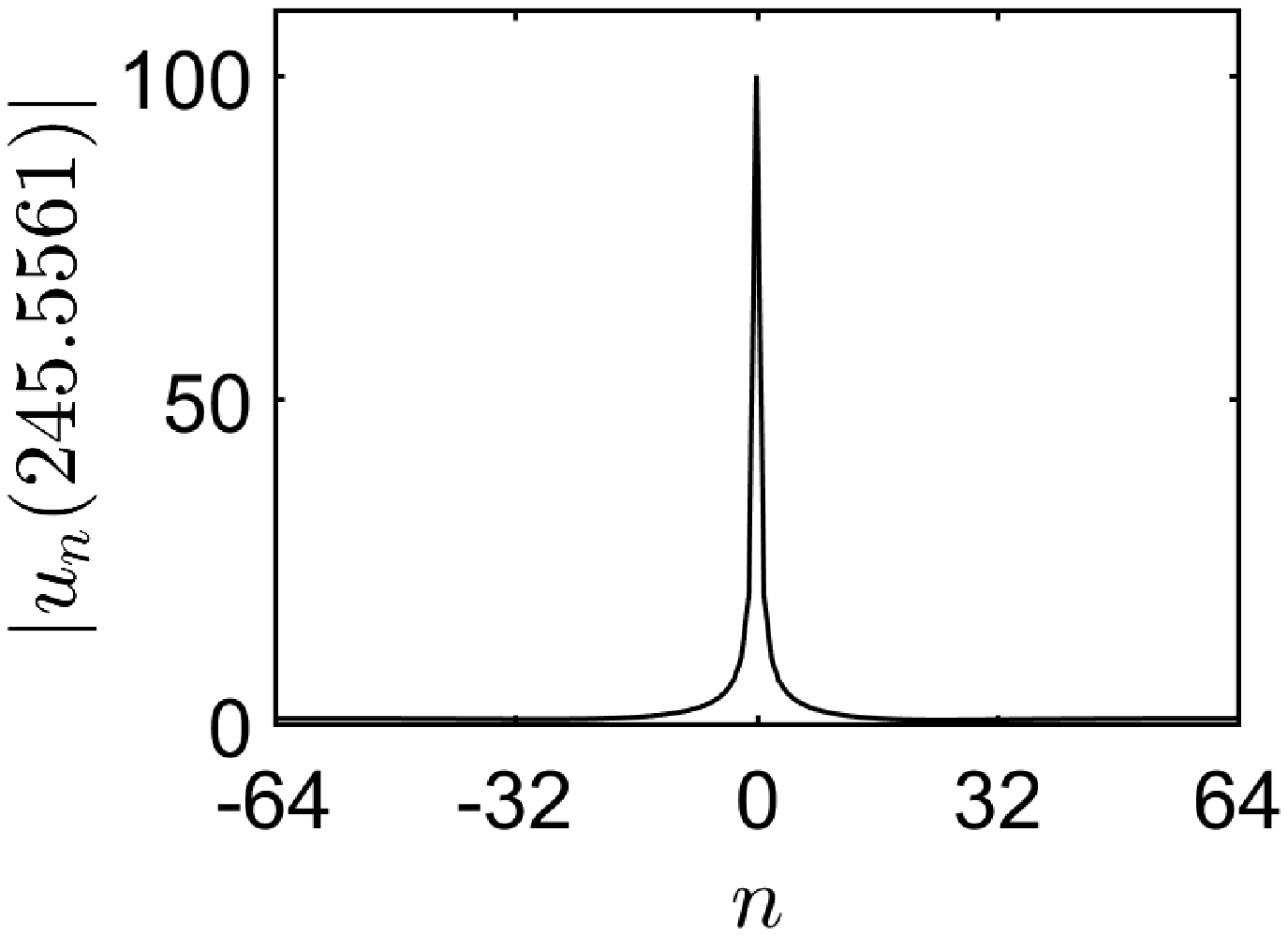
}}~
\subfigure[$N=128$, trajectory of $u_{0}\left(t\right)$, $0\leq t\leq245.5561$]{\includegraphics[width=0.2871\textwidth]{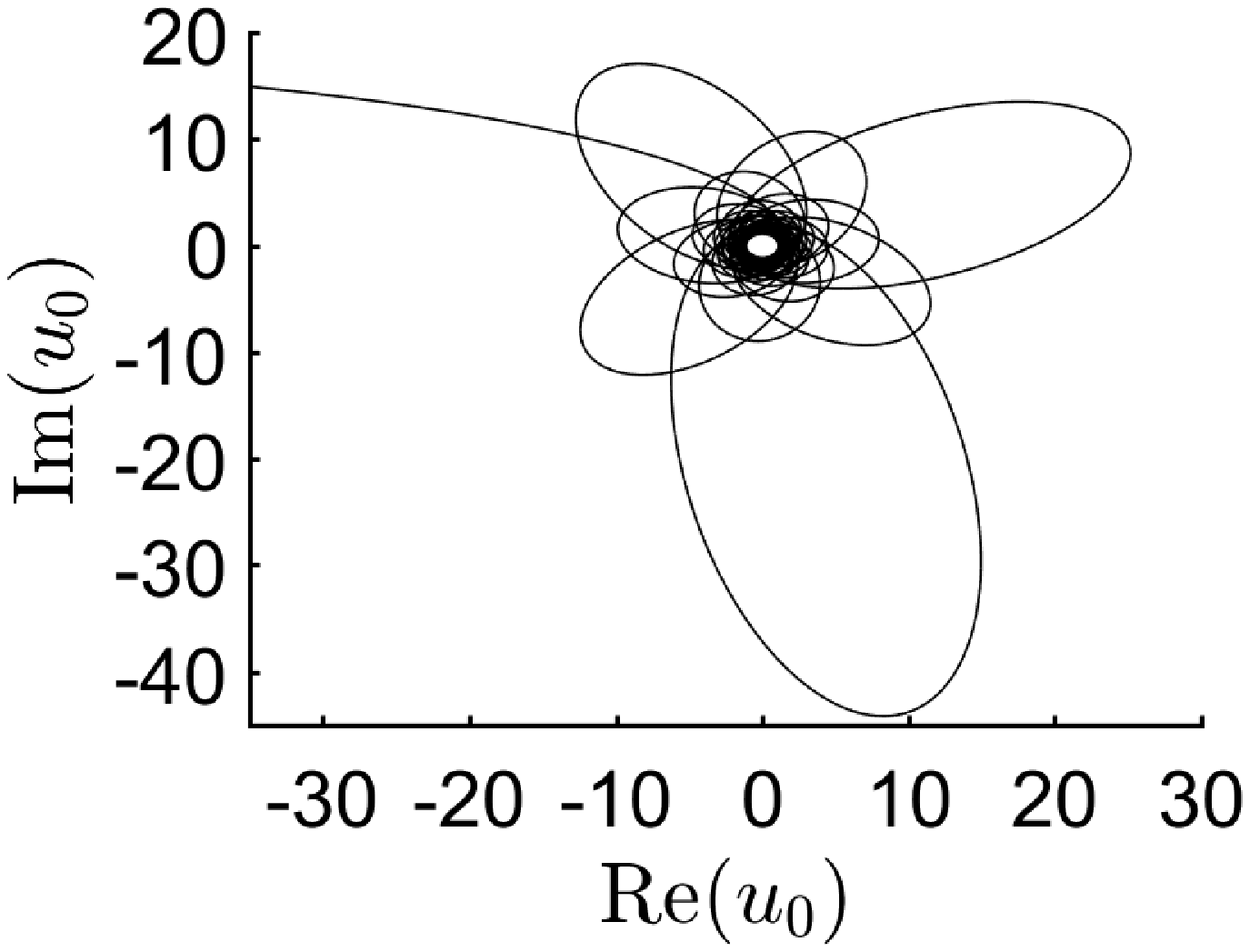
}}
\caption{Numerical solutions of Cauchy problem \eqref{eq:CPDNI} with $\varphi_{0}=\frac{1}{256}\pi$ and $N=32,\,64,\,128$.\label{fig:keji32}}
\end{figure}

We hope to understand different properties for the solutions to Cauchy problems \eqref{eq:CPDNN} and \eqref{eq:CPDNI}. In Fig.\ref{fig:keji32}, we fix $\varphi_{0}=\frac{\pi}{256}$ and set $N=32$, 64 and 128 in Cauchy problem \eqref{eq:CPDNI}. Different from Cauchy problem \eqref{eq:CPDNN}, it takes much more time for the maximum module of the solution to reach 100, and the maximum module of the solutions always reach 100 at $n=0$.

\begin{figure}[htbp]
\centering
\subfigure[$\varphi_{0}=\frac{\pi}{8}$, $\left|u_{0}(t)\right|$]{\includegraphics[width=0.25\textwidth]{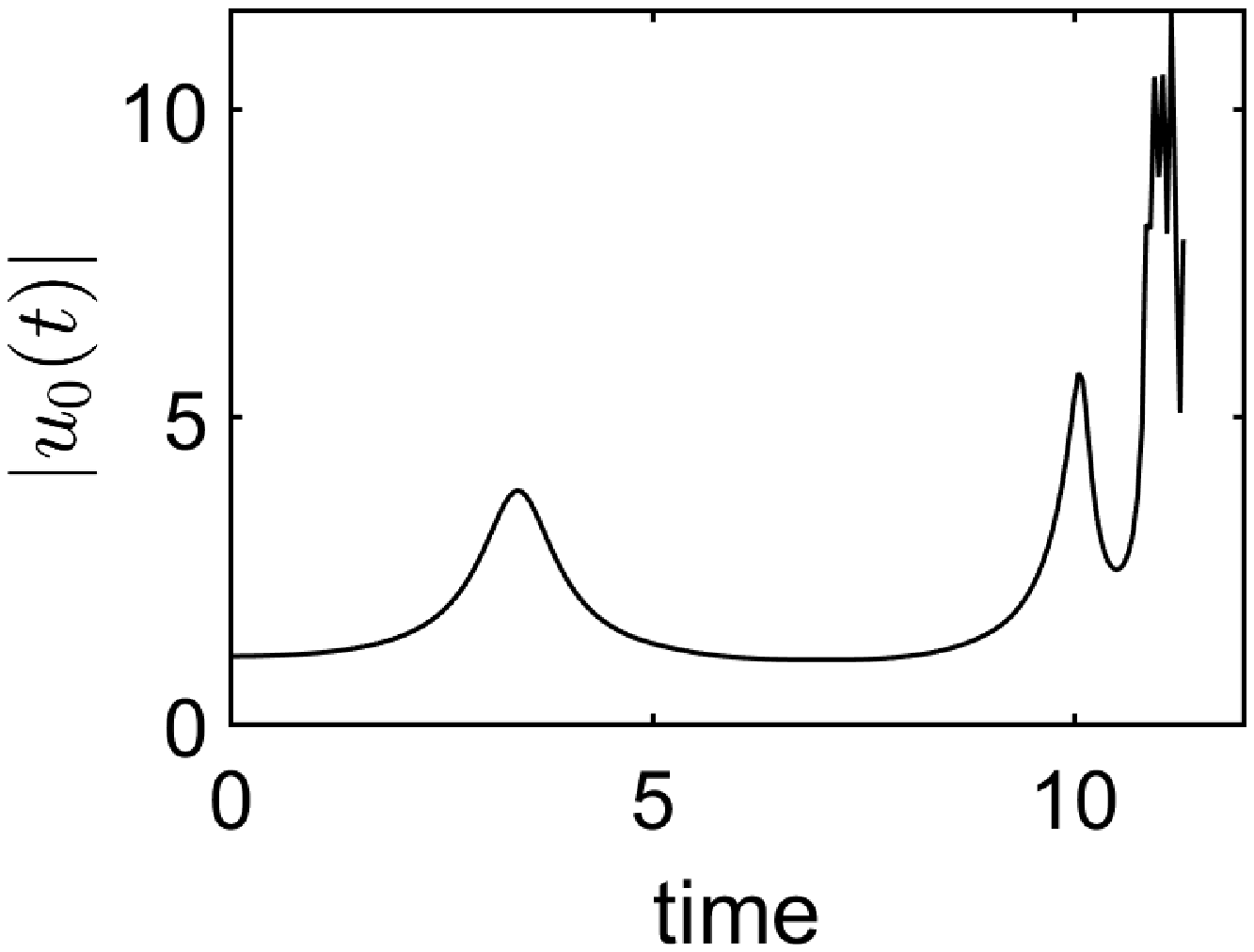
}}\qquad
\subfigure[$\varphi_{0}=\frac{\pi}{4}$, $\left|u_{0}(t)\right|$]{\includegraphics[width=0.25\textwidth]{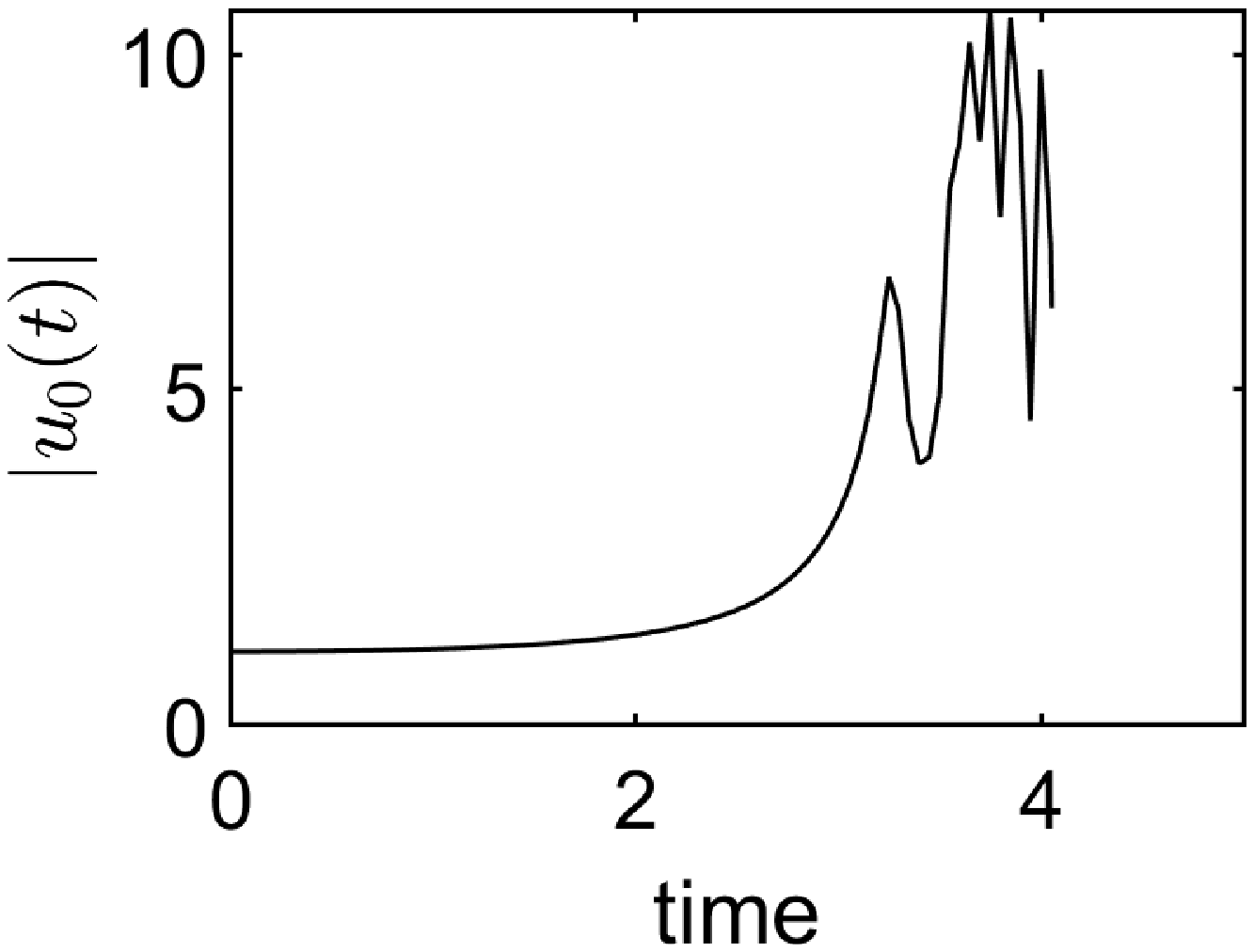
}}\qquad
\subfigure[$\varphi_{0}=\frac{\pi}{2}$, $\left|u_{0}(t)\right|$]{\includegraphics[width=0.25\textwidth]{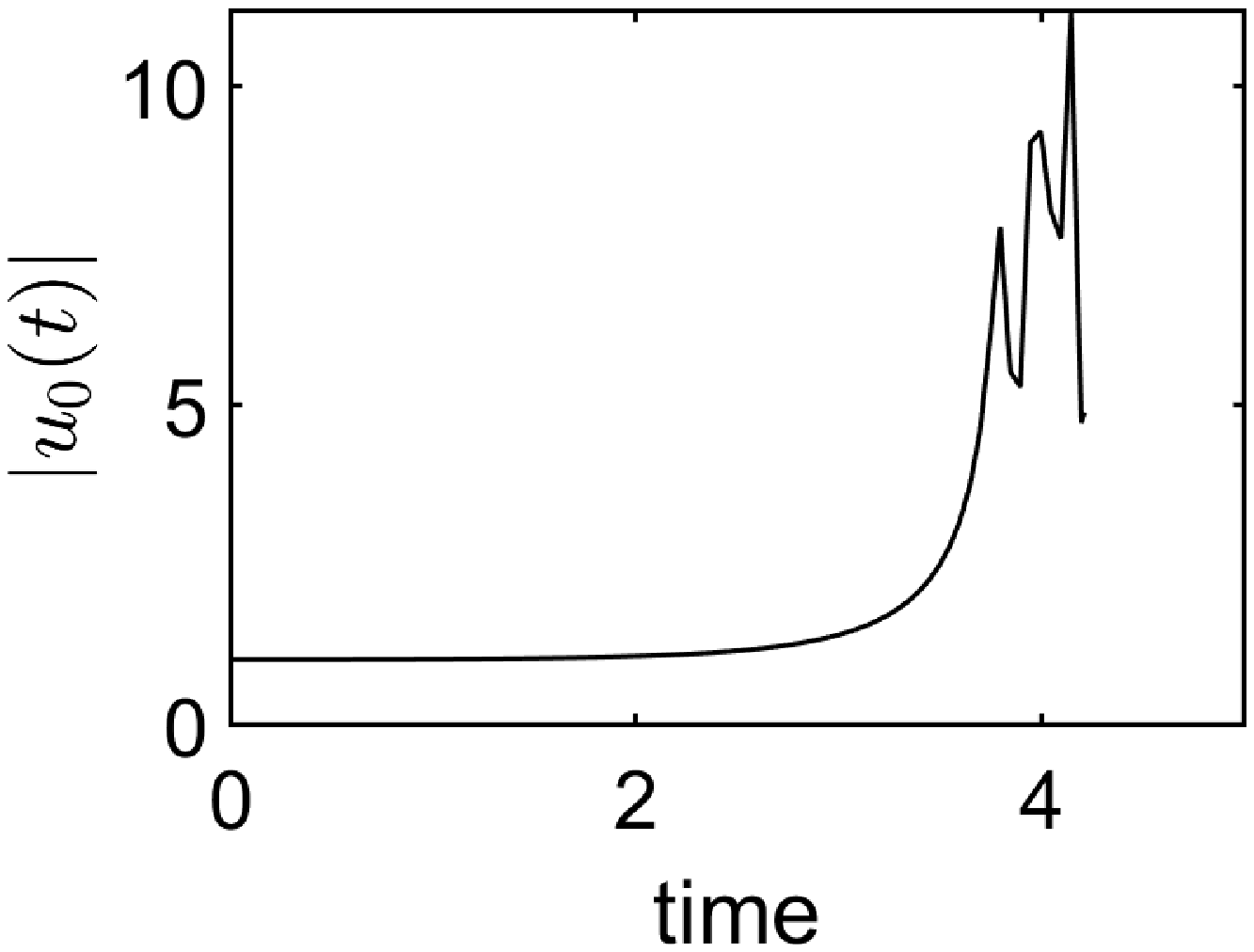
}}\\
\subfigure[$\varphi_{0}=\frac{\pi}{8}$, $\left|u_{n}\left(11.29\right)\right|$]{\includegraphics[width=0.25\textwidth]{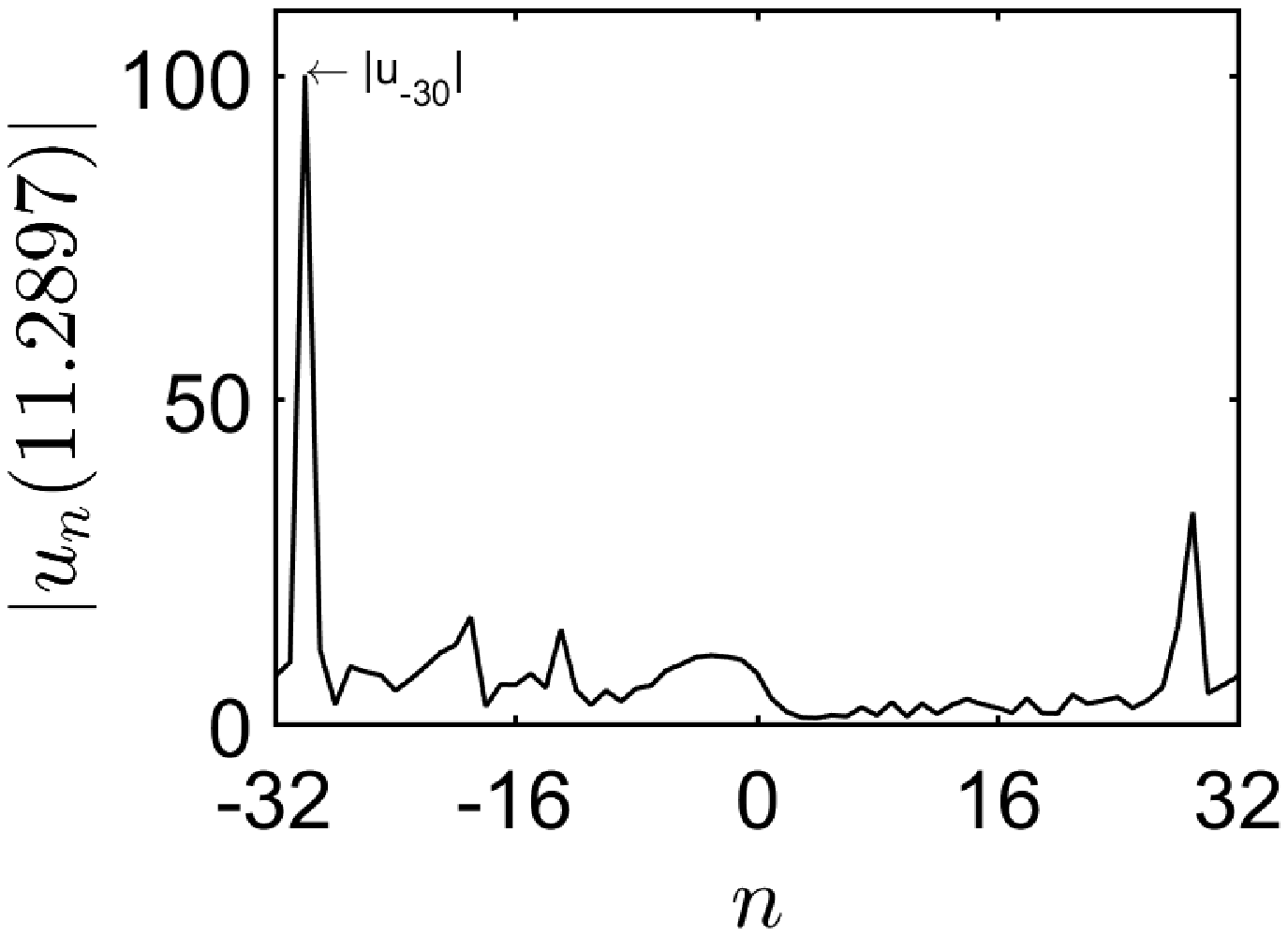
}}\qquad
\subfigure[$\varphi_{0}=\frac{\pi}{4}$, $\left|u_{n}\left(4.05\right)\right|$]{\includegraphics[width=0.25\textwidth]{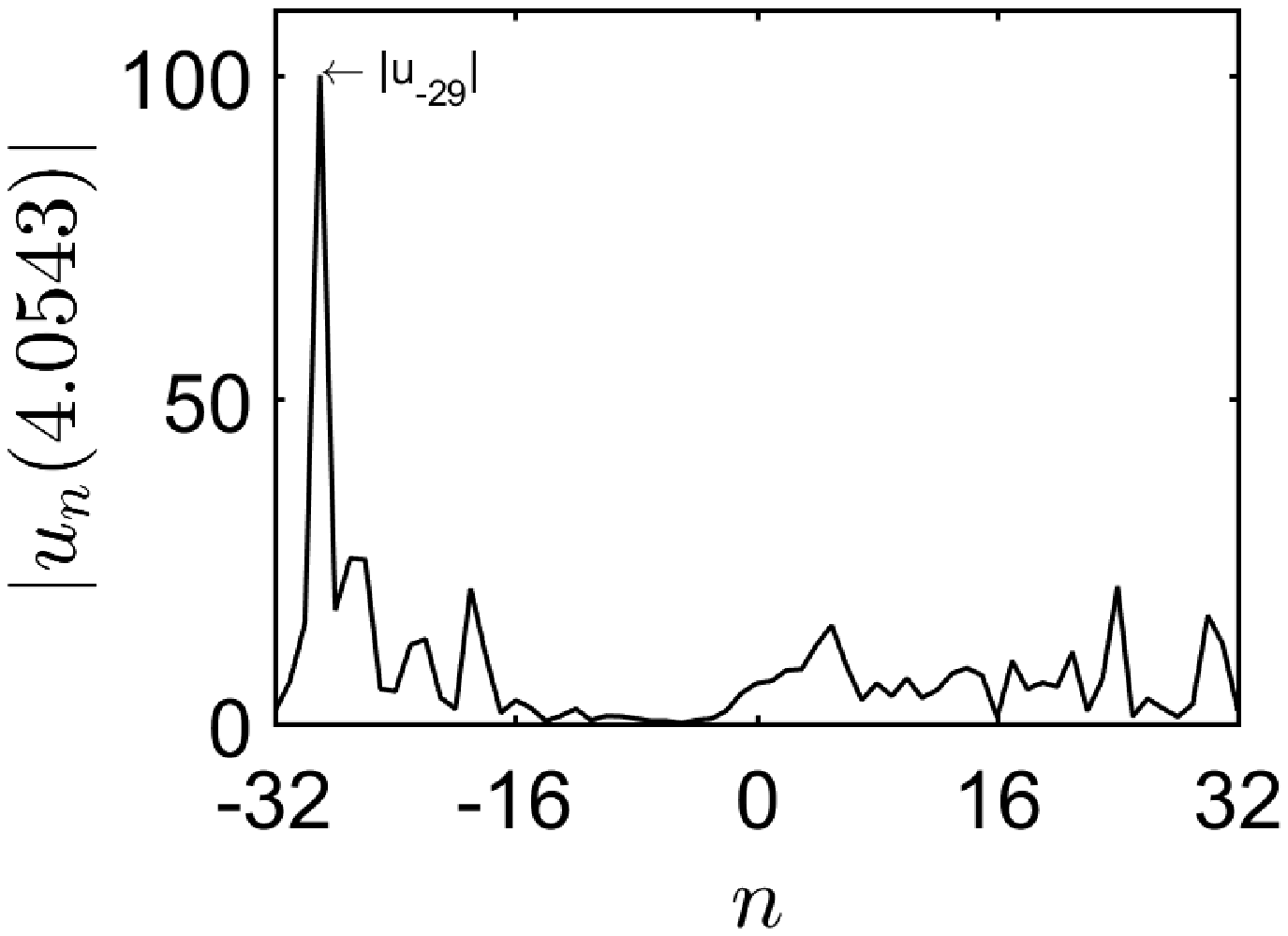
}}\qquad
\subfigure[$\varphi_{0}=\frac{\pi}{2}$, $\left|u_{n}\left(4.22\right)\right|$]{\includegraphics[width=0.25\textwidth]{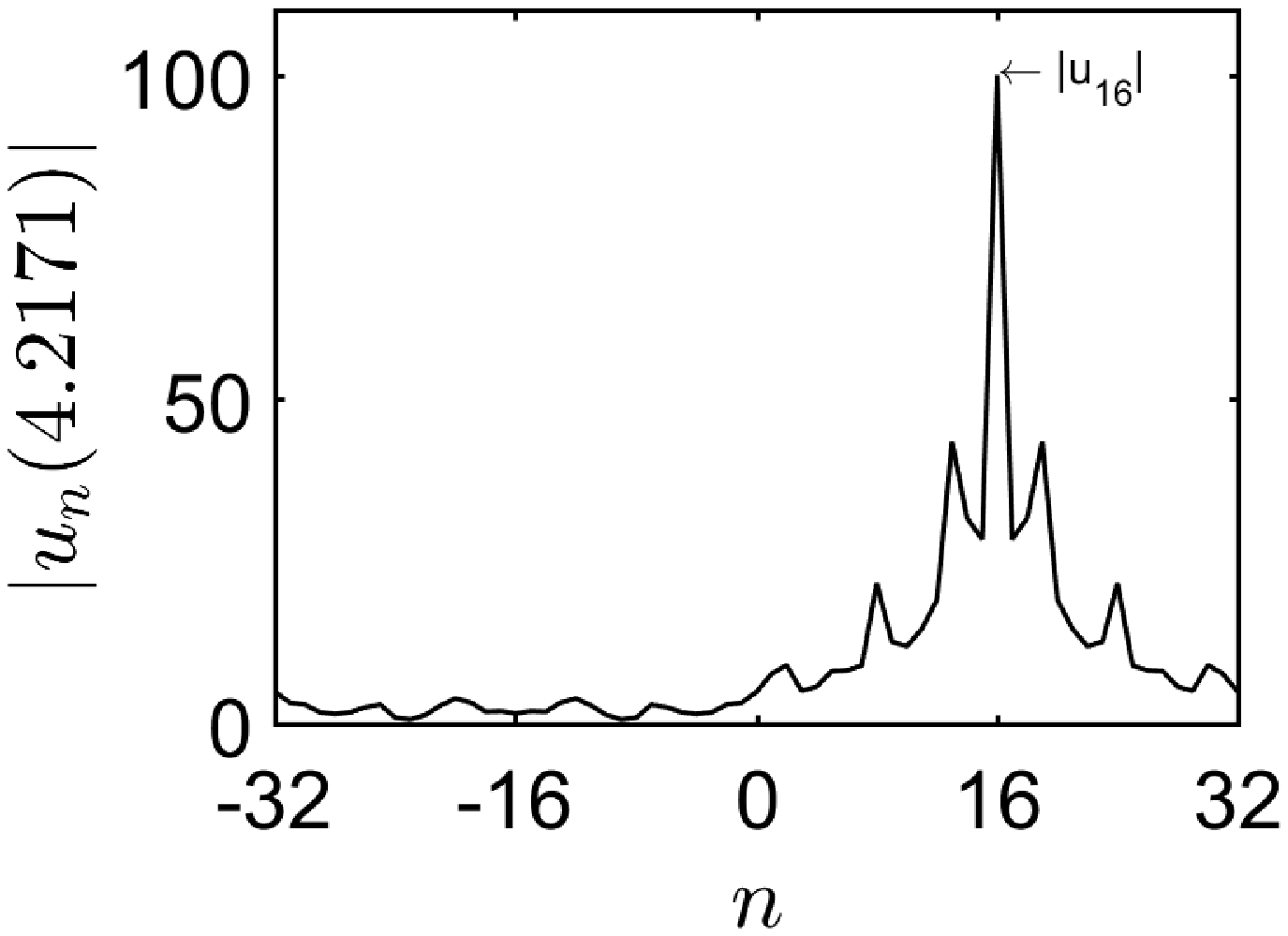
}}
\centering{}\caption{The numerical solution of nonlocal nonintegrable scheme (\ref{eq:ddnoke}) for $N=64$ and $\varphi_0=\frac{\pi}{8},\,\frac{\pi}{4},\,\frac{\pi}{2}$.\label{fig:nokech}}
\end{figure}

\begin{figure}[htbp]
\centering
\subfigure[$\varphi_{0}=\frac{\pi}{8}\thinspace,\left|u_{0}(t)\right|$]{\includegraphics[width=0.25\textwidth]{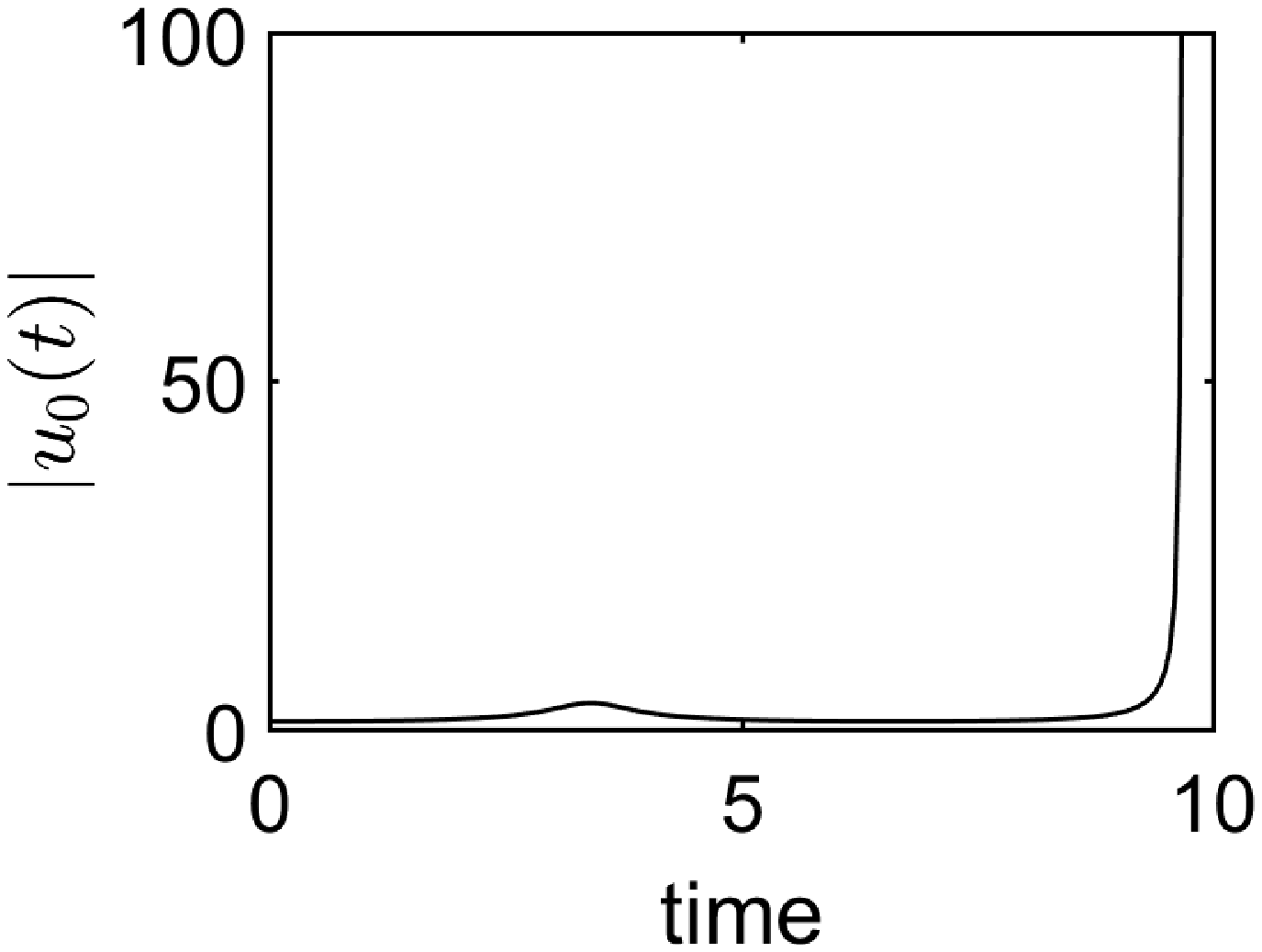
}}\qquad
\subfigure[$\varphi_{0}=\frac{\pi}{4}\thinspace,\left|u_{0}(t)\right|$]{\includegraphics[width=0.25\textwidth]{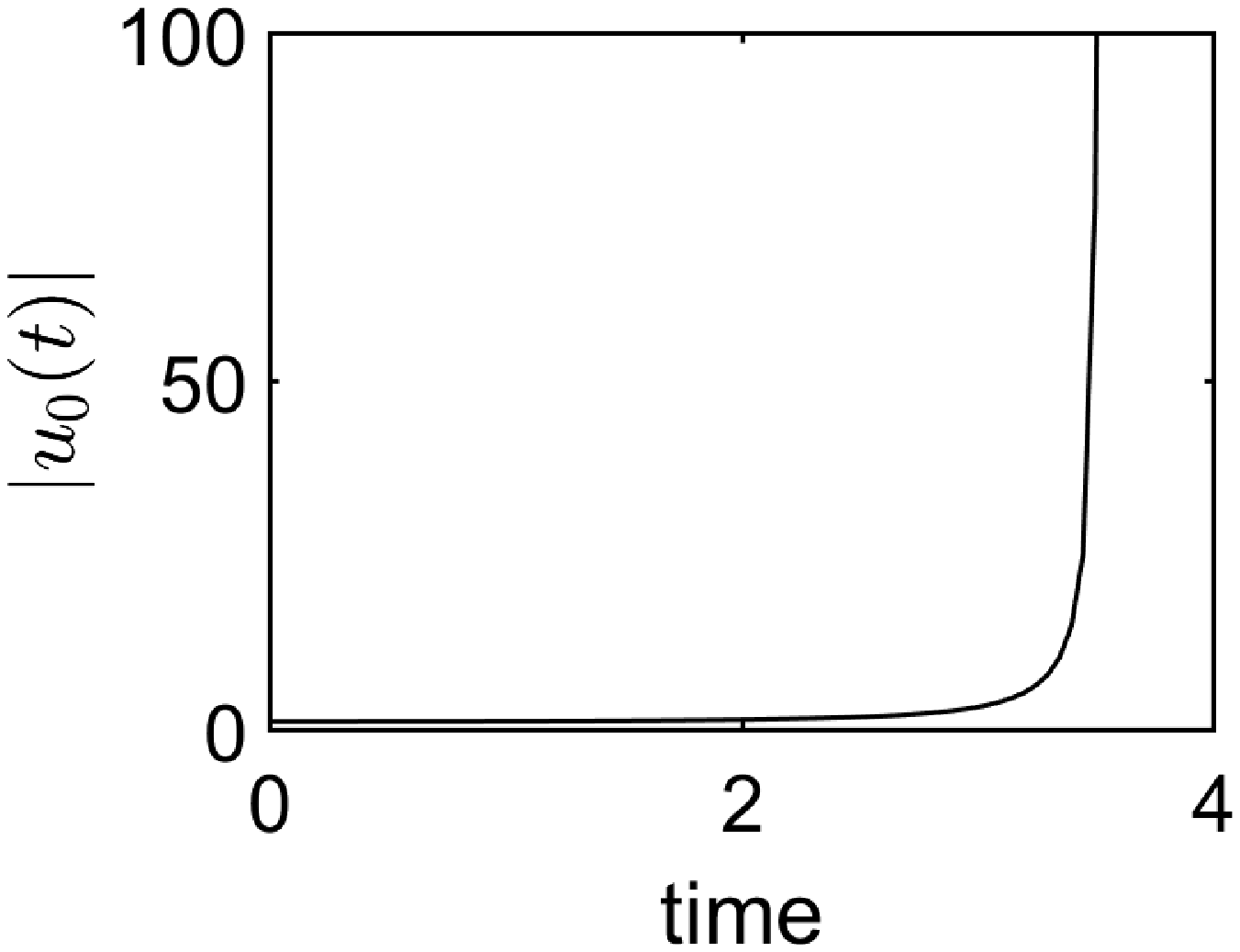
}}\qquad
\subfigure[$\varphi_{0}=\frac{\pi}{2}\thinspace,\left|u_{0}(t)\right|$]{\includegraphics[width=0.25\textwidth]{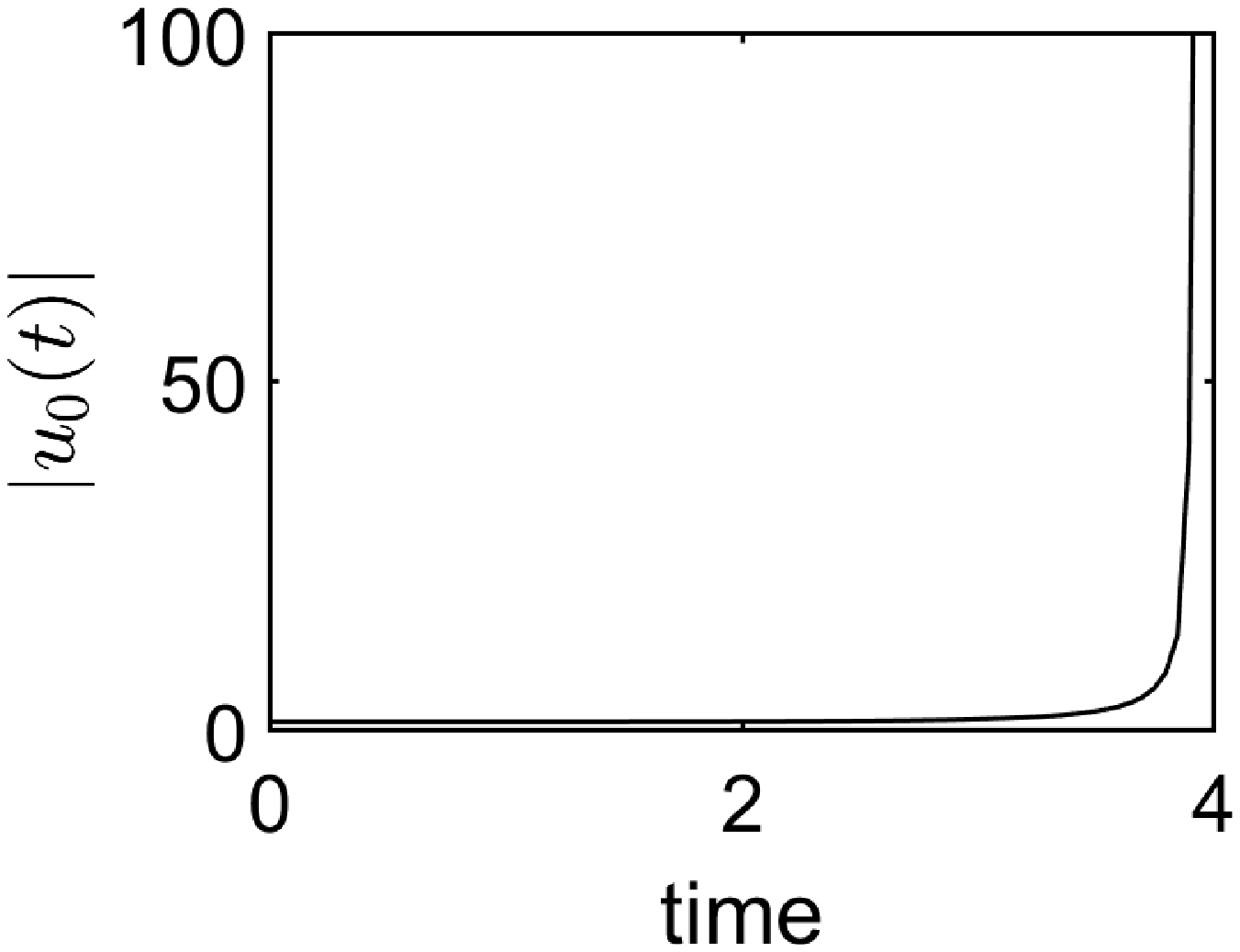
}}\\
\subfigure[$\varphi_{0}=\frac{\pi}{8}\thinspace,\left|u_{n}\left(9.67\right)\right|$]{\includegraphics[width=0.25\textwidth]{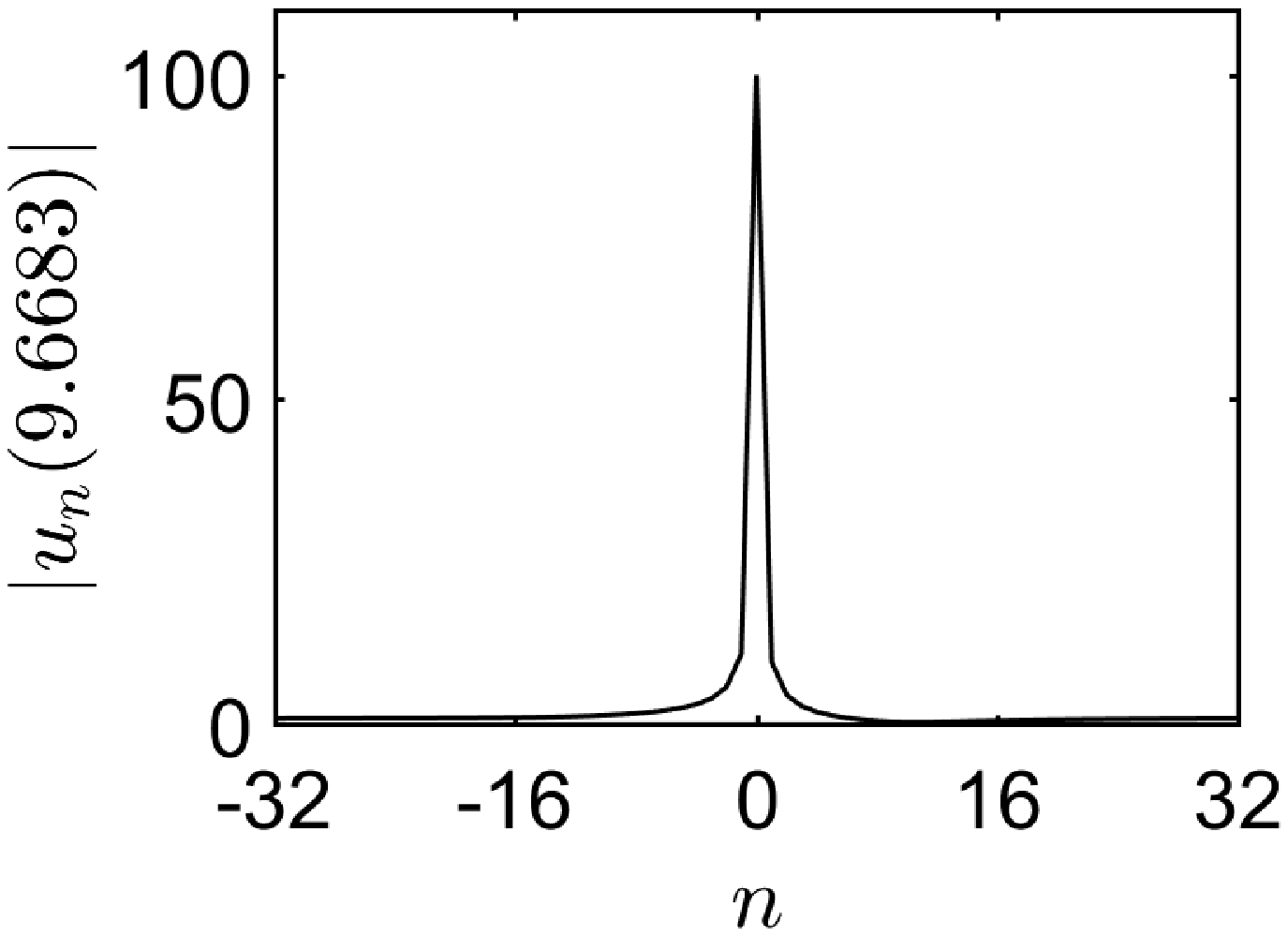
}}\qquad
\subfigure[$\varphi_{0}=\frac{\pi}{4}\thinspace,\left|u_{n}\left(3.51\right)\right|$]{\includegraphics[width=0.25\textwidth]{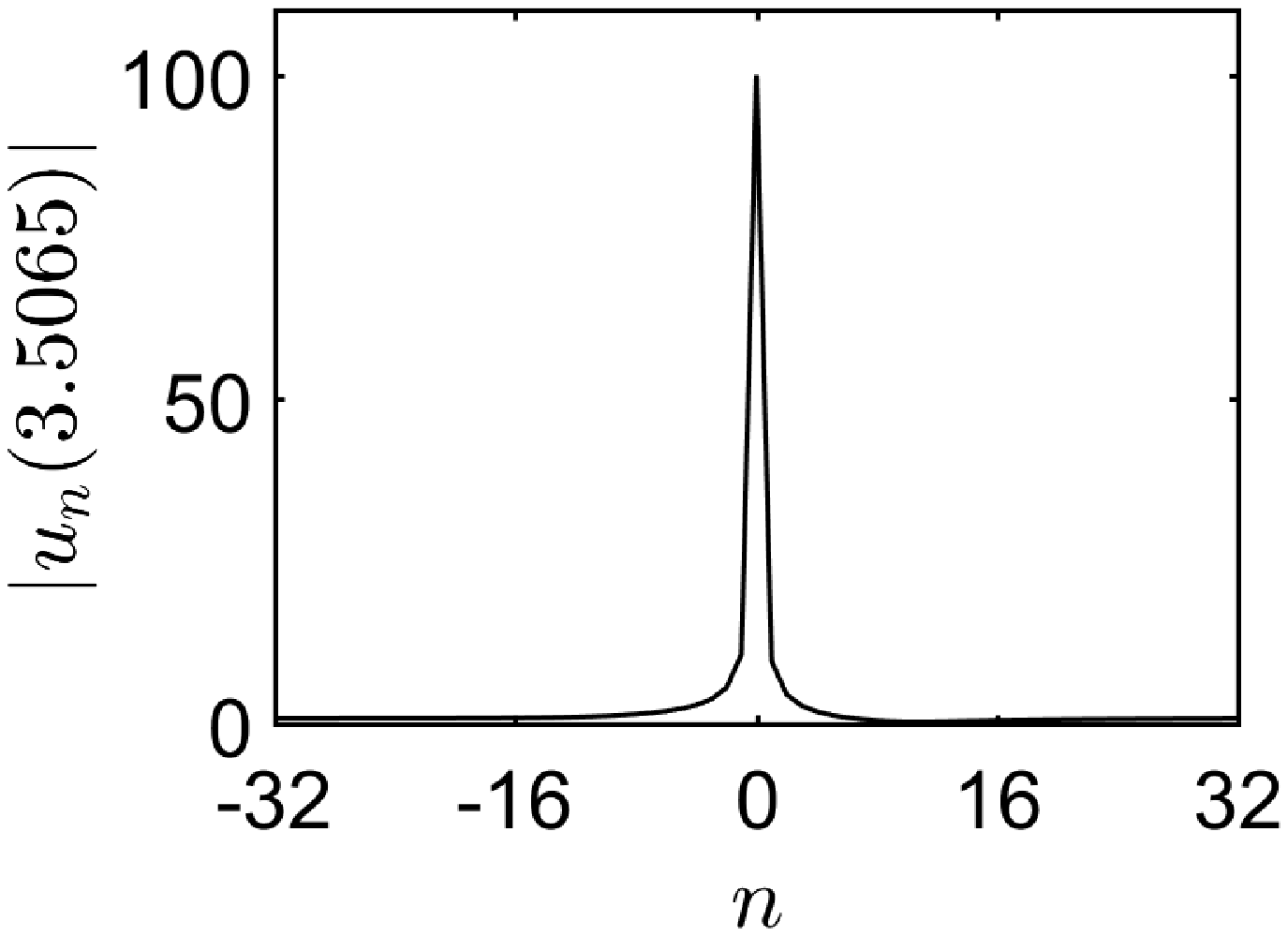
}}\qquad
\subfigure[$\varphi_{0}=\frac{\pi}{2}\thinspace,\left|u_{n}\left(3.91\right)\right|$]{\includegraphics[width=0.25\textwidth]{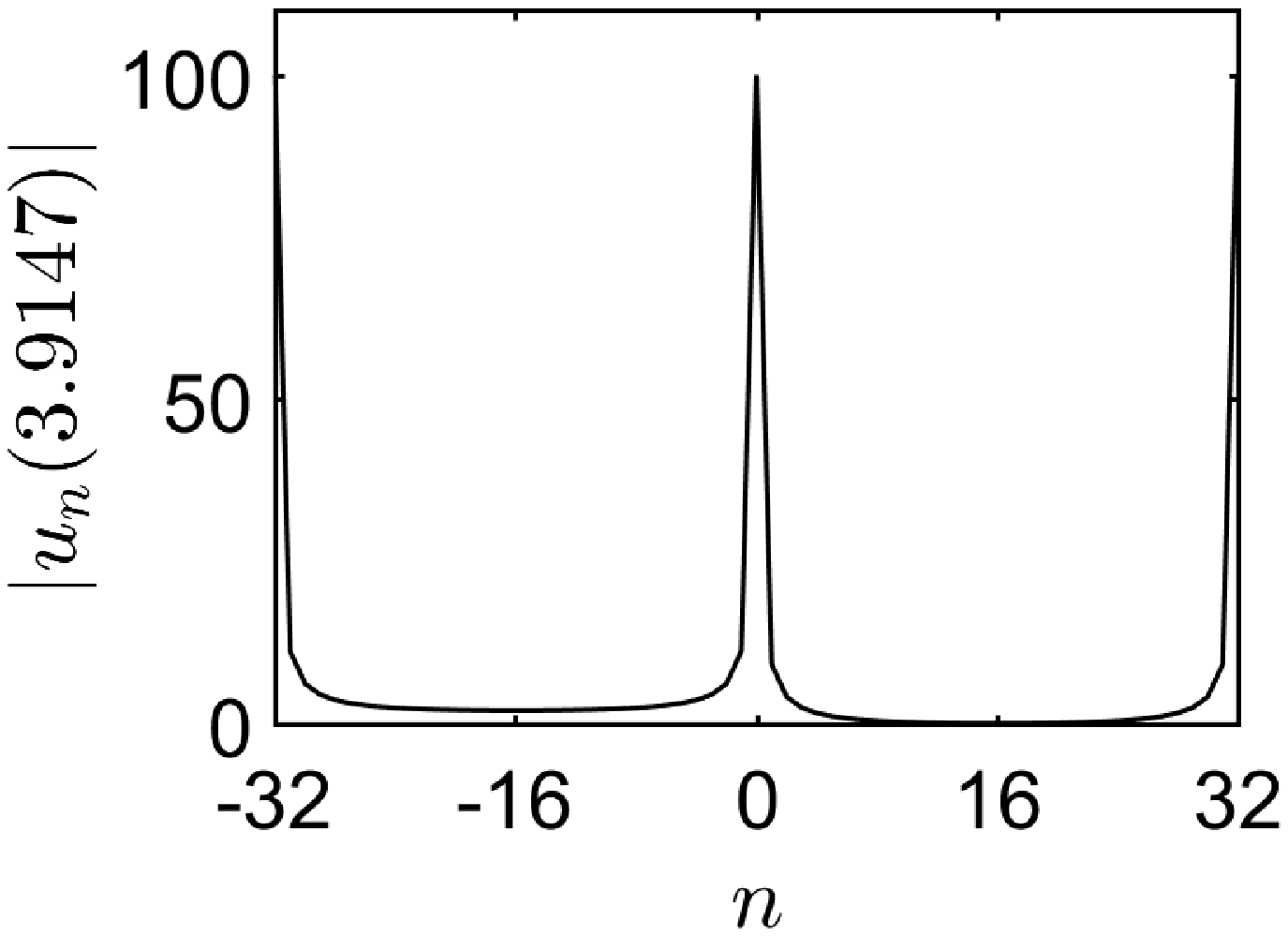
}}
\caption{The numerical solution of nonlocal integrable scheme (\ref{eq:ddkeji}) for $N=64$ and $\varphi_0=\frac{\pi}{8},\,\frac{\pi}{4},\,\frac{\pi}{2}$.\label{fig:nokech-3}}
\end{figure}

Then, let us discuss what would happen when the initial condition changes while space step $h$ is fixed. As we know, NLS equation \eqref{eq:LCNLS} is Galilean invariant. But this property does not hold for nonlocal NLS equation \eqref{eq:NNLS}. By numerical simulation, we can see that the change of the initial data has great impact on the solution to the Cauchy problems \eqref{eq:CPDNN} and \eqref{eq:CPDNI}, which is different from the case of the Cauchy problem \eqref{eq:CPDLN}. We fix $N=64$ and choose $\varphi_{0}=\frac{\pi}{8}$, $\frac{\pi}{4}$ and $\frac{\pi}{2}$ in Cauchy problems \eqref{eq:CPDNN} and \eqref{eq:CPDNI}, respectively. The corresponding computational results are shown in Fig.\ref{fig:nokech} and Fig.\ref{fig:nokech-3}. Different from Fig.\ref{fig:noke32-1} and Fig.\ref{fig:keji32}(d-f), the quasiperiodic phenomenon no longer occurs and 'blow up' comes more quickly. The points at which the modules of the solutions first reaching 100 are variable in Cauchy problem \eqref{eq:CPDNN}.
However, for Cauchy problem \eqref{eq:CPDNI}, the point at which the maximum module of the solution first reaching 100 only locates at $n=0$ or $n=\frac{L}{2h}$. Specifically, for $\varphi_{0}=\frac{\pi}{2}$, both $n=0$ and $n=\frac{L}{2h}$ grow to 100 at $t=4.9147$. For other $\varphi_{0}$, we find while $\varphi_{0}\in\left(-\frac{\pi}{2},\frac{\pi}{2}\right)$, the point at which the maximum module of the solution first to reach 100 is $n=0$. For $\varphi_{0}=\pm\frac{\pi}{2}$, both $n=0$ and $n=\frac{L}{2h}$ grow up and their modules reach 100 ahead of other points. For $\varphi_{0}\in\left(\frac{\pi}{2},\frac{3\pi}{2}\right)$, the point at which the maximum module of the solution first reaching 100 is $n=\frac{L}{2h}$. This property is different from the Cauchy problem (34) of nonintegrable discrete nonlocal NLS.

\begin{figure}[htbp]
\begin{centering}
\includegraphics[width=0.6\textwidth]{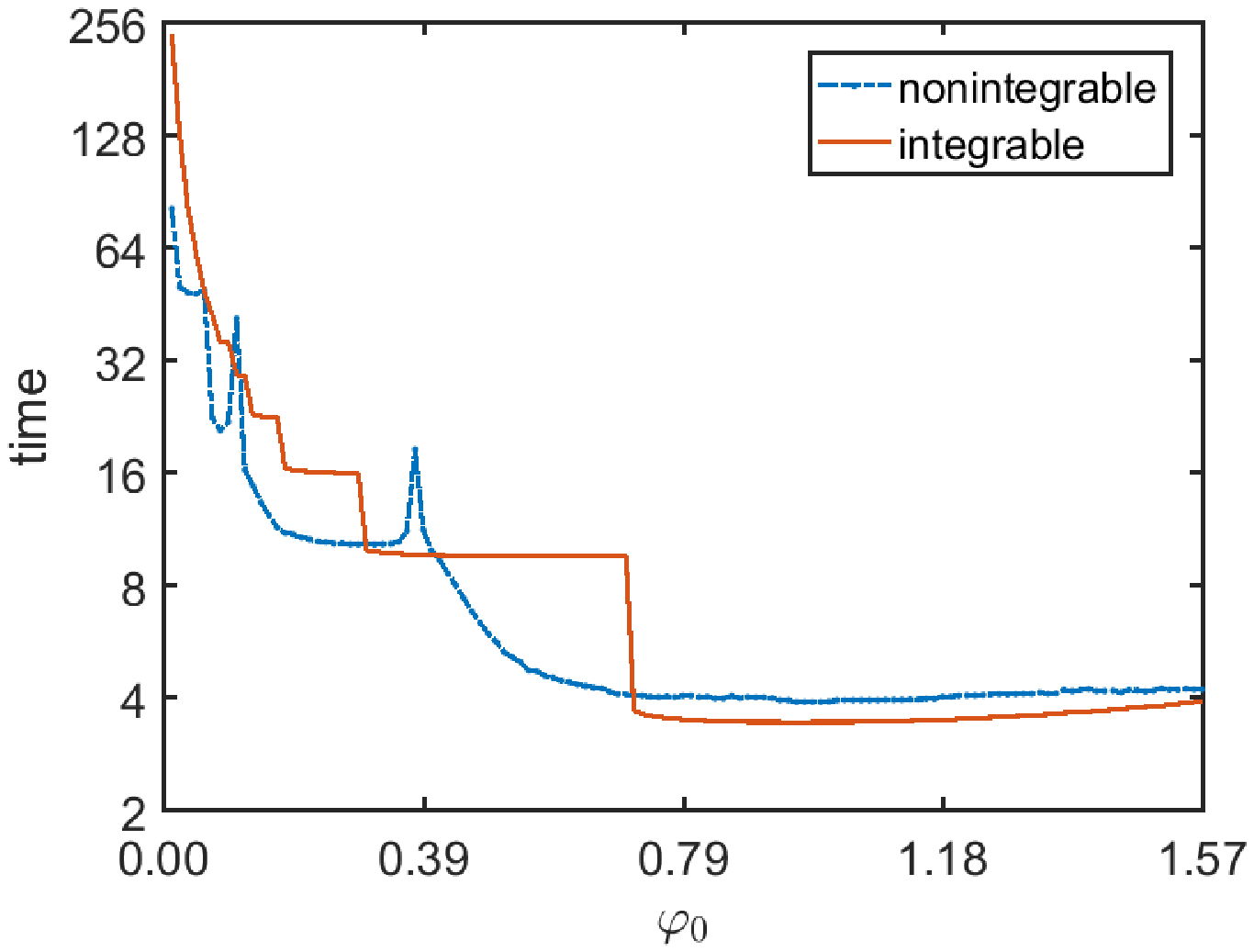
}\caption{The time it takes to reach 100 for maximum norm of numerical solution of Eq.(\ref{eq:ddnoke})
and Eq.(\ref{eq:ddkeji}) to different $\varphi_{0}=\frac{k\pi}{256}$, $k=1,2,\cdots,128$
when $N=64$.\label{fig:Blowup_Time}}
\end{centering}
\end{figure}

Fig.\ref{fig:Blowup_Time} shows the relation between $\varphi_{0}$ and the time at which $\max\left|u_{n}\right|$ first reaches 100 for both Cauchy prolems \eqref{eq:CPDNN} and \eqref{eq:CPDNI} when $N=64$.
It can be seen that the time expends for the maximum module of numerical solution to reach 100 generally increases as $\varphi_0$ grows from 0 to $\frac\pi2$.

\section{Conclusions}
In this paper, we have investigated the nonintegrable spatial discrete nonlocal NLS equation \eqref{eq:ddnoke}. By applying the discrete Fourier transform method, we have obtained its stationary solitary wave solutions. Linear stability analysis of the stationary-solitary-wave solutions is discussed. It has been shown that stationary-solitary-wave solutions to discrete nonlocal NLS equation \eqref{eq:ddnoke} are linear unstable, while stationary-solitary-wave solutions to integrable discrete NLS equation \eqref{eq:LINLS} are linear stable. Numerical simulations for Cauchy problem of nonlocal NLS equation \eqref{eq:NNLS} with periodic boundary conditions are performed, in which we use the integrable scheme \eqref{eq:ddkeji} and the nonintegrable scheme \eqref{eq:ddnoke} as a discrete model of nonlocal NLS equation \eqref{eq:NNLS}. The comparison of numerical simulations between the Cauchy problem of NLS equation \eqref{eq:LCNLS} and the Cauchy problem of nonlocal NLS equation \eqref{eq:NNLS} is given. It is found that the solutions of Cauchy problem of nonlocal NLS equation \eqref{eq:NNLS} are very sensitive to the initial data, and the nonintegrable discrete nonlocal NLS equation \eqref{eq:ddnoke} possesses more new properties than the nonintegrable discrete NLS equation \eqref{eq:LNNLS}.
\section*{Acknowledgements}
The work of ZNZ is supported by the National Natural Science
Foundation of China under grant 11671255 and
by the Ministry of Economy and Competitiveness of Spain under
contract MTM2016-80276-P (AEI/FEDER,EU).

\end{document}